\documentclass[
reprint,
superscriptaddress,
showpacs,showkeys,floatfix,
amsmath,amssymb,
aps,
jcp
]{revtex4-1}

\usepackage{graphicx}
\usepackage[ colorlinks,
    linkcolor={blue},
    citecolor={blue},
    urlcolor={blue}]{hyperref}

\usepackage{amsmath}
\usepackage{amssymb}
\usepackage{dcolumn}
\usepackage{bm}

\usepackage[english]{babel}
\usepackage{microtype}
\usepackage{indentfirst}
\usepackage{siunitx, booktabs}
\usepackage{tabularx}
\usepackage[caption=false]{subfig}
\usepackage{array}
\usepackage{setspace}
\usepackage{braket}
\usepackage{xcolor}
\usepackage{placeins}

\usepackage{amsfonts}
\usepackage{algorithmic}
\usepackage{textcomp}
\usepackage{xcolor}
\usepackage{xspace}
\usepackage{url}
\usepackage{listings}

\newcommand{\CPTWOK}{{\tt CP2K}\xspace}
\newcommand{\DBCSR}{{\tt DBCSR}\xspace}
\newcommand{\LIBXSMM}{{\tt LIBXSMM}\xspace}
\newcommand{\LIBCUSMM}{{\tt LIBCUSMM}\xspace}
\newcommand{\LIBSMMACC}{{\tt LIBSMM\_ACC}\xspace}

\newcommand{\eg}{{e.\,g.}\xspace}

\newcommand{\DM}{density matrix}
\newcommand{\LS}{$\mathcal{O}$($N$)}
\newcommand{\CPXK}{CP2K}

\newcommand*{\imi}{i} 

\newcommand{\ketbra}[2]{\ensuremath{\vert #1 \rangle \langle #2 \vert}} 

\makeatletter
\newcommand\xleftrightarrow[2][]{%
  \ext@arrow 9999{\longleftrightarrowfill@}{#1}{#2}}
\newcommand\longleftrightarrowfill@{%
  \arrowfill@\leftarrow\relbar\rightarrow}
\makeatother

\newcommand{\cp}{{\sc CP2K}}
\newcommand{\qs}{{\sc Quickstep}}
\newcommand{\cpqs}{{\sc CP2K/Quickstep}}
\newcommand{\I}{{\rm i}}

\newcommand{\E}{{\rm e}}

\newcommand{\ba}{{\bm{a}}}
\newcommand{\bA}{{\bm{A}}}
\newcommand{\bB}{{\bm{B}}}
\newcommand{\bb}{{\bm{b}}}
\newcommand{\bC}{{\bm{C}}}

\newcommand{\be}{{\bm{e}}}

\newcommand{\bg}{{\bm{g}}}
\newcommand{\bG}{{\bm{G}}}

\newcommand{\bh}{{\bm{h}}}

\newcommand{\bK}{{\bm{K}}}

\newcommand{\bN}{{\bm{N}}}

\newcommand{\bP}{{\bm{P}}}
\newcommand{\bq}{{\bm{q}}}

\newcommand{\br}{{\bm{r}}}
\newcommand{\bR}{{\bm{R}}}

\newcommand{\bS}{{\bm{S}}}

\newcommand{\bU}{{\bm{U}}}

\newcommand{\op}[1]{{\hat{#1}}}

\DeclareMathOperator{\Tr}{Tr}
\DeclareMathOperator{\erf}{erf}
\DeclareMathOperator{\erfc}{erfc}
\DeclareMathOperator{\sign}{sign}

\newcommand{\ext}{{\rm ext}}

\newcommand{\XC}{{\rm XC}}

\newcommand{\pp}{{\rm pp}}
\newcommand{\loc}{{\rm loc}}
\newcommand{\nl}{{\rm nl}}

\newcommand{\rr}{\mathbf{r}}

\newcommand{\Rmm}{\mathbf{R}_a}

\newcommand{\Rqm}{\mathbf{R}_{\alpha}}
\newcommand{\kk}{\mathbf{k}}
\newcommand{\nn}{\mathbf{L}}
\newcommand{\rri}{\mathbf{r}_i}

\newcommand{\quickstep}{{\sc Quickstep} }

\begin{document}

\title{CP2K: An Electronic Structure and Molecular Dynamics Software Package \\ \textsc{Quickstep}: Efficient and Accurate Electronic Structure Calculations}



\author{Thomas D. K\"uhne}
\affiliation{Dynamics of Condensed Matter and Center for Sustainable Systems Design, Chair of Theoretical Chemistry, Paderborn University, Warburger Str. 100, D-33098 Paderborn, Germany}%
\author{Marcella Iannuzzi}
\affiliation{Department of Chemistry, University of Zurich, Winterthurerstrasse 190, CH-8057 Z\"urich, Switzerland}
\author{Mauro Del Ben}
\affiliation{Computational Research Division, Lawrence Berkeley National Laboratory, Berkeley, California 94720, USA}
\author{Vladimir V. Rybkin}
\affiliation{Department of Chemistry, University of Zurich, Winterthurerstrasse 190, CH-8057 Z\"urich, Switzerland} 
\author{Patrick Seewald}
\affiliation{Department of Chemistry, University of Zurich, Winterthurerstrasse 190, CH-8057 Z\"urich, Switzerland} 
\author{Frederick Stein}
\affiliation{Department of Chemistry, University of Zurich, Winterthurerstrasse 190, CH-8057 Z\"urich, Switzerland} 
\author{Teodoro Laino}
\affiliation{IBM Research - Zurich, S\"{a}umerstrasse 4, CH-8803 R\"{u}schlikon, Switzerland}
\author{Rustam Z. Khaliullin}
\affiliation{Department of Chemistry, McGill University, CH-801 Sherbrooke St. West, Montreal, QC H3A 0B8, Canada}
\author{Ole Sch\"utt}
\affiliation{Department of Materials, ETH Z\"{u}rich, CH-8092 Z\"{u}rich, Switzerland}
\author{Florian Schiffmann}
\affiliation{Centre of Policy Studies, Victoria University, Melbourne, Australia}
\author{Dorothea Golze}
\affiliation{Department of Applied Physics, Aalto University, Otakaari 1, FI-02150 Espoo, Finland}
\author{Jan Wilhelm}
\affiliation{Institute of Theoretical Physics, University of Regensburg, Universit\"atsstra\ss e 31, D-93053 Regensburg, Germany}  
\author{Sergey Chulkov}
\affiliation{School of Mathematics and Physics, University of Lincoln, Brayford Pool, Lincoln, United Kingdom}
\author{Mohammad Hossein \surname{Bani-Hashemian}}
\affiliation{Integrated Systems Laboratory, ETH Z\"{u}rich, CH-8092 Z\"{u}rich, Switzerland}
\author{Val\'ery Weber}
\affiliation{IBM Research - Zurich, S\"{a}umerstrasse 4, CH-8803 R\"{u}schlikon, Switzerland}%
\author{Urban Bor{\v s}tnik}
\affiliation{Scientific IT Services, ETH Z\"urich, 
Z\"urich, Switzerland}
\author{Mathieu Taillefumier}
\affiliation{Swiss National Supercomputing Centre (CSCS), ETH Z\"urich}
\author{Alice Shoshana Jakobovits}
\affiliation{Swiss National Supercomputing Centre (CSCS), ETH Z\"urich}
\author{Alfio Lazzaro}
\affiliation{HPE Switzerland GmbH, Basel, Switzerland}
\author{Hans Pabst}
\affiliation{Intel Extreme Computing, Software and Systems, Z\"urich, Switzerland}
\author{Tiziano M\"uller}
\affiliation{Department of Chemistry, University of Zurich, Winterthurerstrasse 190, CH-8057 Z\"urich, Switzerland}
\author{Robert Schade}
\affiliation{Department of Computer Science and  Paderborn Center for Parallel Computing, Paderborn University, Warburger Str. 100, D-33098 Paderborn, Germany}
\author{Manuel Guidon}
\affiliation{Department of Chemistry, University of Zurich, Winterthurerstrasse 190, CH-8057 Z\"urich, Switzerland}
\author{Samuel Andermatt}
\affiliation{Integrated Systems Laboratory, ETH Z\"{u}rich, CH-8092 Z\"{u}rich, Switzerland}
\author{Nico Holmberg}
\affiliation{Department of Chemistry and Materials Science, Aalto University, P.O. Box 16100, 00076 Aalto, Finland}
\author{Gregory K. Schenter}
\affiliation{Physical Science Division, Pacific Northwest National Laboratory, P. O. Box 999, Richland, Washington 99352, USA}
\author{Anna Hehn}
\affiliation{Department of Chemistry, University of Zurich, Winterthurerstrasse 190, CH-8057 Z\"urich, Switzerland}
\author{Augustin Bussy}
\affiliation{Department of Chemistry, University of Zurich, Winterthurerstrasse 190, CH-8057 Z\"urich, Switzerland}
\author{Fabian Belleflamme}
\affiliation{Department of Chemistry, University of Zurich, Winterthurerstrasse 190, CH-8057 Z\"urich, Switzerland}
\author{Gloria Tabacchi}
\affiliation{Department  of  Science  and  High  Technology,  University  of  Insubria  and  INSTM,  via  Valleggio  9,  I-22100 Como,  Italy}
\author{Andreas Gl\"{o}{\ss}}
\affiliation{BASF SE, Carl-Bosch-Stra{\ss}e 38, D-67056 Ludwigshafen am Rhein, Germany}
\author{Michael Lass}
\affiliation{Department of Computer Science and Paderborn Center for Parallel Computing, Paderborn University, Warburger Str. 100, D-33098 Paderborn, Germany}
\author{Iain Bethune}
\affiliation{Hartree Centre, Science and Technology Facilities Council, United Kingdom}
\author{Christopher J. Mundy}
\affiliation{Physical Science Division, Pacific Northwest National Laboratory, P. O. Box 999, Richland, Washington 99352, USA}
\author{Christian Plessl}
\affiliation{Department of Computer Science and Paderborn Center for Parallel Computing, Paderborn University, Warburger Str. 100, D-33098 Paderborn, Germany}
\author{Matt Watkins}
\affiliation{School of Mathematics and Physics, University of Lincoln, Brayford Pool, Lincoln, United Kingdom}
\author{Joost VandeVondele}
\affiliation{Swiss National Supercomputing Centre (CSCS), ETH Z\"urich}
\author{Matthias Krack}
\email{matthias.krack@psi.ch}
\affiliation{Laboratory for Scientific Computing and Modelling, Paul Scherrer Institute, CH-5232 Villigen PSI, Switzerland}
\author{J\"urg Hutter}
\altaffiliation{}
\affiliation{Department of Chemistry, University of Zurich, Winterthurerstrasse 190, CH-8057 Z\"urich, Switzerland}

\date{\today}

\verb

\begin{abstract}
\cp\ is an open source electronic structure and molecular dynamics software package to perform atomistic simulations of solid-state, liquid, molecular and biological systems. It is especially aimed at massively-parallel and linear-scaling electronic structure methods and state-of-the-art ab-initio molecular dynamics simulations. Excellent performance for electronic structure calculations is achieved using novel algorithms implemented for modern high-performance computing systems. This review revisits the main capabilities of \cp\ to perform efficient and accurate electronic structure simulations. The emphasis is put on density functional theory and multiple post-Hartree-Fock methods using the Gaussian and plane wave approach and its augmented all-electron extension.
\end{abstract}

\maketitle 

%

\section{Introduction}
\label{section_intro}
The geometric increase in the performance of computers over the last few decades, together with advances in theoretical methods and applied mathematics, has established computational science as an indispensable technique in chemistry, physics, life and materials sciences. In fact, computer simulations have been very successful in explaining a large variety of new scientific phenomena, interpret experimental measurements, predict materials properties and even rationally design new systems. Therefore, conducting experiments \textit{in silico} permits to investigate systems atomistically that otherwise would be too difficult, expensive or simply impossible to perform. However, the by far most rewarding outcome of such simulations is the invaluable insights they provide into the atomistic behavior and the dynamics. Therefore, electronic structure theory based \textit{ab-initio} molecular dynamics (AIMD) can be sought of as a computational microscope~\cite{Marx_Hutter_2009,Hutter2011,Kuehne2014}. 

The open source electronic structure and molecular dynamics (MD) software package \cp\ aims at providing a broad range of computational methods and simulation approaches, suitable for extended  condensed-phase systems. The latter is made possible by combining efficient algorithms with excellent parallel scalability to exploit modern high-performance computing architectures. However, beside conducting efficient large-scale AIMD simulations, \cp\ provides a much broader range of capabilities, which includes the possibility of choosing the most adequate approach for a given problem and the flexibility of combining computational methods.

The remaining of the manuscript is organized as follows. The Gaussian and plane wave (GPW) approach to density functional theory (DFT) is reviewed in section~\ref{section_gpw}, before Hartree-Fock and beyond Hartree-Fock methods are covered in sections~\ref{section_hfx} and \ref{section_Post-HF}, respectively. Thereafter, density functional perturbation theory (DFPT) and time-dependent DFT (TD-DFT) are described in  sections~\ref{section_DFPT} and \ref{section_tddft}. Sections~\ref{section_eigensolver} and \ref{sec:DBCSR} are devoted to low-scaling eigenvalue solver based on sparse matrix linear algebra using the \DBCSR library. Conventional orthogonal localized orbitals, non-orthogonal localized orbitals (NLMO), absolutely localized molecular orbitals (ALMO) and compact localized molecular orbtials (CLMO) are discussed in section~\ref{section_ALMO} to facilitate linear-scaling AIMD, whose key concepts are detailed in section~\ref{section_AIMD}.  
Energy decomposition and spectroscopic analysis methods are presented in section~\ref{section_analysis}, followed by various embedding techniques, which are summarized in section~\ref{section_embedding}. Interfaces to other programs and technical aspects of \cp\ are specified in sections~\ref{section_interfaces} and \ref{section_technical}, respectively. 
%

\section{Gaussian and plane waves method}
\label{section_gpw}
%
%
The electronic structure module \qs~\cite{Krack2004,VandeVondele2005} 
in \cp\ can handle a wide spectrum of methods and approaches. 
Semi-empirical (SE) and tight-binding (TB) methods, orbital-free and Kohn--Sham DFT (KS-DFT) and wavefunction-based correlation methods (e.g. MP2, dRPA, GW) all
make use of the same infrastructure of integral routines and optimization algorithms. 
In this section, we give a brief overview of the methodology that
sets \cp\ apart from most other electronic structure programs, namely its 
use of a plane wave (PW) auxiliary basis set within a Gaussian orbital scheme.
As many other programs, \cp\ uses contracted Gaussian basis sets $g(\br)$
to expand orbital functions
\begin{equation}
             \varphi(\br) = \sum_u d_u \, g_u(\br),
\end{equation}
where the contraction coefficients $d_u$ are fixed and the primitive Gaussians 
\begin{equation}
           g(\br) = r^l \exp[-\alpha (\br - \bA)^2] \, Y_{lm}(\br - \bA)
\end{equation}
are
centered at atomic positions. 
These functions are defined by the exponent $\alpha$, the spherical harmonics $Y_{lm}$ with
angular momentum $(l,m)$ and the coordinates of its center $\bA$.
The unique properties of Gaussians, e.g. analytic integration or the product theorem,
are exploited in many programs.
In \cp\ we make use of an additional property of Gaussians, namely, that their
Fourier transform is again a Gaussian function, i.e. 
\begin{equation}
          \int{ \! \exp[-\alpha \br^2] \exp[-\I \bG \cdot \br]} \, d\br = \exp \left[ -\frac{\bG^2}{4\alpha} \right].
\end{equation}
This property is directly connected with the fact, that integration of Gaussian functions
on equidistant grids shows exponential convergence with grid spacing. In order to take advantage of this property we define, within a computational box or periodic unit cell, a set of equidistant grid points
\begin{equation}
          \bR = \bh \, \bN \, \bq.
\end{equation}
The three vectors $\ba_1, \ba_2$, and $\ba_3$ define a computational box with
a $3 \times 3$ matrix $\bh = [\ba_1, \ba_2, \ba_3]$ and a volume $\Omega = \det {\bh}$.
Furthermore, $\bN$ is a diagonal matrix with entries $1/N_s$, where $N_s$ is the number of grid points along vector $s=1,2,3$, whereas  $\bq$ is a vector of integers ranging from $0$ to $N_s - 1$. Reciprocal lattice vectors $\bb_i$, defined by
$\bb_i \cdot \ba_j = 2\pi \delta_{ij}$
can also be arranged into a $3 \times 3$ matrix
$ [\bb_1, \bb_2, \bb_3] = 2 \pi (\bh^t)^{-1} $
which allows us to define reciprocal space vectors
\begin{equation}
          \bG = 2 \pi (\bh^t)^{-1} \bg,
\end{equation}
where $\bg$ is a vector of integer values.
Any function with periodicity given by the lattice vectors and defined on the
real-space points $\bR$ can be transformed into a reciprocal space 
representation by the Fourier transform
\begin{equation}
          f(\bG) = \sum_{\bR} f(\bR) \exp[\I \bG \cdot \bR].
\end{equation}
The accuracy of this expansion is given by the grid spacings or the PW cutoff
defining the largest vector $\bG$ included in the sum.

In the GPW method~\cite{Lippert1997}, the equidistant grid, or equivalently
the PW expansion within the computational box, is used for an alternative
representation of the electron density.
In the KS method, the electron density is defined by
\begin{equation}
          n(\br) = \sum_{\mu \nu} P_{\mu \nu} \varphi_\mu(\br) \varphi_\nu(\br),
\end{equation}
where the density matrix $P_{\mu \nu} = \sum_i f_i c_{\mu i} c_{\nu i}$ is calculated from the
orbital occupations $f_i$ and the 
orbital expansion coefficients $c_{\mu i}$ of the common linear combination of atomic orbitals $\Phi_i(\br) = \sum_\mu c_{\mu i} \varphi_\mu(\br)$. Therein, $\Phi_i(\br)$ are the so-called molecular orbitals (MOs) and $\varphi_\mu(\br)$ the atomic orbitals (AOs). 
In the PW expansion, however, the density is given by
\begin{equation}
          n(\br) = \sum_{\bG} n(\bG) \exp[\I \bG \cdot \br].
\end{equation}
The definitions given above allow us to calculate the expansion coefficients $n(\bG)$
from the density matrix $P_{\mu \nu}$ and the basis functions $\varphi_\mu(\br)$.
The dual representation of the density is used in the definition of the GPW-based KS
energy expression (see section \ref{subsection_ks}) to facilitate efficient and accurate algorithms for
the electrostatic, as well as exchange and correlation energies.
The efficient mapping $P_{\mu \nu} \rightarrow n(\bG)$ is achieved by using multigrid
methods, optimal screening in real-space, and the separability of Cartesian Gaussian functions in
orthogonal coordinates. Details of these algorithms that result in a linear scaling
algorithm with small prefactor for the mapping is described elsewhere~\cite{Lippert1997,Krack2004,VandeVondele2005}.

\subsection{Kohn--Sham energy, forces, and stress tensor}
\label{subsection_ks}
The KS electronic energy functional~\cite{Kohn_1965_a} for a molecular or crystalline system 
within the supercell or $\Gamma$-point approximation in
the GPW framework~\cite{Lippert1997,Krack2004,VandeVondele2005} is defined as
\begin{subequations}
\begin{eqnarray}
 E^{\rm el}[P] &=& E^{\rm kin}[P] + E^{\ext}[P] + E^{\rm ES}[n_\bG] + E^{\XC}[n_\bG]\\
               &=& \sum_{\mu\nu} P_{\mu\nu}
                    \langle
                     \varphi_\mu(\br)
                     \mid -\frac{1}{2}\nabla^2 +  V^{\ext}(\br) \mid
                     \varphi_\nu(\br)
                    \rangle \nonumber\\
               \label{Eq:GPW-energy}
               &+& 4\pi\,\Omega \sum_{\bG}
                    \frac{|n_{\rm tot}(\bG)|^2}{\bG^2} + E_{\rm ovrl} - E_{\rm self}\\ 
               &+& \frac{\Omega}{N_1 N_2 N_3} \sum_{\bR} F_{\XC}[n_\bG](\bR), \nonumber
\end{eqnarray}
\end{subequations}
where $E^{\rm kin}$ is the kinetic energy, $E^{\ext}$ is the electronic
interaction with the ionic cores (see section \ref{subsection_pp}), $E^{\rm ES}$ is the total electrostatic
(Coulomb) energy and $E^{\XC}$ is the exchange-correlation (XC) energy.
An extension of Eq.~(\ref{Eq:GPW-energy}) to include a k-point sampling within the first Brillouin zone
is also available in \cp. This implementation follows the methods from Pisani~\cite{Roetti1996}, 
Scuseria~\cite{Kudin_CPL_1998} and coworkers.
The electrostatic energy is calculated using an Ewald method~\cite{Marx2000}. A total charge density is defined by adding Gaussian charge distributions of the form
\begin{equation}
         n^A(\br) = -\frac{Z_A}{(R_A^c)^3} \pi^{-3/2} \exp{ \left[ - \left( \frac{\br - \bA}{R_A^c} \right)^2 \right] }
\end{equation}
to the electronic charge distribution $n(\br)$. Self and overlap terms 
\begin{subequations}
\begin{eqnarray}
    E_{\rm self} &=& \sum_A \frac{1}{\sqrt{2 \pi}} \frac{Z_A^2}{R_A^c} \enspace \\
    E_{\rm ovrl} &=& \sum_{A,B} \frac{Z_A Z_B}{|\bA - \bB|} 
                    \erfc \left[ \frac{|\bA - \bB|}{\sqrt{{R_A^c}^2 + {R_B^c}^2}} \right],
\end{eqnarray}
\end{subequations}
as generated by this Gaussian distributions, have to be compensated. 
The double sum for $E_{\rm ovrl}$ runs over unique atom pairs and has to be extended, if necessary,\
beyond the minimum image convention.
The electrostatic term also includes an interaction of the compensation charge with the electronic charge density that
has to be subtracted from the external potential energy. 
The correction potential is of the form
\begin{equation}
    V_{\rm core}^A = \int \; d\br' \frac{n^A(\br')}{|\br - \br'|} = - \frac{Z_A}{|\br - \bA|} 
                     \erf \left[ \frac{|\br - \bA|}{R_A^c} \right].
\end{equation}
The treatment of the electrostatic energy terms and the XC energy is the same as in 
PW codes~\cite{Marx2000}. 
This means that the same methods that are used in those codes to adapt Eq.~(\ref{Eq:GPW-energy})
for cluster boundary conditions can be used here. 
This includes analytic Green's function methods~\cite{Marx2000}, the method of
Eastwood and Brownrigg~\cite{Eastwood1979}, the methods of Martyna and Tuckerman~\cite{Martyna1999,Minary2002}, 
and wavelet approaches~\cite{Genovese_JCP_2006,Genovese_JCP_2007}.
Starting from $n(\bR)$, using Fourier transformations we can calculate the combined potential for the electrostatic and XC energies. 
This includes the calculation of the gradient of the charge density needed for generalized gradient approximation (GGA) functionals. 
For non-local van der Waals functionals~\cite{Dion_PRL_2004}, we use the Fourier transform based algorithm of Soler et al.~\cite{Roman-Perez_PRL_2009}. The combined potential calculated on the grid points $V^\XC(\bR)$ is then the starting point to calculate the KS matrix elements
\begin{equation}
         H^{\XC}_{\mu \nu} = \frac{\Omega}{N_1 N_2 N_3} \sum_{\bR} V^\XC(\bR) \varphi_\mu(\bR) \varphi_\nu(\bR).
\end{equation}
This inverse mapping $V^\XC(\bR) \rightarrow H^{\XC}_{\mu \nu}$ is using the same methods that have been used for the charge density. Also, in this case we can achieve linear scaling with small prefactors.

A consistent calculation of nuclear forces within the GPW method can be done easily. 
Pulay terms, i.e. the dependence
of the basis functions on nuclear positions, require the integration of potentials given on the real-space grid
with derivatives of the basis functions~\cite{PulayForces}. 
However, the basic routines work with Cartesian Gaussian functions and
their derivatives are again functions of the same type, so the same routine can be used.
Similarly, for XC functionals including the kinetic energy density, we can calculate $\tau(\br)$, the corresponding potential
and matrix representation using the same mapping functions.

For the internal stress tensor
\begin{equation}
     \Pi_{uv} = -\frac{1}{\Omega} \sum_s \frac{\partial E_{\rm el}}{\partial \bh_{us}} \bh^t_{sv},
\end{equation}
we use again the Fourier transform framework for the $E^\XC$ terms and the simple virial for all pair forces.
This applies to all Pulay terms and only for the XC energy contributions from GGA functionals special care is needed
due to the cell dependent grid integration~\cite{Corso_PRB_1994,Balbas_PRB_2001,Schmidt_JPCB_2009}. 

\subsection{Dual-space pseudopotentials}
\label{subsection_pp}
The accuracy of the PW expansion of the electron density in the GPW method
is controlled by cutoff value $E_{\rm cut}$ restricting the maximal allowed value
of $|\bG|^2$. 
In case of a Gaussian basis sets the cutoff needed to get a given 
accuracy is proportional to the largest exponent. 
As can been easily seen by inspecting
common Gaussian basis sets for elements of different rows in the periodic table, the
value of the largest exponent rapidly increases with atomic number. 
Therefore, the prefactor in GPW calculations will increase similarly. 
In order to avoid this, we can either resort to
a pseudopotential description of inner shells, or use a dual representation as
described in Bl\"ochl's projector augmented-wave method (PAW) \cite{Bloechl1994}.
The pseudopotentials used together with PW basis sets are constructed
to generate nodeless atomic valence functions. Fully non-local forms are
computationally more efficient and easier to implement.
Dual-space pseudopotentials are of this form and are, because of their analytic
form consistent of Gaussian functions, 
easily applied together with Gaussian basis sets~\cite{Hartwigsen1998,Goedecker1996,Krack2005}.

The pseudopotentials are given in real-space as
%
\begin{eqnarray}
     V^{\pp}_{\loc}(r) & = & -\frac{Z_{\rm ion}}{r} \erf \left( \alpha^{\pp} r \right) +
                             \sum_{i=1}^4 C^{\pp}_i \left( \sqrt{2} \alpha^{\pp} r \right)^{2i-2} \nonumber \\
    &\times& \exp\left[ -\left( \alpha^{\pp} r \right)^2 \right], \, 
     \mbox{with} \ \alpha^{\pp}
                      = \frac{1}{\sqrt{2}r^{\pp}_{\loc}} 
\end{eqnarray}
and a non-local part
\begin{equation}
     V^{\pp}_{\nl}({\br},{\br'}) =  \sum\limits_{lm} \sum\limits_{ij} \langle\,{\br}\,|\,p_i^{lm} \,\rangle
                                       \,h_{ij}^{l}\,\langle\,p_j^{lm} \,|\,{\br'} \,\rangle, 
                \\
\end{equation}
where
\begin{equation}
        \langle\, {\br}\,|\,p_i^{lm}\,\rangle  = N_i^{l} Y^{lm}(\hat r) r^{l+2i-2}
                                                    \exp\left[-\frac{1}{2}\left(\frac{r}{r_l}\right)^2\right] 
\end{equation}
are Gaussian-type projectors 
%
resulting in a fully analytical formulation that requires only the definition of a small parameter set for each element. 
Moreover, the pseudopotentials are transferable and norm-conserving.
The pseudopotential parameters are optimized with respect
to atomic all-electron wavefunction obtained from
relativistic density functional calculations using a numerical
atom code, which is also part of \cp.
A database with many pseudopotential parameter sets
optimized for different XC potentials
is available together with the distribution of the \cp\ program.

\subsection{Basis sets}
\label{subsection_basis}
The use of pseudopotentials in the GPW method also requires the
use of correspondingly adapted basis sets. In principle, the same
strategies that have been used to generate molecular or
solid-state Gaussian basis sets could be used. 
It is always possible to generate specific basis sets for an application type, e.g.
for metals or molecular crystals, but for ease of use a general basis set is desirable.
Such a general basis should fulfill the following requirements.
High accuracy for smaller basis sets and a route for systematic improvements. 
One and the same basis set should perform in various environments from isolated molecules 
to condensed phase systems. Ideally, the 
basis sets should lead for all systems to well conditioned overlap matrices and be 
therefore well suited for linear scaling algorithms. To fulfill all the above requirements, generally contracted basis sets with shared exponents for all angular momentum states were proposed~\cite{VandeVondele_JCP_2007}. 
In particular, a full contraction over all primitive functions is used for both 
valence and polarization functions. 
The set of primitive functions includes diffuse functions with small exponents that are mandatory for the description of weak interactions.
However, in contrast to the practice used in augmented basis sets, 
these primitive functions are always part of a contraction with tighter functions.
Basis sets of this type were generate according to a recipe that includes global optimization
of all parameters with respect to the total energy of a small set of 
reference molecules~\cite{VandeVondele_JCP_2007}.
The target function was augmented with a penalty function that includes the
overlap condition number.
These basis sets of type {\em molopt} have been created for the first two rows of the periodic table.
They show good performance for molecular systems, liquids and dense solids. 
Results for the Delta test are shown in Fig.~\ref{fig:deltatest_histogram}.
The basis sets have 7 primitive functions with a smallest exponent of $\approx 0.047$ for oxygen.
This is very similar to the smallest exponent found in augmented basis sets of the 
correlation consistent type, e.g. $\approx 0.060$ for aug-cc-pVTZ.
The performance of the grid mapping routines depends strongly on this most diffuse function in the basis sets.
We have therefore optimized a second set of basis sets of the {\em molopt} type, where the number
of primitives has been reduced (e.g. 5 for first row elements) which also leads to less diffuse functions.
The smallest exponent for oxygen in this basis set is now $\approx 0.162$.
These basis sets still show good performance in many different environments and are especially suited
for all types of condensed matter systems (e.g. see Delta test results in section~\ref{subsection_deltatest}).
The reduction of Gaussian primitives and the removal of very diffuse functions leads to a 10-fold time reduction
for the mapping routines for liquid water calculations using default settings.

\subsection{Local density fitting approach}
\label{subsection_lri}
Within the GPW approach, the mapping of the electronic density on the grid is often the
time determining step. 
With an appropriate screening this is a linear scaling step with a prefactor determined by the 
number of overlapping basis functions.
Especially in condensed phase systems, the number of atom pairs that have to be included 
can be very large.
For such cases it can be beneficial, to add an intermediary step in the density mapping.
In this step, the charge density is approximated by another auxiliary Gaussian basis.
The expansion coefficients are determined using the local density fitting approach by
Baerends et al.~\cite{Baerends_CP_1973}.
They introduced a local metric, where the electron density is decomposed into pair-atomic densities, 
which are approximated as a linear combination of auxiliary functions localized at atoms A and B. 
The expansion coefficients are obtained by employing an overlap metric.
This local resolution-of-the-identity (LRI) method combined with GPW is available in \cp\ as the
LRIGPW approach~\cite{Golze_JCTC_2017}. For details how to set-up such a calculation see Ref.~\onlinecite{lrigpw_tutorial}. 

The atomic pair densities $n^{AB}$ are approximated by an expansion in a set of
Gaussian fit functions ${f(\br)}$ centered at atoms $A$ and $B$, respectively.
The expansion coefficients are obtained for each pair $AB$ by fitting the exact density 
while keeping the number of electrons fixed.
This leads to a set of linear equations that can be solved easily. 
The total density is then represented by the sum of coefficients of
all pair expansions on the individual atoms.
The total density is now presented as a sum
over the number of atoms, whereas in GPW we have a sum over pairs.
In the case of 64 water molecules in a periodic box,
this means that the fitted density is mapped on the grid by
192 atom terms rather than $\approx 200000$ atom pairs.

LRIGPW requires the calculation of two- and three-index overlap integrals that is computationally demanding for large auxiliary basis sets. To increase the efficiency of the LRIGPW implementation, we developed an integral scheme based on solid harmonic Gaussian functions~\cite{Golze2017}, which is superior to the widely used Cartesian Gaussian-based methods.

An additional increase in efficiency can be achieved by recognizing that most of the
electron density is covered by a very small number of atom pair densities.
The large part of more distant pairs can be approximated by an expansion on
a single atom. 
By using a distance criteria and a switching function, a method with a smooth
potential energy is created.
The single atom expansion reduces memory requirements and computational time considerably.
In the above water box example about 99\% of the pairs can be treated as distant pairs.

\subsection{Gaussian-augmented plane waves approach}
\label{subsection_gapw}
An alternative to pseudopotentials, or a method to allow for smaller cutoffs in 
pseudopotential calculations is provided by the Gaussian-augmented plane waves (GAPW) approach~\cite{Lippert1999,Krack_PCCP_2000}.
The GAPW method uses a dual representation of the electronic density,
where the usual expansion of the density using the density matrix $\bP$
is replaced in the calculation of the Coulomb and XC energy by
\begin{equation}
   n({\br}) = \tilde n({\br}) + \sum_A n_A({\br}) - \sum_A \tilde n_A({\br})
   \label{den_split}.
\end{equation}
The densities $\tilde n({\br}), n_A({\br})$,
and $\tilde n_A({\br})$ are expanded in plane waves and products of primitive
Gaussians centered on atom $A$, respectively, i.e.
\begin{subequations}
\begin{align}
        \tilde n({\br}) &= \frac{1}{\Omega} \sum_{\bG}
                            \tilde n({\bG}) \, e^{i {\bG} \cdot {\br}}
            \label{eq_pw}, \\
        n_A({\br}) &= \sum_{mn \in A} P^A_{mn} \, g_m({\br}) \, g_n^\star({\br}), \\
        \tilde n_A({\br}) &= \sum_{mn \in A} \tilde P^A_{mn} \, g_m({\br}) \, g_n^\star({\br}).
\end{align}
\end{subequations}
In Eq.~(\ref{eq_pw}), $\tilde n({\bG})$ are the Fourier coefficients of the soft density,
as obtained in the GPW method by keeping in the expansion of the
contracted Gaussians only those primitives with exponents smaller than a given threshold.
The expansion coefficients $P^A_{mn}$, and $\tilde P^A_{mn}$ are also functions
of the density matrix $P_{\alpha \beta}$ and can be calculated efficiently.
The separation of the density from Eq.~(\ref{den_split}) is borrowed from the
PAW approach~\cite{Bloechl1994}.
Its special form allows the separation of the smooth parts, characteristic of the
interatomic regions, from the quickly varying parts close to the atoms, while still
expanding integrals over all space.
The sum of the contributions in Eq.~(\ref{den_split}) gives the
correct full density if the following conditions are fulfilled
\begin{subequations}
\begin{align}
    n({\br}) &= n_A({\br}) &
     \tilde n({\br}) &= \tilde n_A({\br}) & \text{close to atom} \ A, \\
    n({\br}) &= \tilde n({\br})
    &   n_A({\br}) &= \tilde n_A({\br})  & \text{far from atom} \ A.
\end{align}
\end{subequations}
The first conditions are exactly satisfied only in the limit of a complete basis set.
However, the approximation introduced in the construction of the local densities,
can be systematically improved by choosing larger basis sets.

For semi-local XC functionals such as the local density approximation (LDA),
GGA or meta functionals using the kinetic energy
density, the XC energy can be simply written as
\begin{equation}
     E_{\rm GAPW}^{\XC}[ n ] = E^{\XC}[ \tilde n ] + \sum_A E^{\XC}[ n_A ]
                       - \sum_A E^{\XC}[ \tilde n_A ]
     \enspace .
\end{equation}
The first term is calculated on the real-space grid defined by the PW expansion and the other two are efficiently and accurately
calculated using atom centered meshes.

Due to the non-local character of the Coulomb operator,
the decomposition for the electrostatic energy is more complex.
In order to distinguish between local and global terms, we need to
introduce atom-dependent screening densities $n_A^0$ 
that generate the same multipole expansion $Q^{lm}_A$
as the local density $n_A - \tilde n_A + n^Z_A$, where $n^Z_A$ is the nuclear charge of
atom $A$, i.e.
\begin{equation}
    n_A^0({\br}) = \sum_{lm} Q^{lm}_A \, g^{lm}_A({\br}).
\end{equation}
The normalized primitive Gaussians $g^{lm}_A({\br})$ are defined
with an exponent such that they are localized within an atomic region.
Since the sum of local densities $n_A - \tilde n_A + n^Z_A -  n_A^0$
has vanishing multiple moments, it does not interact with charges outside the
localization region, and the
corresponding energy contribution can be calculated by one-center integrals.
The final form of the Coulomb energy in the GAPW method then reads as 
\begin{eqnarray}
      E_{\rm GAPW}^{\rm C}[n + n^Z] &=&
            E^{\rm H}[\tilde n + n^0] + \sum_A E^{\rm H}[n_A + n^Z_A] \nonumber \\
            &-& \sum_A E^{\rm H}[\tilde n_A + n^0_A]
      \label{eq_coul},
\end{eqnarray}
where $n^0$ is summed over all atomic contributions,
and $E^{\rm H}[n]$ denotes the Coulomb energy of a charge distribution $n$.
The first term in Eq.~(\ref{eq_coul}) can be calculated efficiently using
fast Fourier transform (FFT) methods using the GPW framework. 
The one-centered terms are calculated on radial atomic grids.

The special form of the GAPW energy functional involves several
additional approximations in addition to a GPW calculation.
The accuracy of the local expansion of the density is controlled by the
flexibility of the product basis of primitive Gaussians.
As we fix this basis to be the primitive Gaussians present in the
original basis we cannot independently vary the accuracy of the expansion. 
Therefore, we have to consider this approximation as
inherent to the primary basis used.

With the GAPW method it is possible to calculate materials properties that
depend on the core electrons. 
This has been used for the simulation of the X-ray scattering in liquid water~\cite{Krack_JCP_2002}.
X-ray absorption spectra are calculated using the 
transition potential method~\cite{Triguero_PRB_1998,Iannuzzi_PCCP_2007,Iannuzzi_JCP_2008,Mueller2019}.
Several nuclear and electronic magnetic properties are also available~\cite{Weber_JCP_2009,Mondal_JCTC_2018}.

\section{Hartree-Fock and hybrid density functional theory methods}
\label{section_hfx}
Even though, semi-local DFT is a cornerstone of much of condensed phase electronic structure modeling, it is also recognized that going beyond GGA-based DFT is necessary to improve the accuracy and reliability of electronic structure methods.
One path forward is to augment DFT by elements of wavefunction theory, or to adopt wavefunction theory itself. 
This is the approach taken in successful hybrid XC functionals such as B3LYP or HSE~\cite{Becke1988,Lee1988}, where part of the exchange functional is replaced
by exact Hartree-Fock exchange (HFX).
The capability to compute HFX was introduced in CP2K by Guidon et al.~\cite{Guidon2008, Guidon2009, Guidon2010}.
The aim at that time was to enable the use of hybrid XC functionals for condensed phase calculations, of relatively large, disordered systems, in the context of AIMD simulations.
This objective motivated a number of implementation choices and developments that will be described in the following.
The capability was particularly important to make progress in the field of first principles electro-chemistry~\cite{Adriaanse2012, Cheng2014},
but is also the foundation for the correlated wavefunction methods such as MP2, RPA and GW that are available in CP2K and will be described in section~\ref{section_Post-HF}.

In the periodic case, HFX can be computed as 
\begin{widetext}
\begin{equation}
\label{Eq:HFXP}
E_X^{\mbox{PBC}}=-\frac{1}{2N_k}\sum_{i,j}\sum_{\mathbf{k},\mathbf{k}'}\int\int \psi_i^{\mathbf{k}}(r_1)\psi_j^{\mathbf{k'}}(r_1)g(|r_1-r_2|)\psi_i^{\mathbf{k}}(r_2)\psi_j^{\mathbf{k'}}(r_2)\, d^3r_1 d^3r_2,
\end{equation}
\end{widetext}
where an explicit sum over the k-points is retained and a generalized Coulomb operator $g(|r_1-r_2|)$ is employed.
The k-point sum is important, as at least for the standard Coulomb operator $\frac{1}{r}$,
the term for $\mathbf{k}=\mathbf{k'}=\mathbf{0}$ (the so-called $\Gamma$-point) is singular,
though the full sum approximates an integrable expression.
If the operator is short-ranged or screened, the $\Gamma$-point term is well-behaved.
CP2K computes HFX at the $\Gamma$-point only and employs a localized atomic basis set,
using an expression where the sum is explicit over the indices of the localized basis ($\lambda\sigma\mu\nu$), as well as the image cells ($\mathbf{a}\mathbf{b}\mathbf{c}$), thus 
\begin{equation}
\label{Eq:HFXbasis}
E_X^{\mbox{$\Gamma$}} = -\frac{1}{2} \sum_{\lambda\sigma\mu\nu} P^{\mu\sigma}P^{\nu\lambda}\sum_{\mathbf{a}\mathbf{b}\mathbf{c}}\left(\mu\nu^{\mathbf{a}}|\lambda^{\mathbf{b}}\sigma^{\mathbf{b}+\mathbf{c}}\right)_{g}.
\end{equation}
%
For an operator with finite range, the sum over the image cells will terminate.
This expression was employed to perform hybrid DFT-based AIMD simulations of liquid water with CP2K~\cite{Guidon2008}.
Several techniques have been introduced to reduce the computational cost.
First, screening based on the overlap of basis functions is employed to reduce the scaling of the calculation from $O(N^4)$ to $O(N^2)$. 
This does not require any assumptions on the sparsity of the density matrix, nor the range of the operator, and makes HFX feasible for fairly large systems.
Second, the HFX implementation in CP2K is optimized for 'in-core' calculations,
where the four center integrals are computed (analytically) only once at the beginning of the SCF procedure, stored in main memory, and reused afterwards.
This is particularly useful in the condensed phase, as the sum over all image cells multiplies the cost of evaluating the integral, relative to gas phase calculations.
To store all computed integrals, the code has been very effectively parallelized using MPI and OMP, yielding super-linear speed-ups as long as added hardware resources provide additional memory to store all integrals.
Furthermore, a specialized compression algorithm is used to store each integral with just as many bits as needed to retain the target accuracy.
Third, a multiple-time-step (MTS) algorithm (see section~\ref{subsection_mts}) is employed to evaluate HFX energies only every few time-steps during an AIMD simulation,
assuming that the difference between the potential energy surface of a GGA and a hybrid XC functional is slowly varying with time. 
This technique has found reuse in correlated wavefunction simulations described in section~\ref{section_Post-HF}.

In Ref.~\onlinecite{Guidon2009}, the initial implementation was revisited, in particular to be able to robustly compute HFX at the $\Gamma$-point for the case where the operator in the exchange term is $\frac{1}{r}$, and not a screened operator as e.g. in HSE~\cite{HSE03,HSE06}.
The solution is to truncate the operator (not any of the image cell sums), with a truncation radius $R_C$ that grows with the cell size.
The advantage of this approach is that the screening of all terms in Eq.~\ref{Eq:HFXbasis} can be performed rigorously,
and that the approach is stable for a proper choice of screening threshold, 
also in the condensed phase with good quality (but non-singular) basis sets.
The value of $R_C$ that yields convergence is system dependent, and large values of $R_C$ might require
the user to explicitly consider multiple unit cells for the simulation cell.
Note that the HFX energy converges exponentially with $R_C$ for typical insulating systems,
and that the same approach was used previously to accelerate k-point convergence~\cite{Spencer2008}.
In Ref.~\onlinecite{Paier2009}, it was demonstrated that two different codes (CP2K and Gaussian), with very different implementations of HFX,
could reach micro-Hartree agreement for the value of the HF total energy of the LiH crystal.
In Ref.~\onlinecite{Guidon2009}, a suitable GGA-type exchange function was derived that can be used as a long range correction (LRC) together with the truncated operator.
This correction functional, in the spirit of the the exchange functional employed in HSE,
effectively allows for model chemistries that employ very short range exchange (e.g. $\approx 2$\AA{}) only.
Important for the many applications of HFX with CP2K is the auxiliary density matrix method (ADMM) method, introduced in Ref.~\onlinecite{Guidon2010}.
This method reduces the cost of HFX significantly, often bringing it to within a few times the cost of conventional GGA-based DFT, by addressing the unfavorable scaling of
the computational cost of Eq.~\ref{Eq:HFXbasis} with respect to basis set size. 
The key ingredient of the ADMM method is the use of an auxiliary density matrix $\mathbf{\hat{P}}$, which approximates the original $\mathbf{P}$,
for which the HFX energy is more cost-effective to compute:
\begin{eqnarray}
\label{Eq:ADMM}
E_X^{{HFX}}[\mathbf{P}]&=&E_X^{{HFX}}[\mathbf{\hat{P}}]+\left(E_X^{{HFX}}[\mathbf{P}]-E_X^{{HFX}}[\mathbf{\hat{P}}]\right)\nonumber\\
&\approx& E_X^{{HFX}}[\mathbf{\hat{P}}]+\left(E_X^{{DFT}}[\mathbf{P}]-E_X^{{DFT}}[\mathbf{\hat{P}}]\right).
\end{eqnarray}
Effectively, $E_X^{{HFX}}[P]$ is replaced with by computationally more efficient  $E_X^{{HFX}}[\hat{P}]$,
and the difference between the two is corrected approximately with a GGA-style exchange functional.
Commonly, the auxiliary $\mathbf{\hat{P}}$ is obtained by projecting the density matrix $\mathbf{P}$ using a smaller, auxiliary basis.
This approximation, including the projection, can be implemented fully self-consistently.
In Ref.~\onlinecite{Ohto2019}, the efficiency of the ADMM method was demonstrated by computing over 300~ps of AIMD trajectory for systems containing 160 water molecules, 
and by computing spin densities for a solvated metallo-protein system containing approximately 3000~atoms~\cite{Guidon2010}.
%

\section{Beyond Hartree-Fock Methods}
\label{section_Post-HF}
In addition, so-called post-Hartree-Fock methods that are even more accurate than the just describe hybrid-DFT approach are also available within \cp. 

\subsection{Second-order M\o{}ller-Plesset perturbation theory}
\label{subsection_MP2}
\subsubsection{Theory}
\label{subsubsection_MP2-Theory}
Second-order M\o{}ller-Plesset perturbation theory (MP2) is the simplest \textit{ab-initio} correlated wavefunction method~\cite{Moller1934}, applied to the Hartree-Fock reference and able to capture most of the dynamic electron correlation~\cite{Helgaker2000}. In the DFT framework, the MP2 formalism gives rise to the doubly-hybrid XC functionals~\cite{Grimme2006}. In the spin-orbital basis, the MP2 correlation energy is given by:
\begin{widetext}
\begin{equation} 
 E^{(2)}= \frac{1}{2} \Bigg\{ \sum_{ij,ab} \frac{(ia|jb) [(ia|jb) - (ib|ja)]}{\Delta_{ij}^{ab}} + 
 \sum_{\overline{ij},\overline{ab}} \frac{(\overline{ia}|\overline{jb}) [(\overline{ia}|\overline{jb}) - (\overline{ib}|\overline{ja})]}{\Delta_{\overline{ij}}^{\overline{ab}}} \Bigg\}
  -\sum_{\overline{i}j, \overline{a}b} \frac{(\overline{ia}|jb)}{\Delta_{\overline{i}j}^{\overline{a}b}}, 
 \label{eq:MP2_canonical_energy}
\end{equation}
\end{widetext}
%
where $i, j, ...$ run over occupied spin-orbitals, $a, b, ...$ run over virtual spin-orbitals (indexes without bar stand for $\alpha$-spin-orbitals, indexes with bar for $\beta$-spin-orbitals),  $\Delta_{ij}^{ab} = \epsilon_a + \epsilon_b - \epsilon_i - \epsilon_j $  ($\epsilon_a$ and $\epsilon_i$ are orbital energies) and $(ia|jb)$ are electron repulsion integrals in the Mulliken notation.

In a canonical MP2 energy algorithm, the time limiting step is the computation of the $(ia|jb)$ integrals obtained
from the electron repulsion integrals over AOs $(\mu \nu | \lambda \sigma)$ via four consecutive index integral transformations.
The application of the resolution-of-identity (RI) approximation to MP2~\cite{Feyereisen1993}, which consists of replacing $(ia|jb)$ integrals with the approximated $(ia|jb)_{RI}$, is given by 
\begin{equation} 
 (ia|jb)_{RI} = \sum_{P} B^{ia}_P B^{jb}_P, B^{ia}_{P} = \sum_{R} (ia|R) L^{-1}_{PR},
 \label{eq:RI_contraction}
\end{equation}
where $P, R, ...$ (the total number of them is $N_a$) are auxiliary basis functions and $\mathbf{L}$ are two-center integrals over them.
The RI-MP2 method is also scaling $O(N^5)$ with a lower prefactor: the main reason for the speed-up in RI-MP2 lies in the strongly reduced integral computation cost.

As the MP2 is non-variational with respect to wavefunction parameters, analytical expressions for geometric energy derivatives of RI-MP2 energies are complicated, since its calculation requires the solution of Z-vector equations~\cite{Weigend1997}. 

\subsubsection{Scaled opposite-spin MP2}\label{subsubsection_SOS-MP2}
The scaled opposite-spin MP2 (SOS-MP2) method is a simplified variant of MP2~\cite{Jung2004}. Starting from Eq.~\ref{eq:MP2_canonical_energy}, we neglect the same spin term in curly brackets and scale the remaining opposite spin term to account  for the introduced error. We can rewrite the energy term with the RI approximation and the Laplace transform
\begin{equation}
    \frac{1}{x}=\int_0^\infty dt\ e^{-x t}.
\end{equation}
When we exploit a numerical integration, we obtain the working equation for SOS-MP2, which reads as 
%
\begin{equation}
    E^{SOS-MP2}=-\sum_{q}\sum_{PQ}w_{q}Q_{PQ}^{\alpha}(t_q)Q_{QP}^{\beta}(t_{q})\label{eq:E_SOS},
\end{equation}
with
\begin{equation}
    Q_{PQ}^{\alpha}(t_{q})=\sum_{ia}B_{P}^{i a}B_{Q}^{i a}e^{(\epsilon_{a}-\epsilon_{i})t_q}\label{eq:Q_SOS}
\end{equation}
%
and similarly for the beta spin. The integration weights $w_{q}$ and abscissa $t_q$ are determined by a minimax procedure~\cite{Takatsuka2008}. In practice, $\simeq7$ quadrature points are needed for $\mu$Hartree accuracy~\cite{Jung2004}.

The most expensive step is the calculation of the matrix elements $Q_{PQ}^{\alpha}$ which scales like $\mathcal{O}(N^4)$. Due to similarities with the random phase approximation (RPA), we will discuss the implementation of this method in section~\ref{subsection_RPA}.

\subsubsection{Implementation}
\label{subsubsection_MP2-Implementation}
CP2K features $\Gamma$-point implementations of canonical MP2, RI-MP2, Laplace-transformed MP2 and SOS-MP2 energies~\cite{DelBen2012, DelBen2013}. For RI-MP2, analytical gradients and stress tensors are available~\cite{DelBen2015a}, for both closed and open electronic shells \cite{Rybkin2016}. Two- and three-center integrals can be calculated by the GPW method or analytically.

The implementation is massively parallel, takes the advantage of sparse matrix algebra and allows for GPU acceleration of large matrix multiplies. The evaluation of the gradients of the RI-MP2 energy can be performed within a few minutes for systems containing hundreds of atoms and thousands of basis functions on thousands of CPU cores, allowing for MP2-based structure relaxation and even AIMD simulations on HPC facilities. The cost of the gradient calculation is 4-5 times larger than the energy evaluation and open-shell MP2 calculation is typically 3-4 times more expensive than the closed-shell calculation.

\subsubsection{Applications}
\label{subsubsection_MP2-Applications}
The RI-MP2 implementation, which is the computationally most efficient MP2 variant available in CP2K, has been successfully applied to study a number of aqueous systems. In fact, \textit{ab-initio} Monte Carlo (MC) and AIMD simulations of bulk liquid water (with simulation cells containing 64 water molecules) predicted the correct density, structure and IR spectrum~\cite{DelBen2013a,DelBen2013a_corr,DelBen2015}. Other applications include structure refinements of ice XV~\cite{DelBen_JPCL_2014}, AIMD simulations of the bulk hydrated electron (with simulation cells containing 47 water molecules)~\cite{Wilhelm2019}, as well as the first AIMD simulation of a radical in the condensed phase using wavefunction theory. 

\subsection{Random Phase Approximation  Correlation Energy Method}
\label{subsection_RPA}
Total energy methods based on the RPA correlation energy have emerged in the recent years as promising approaches to include non-local dynamical electron correlation effects at the fifth rung on the Jacob's ladder of density functional approximations~\cite{doi:10.1063/1.1390175}.
In this context, there are numerous ways to express the RPA correlation energy depending on the theoretical framework and approximations employed to derive the working equations~\cite{PhysRevB.64.195120, PhysRevB.65.235109,
PhysRevB.66.245103, Furche2005, doi:10.1063/1.1858371, PhysRevB.77.045136,  Lu2009, PhysRevA.85.012517, PhysRevLett.102.096404, PhysRevLett.103.056401, Ren2012, molphys_gorling, doi:10.1080/00268970903476662, doi:10.1063/1.3317437, doi:10.1063/1.3250347, Furche2008, doi:10.1063/1.4849416, doi:10.1063/1.3090814, doi:10.1021/acs.jctc.5b01129}.  
Our implementation uses the approach introduced by Eshuis \emph{et al.}~\cite{Eshuis2010}, which can be referred to as based on the dielectric matrix formulation, involving the numerical integral over frequency of a logarithmic expression including the dynamical dielectric function, expressed in an Gaussian RI auxiliary basis.
%
%
Within this approach, the direct-RPA (sometimes referred as dRPA) correlation energy, which is a RPA excluding exchange contributions~\cite{Eshuis2010}, is formulated as a frequency integral 
\begin{equation}
  E_{c}^{\text{RPA}} = \frac{1}{2} \int_{-\infty}^{+\infty}{ \frac{d \omega}{2 \pi} 
                           \text{Tr}(\ln(\mathbf{1} + \mathbf{Q}(\omega) ) - \mathbf{Q}(\omega))},
  \label{eq:integral_imaginary_freq}
\end{equation}
with the frequency dependent matrix $\mathbf{Q}(\omega)$, expressed in the RI basis, determined by 
\begin{equation}
Q_{PQ}(\omega) = 2\sum_{i}^o\sum_{a}^v B_P^{ia} \,\frac{\varepsilon_a-\varepsilon_i}{\omega^2 + (\varepsilon_a-\varepsilon_i)^2}\,B_Q^{ia}.\label{eq:Qmat_RPA}
\end{equation}
For a given $\omega$, the computation of the integrand function in Eq.~\ref{eq:integral_imaginary_freq} and using Eq.~\ref{eq:Qmat_RPA} requires $O(N^4)$ operations.
The integral of Eq.~\ref{eq:integral_imaginary_freq} can be efficiently calculated by a minimax quadrature requiring only $\simeq10$ integration points to achieve $\mu$Hartree accuracy~\cite{DelBen2015120,doi:10.1021/ct5001268}.
Thus, the introduction of the RI approximation and the frequency integration techniques for computing $E_{c}^{\text{RPA}}$
leads to a computational cost of $O(N^4 N_q)$ and $O(N^3)$ storage, where $N_q$ is the number of quadrature points~\cite{DelBen2013}.

We note here that the conventional RPA correlation energy methods in CP2K are based on the the exact exchange (EXX) and RPA correlation energy formalism (EXX/RPA), which has extensively been applied to a large variety of systems including molecules, systems with reduced dimensionality and solids~\cite{PhysRevB.64.195120, PhysRevB.65.235109, PhysRevLett.88.166401, Furche2005, PhysRevB.79.205114, PhysRevLett.102.096404, molphys_gorling, Eshuis2012, new_j_phys_scheffler, Ren2012, DelBen_JPCL_2014, PhysRevB.66.245103, doi:10.1021/jp0746998, PhysRevLett.103.056401, PhysRevB.81.115126, PhysRevB.86.094109, PhysRevLett.103.056401, PhysRevLett.101.266106, PhysRevB.80.045402, PhysRevLett.96.136404, PhysRevB.84.201401, PhysRevB.77.045136, Lu2009, doi:10.1021/jp9095425}.
Within the framework of the EXX/RPA formalism, the total energy is given as
\begin{align}
  E_{\text{tot}}^{\text{EXX/RPA}} & = E_{\text{tot}}^{\text{HF}} + E_{\text{C}}^{\text{RPA}} \notag \\
                                  & = \left( E_{\text{tot}}^{\text{DFT}}-E_{\text{XC}}^{\text{DFT}} \right) + E_{\text{X}}^{\text{EXX}} + E_{\text{C}}^{\text{RPA}},
  \label{eq:rigpw:EXX_RPA_formula}
\end{align}
where the right-hand side terms of the last equation are: the DFT total energy, the DFT xc energy, the EXX energy and the (direct) RPA correlation energy, respectively.
The sum of the first three terms is identical to the Hartree-Fock energy as calculated employing DFT orbitals, which is usually denoted as HF@DFT.
The last term corresponds to the RPA correlation energy as computed using DFT orbitals and orbital energies and is often referred to as RPA@DFT.
The calculation of the $E_{\text{tot}}^{\text{EXX/RPA}}$ thus requires a ground state calculation with a given DFT functional, followed by an EXX energy evaluation and a RPA correlation energy evaluation employing DFT ground state wavefunctions and orbital energies.

\subsubsection{Implementation of the quartic scaling RPA and SOS-MP2 methods}
\label{subsubsection_RPA-Implementation}
We summarize here the implementation in CP2K of the quartic scaling computation of the RPA and SOS-MP2 correlation energies. The reason to describe the two implementations here is due to the fact that the two approaches share several components. In fact, it can be shown that the direct MP2 energy can be obtained by truncating at the first non-vanishing term of the Taylor expansion to compute the logarithm in Eq.~\ref{eq:integral_imaginary_freq} and integrating over frequencies~\cite{Eshuis2010}.

After the three center RI integral matrix $B_{P}^{ia}$ is made available (via $i.e.$ RI-GPW~\cite{DelBen2013}), the key component of both methods is the evaluation of the frequency dependent $Q_{PQ}(\omega)$ for the RPA and the $\tau$ dependent $Q_{PQ}(\tau)$ for the Laplace transformed SOS-MP2 method (see section~\ref{subsubsection_SOS-MP2}). The matrices $Q_{PQ}(\omega)$ and $Q_{PQ}(\tau)$ are given by the contractions in Eqs.~\ref{eq:Qmat_RPA} and \ref{eq:Q_SOS}, respectively. 
Their computation entails, as a basic algorithmic motif, a large distributed matrix multiplication between tall and skinny matrices for each quadrature point. Fortunately, the required operations at each quadrature point are independent of each other. The parallel implementation in CP2K exploits this fact by distributing the workload for the evaluation of $Q_{PQ}(\omega)$ and $Q_{PQ}(\tau)$ over pools of processes, where each pool is working independently on a subset of quadrature points. Furthermore, the operations necessary for each quadrature point are performed in parallel within all members of the pool. In this way, the $O(N^4)$ bottleneck of the computation displays an embarrassingly parallel distribution of the workload and in fact, it shows excellent parallel scalability to several thousand nodes~\cite{DelBen2013,DelBen2015120}.
Additionally, since at the pool level the distributed matrix multiplication employs the widely adopted data layout of the parallel BLAS library, minimal modifications are required to exploit accelerators (such as graphics processing units (GPUs) and field-progammable gate arrays (FPGAs), see section~\ref{subsection_acceleration} for details), as interfaces to the corresponding accelerated libraries are made available~\cite{DelBen2015120}.


Finally, the main difference between RPA and SOS-MP2 is the postprocessing. After the contraction step to obtain the $Q_{PQ}$ matrix, the sum in Eq.~\ref{eq:E_SOS} is performed with computational costs of $\mathcal{O}(N_a^2N_q)$ for SOS-MP2 instead of $\mathcal{O}(N^3_aN_q)$, which is associated with the evaluation of the matrix logarithm of Eq.~\ref{eq:integral_imaginary_freq} for the RPA (in CP2K this operation is performed using the identity $\text{Tr} [\ln{\mathbf{A}}] = \ln(\text{det}[\mathbf{A}])$, where a Cholesky decomposition is used to efficiently calculate the matrix determinant). Therefore, the computational costs of the quartic-scaling RPA and SOS-MP2 are the same for large systems assuming the same number of quadrature points is used for the numerical integration. 

\subsubsection{Cubic Scaling RPA and SOS-MP2 method}
\label{subsubsection_Cubic-Scaling-RPA}
The scaling of RPA and SOS-MP2 can be reduced from $O(N^4)$ to $O(N^3)$, or even better by alternative analytical formulations of the methods \cite{doi:10.1063/1.4939841, cubic-rpa}. Here we describe the CP2K-specific cubic scaling RPA/SOS-MP2 implementation and demonstrate the applicability to systems containing thousands of atoms. 

For cubic scaling RPA calculations, the matrix $Q_{PQ}(\omega)$ from Eq.~\ref{eq:Qmat_RPA} is transformed to imaginary-time $Q_{PQ}(\tau)$~\cite{doi:10.1021/ct5001268}, as it is already present in SOS-MP2. The tensor $B_P^{ia}$ is transformed from occupied-virtual MO pairs $ia$ to pairs $\mu \nu$ of AO basis set functions. This decouples the sum over occupied and virtual orbitals and thereby reduces the formal scaling from quartic to cubic. Further requirements for a cubic scaling behavior are the use of localized atomic Gaussian basis functions and the localized overlap RI metric such that the occurring 3-center integrals are sparse. A sparse representation of the density matrix is not a requirement for our cubic scaling implementation but it reduces the effective scaling of the usually dominant $O(N^2)$ sparse tensor contraction steps to $O(N)$~\cite{doi:10.1063/1.4939841}.


The operations performed for the evaluation of $Q_{PQ}(\tau)$ are generally speaking contractions of sparse tensors of ranks 2 and 3 - starting from the 3-center overlap integrals and the density matrix. Consequently, sparse linear algebra is the key to good performance, as opposed to the quartic scaling implementation that relies mostly on parallel BLAS for dense matrix operations.

The cubic scaling RPA/SOS-MP2 implementation is based on the  distributed block compressed sparse row (DBCSR) library~\cite{Borstnik2014}, which is described in detail in section \ref{sec:DBCSR} and was originally co-developed with CP2K,  to enable linear scaling DFT~\cite{VandeVondele2012}. The library was extended with a tensor API in a recent effort to make it more easily applicable to algorithms involving contractions of large sparse multi-dimensional tensors. Block-sparse tensors are internally represented as DBCSR matrices, whereas tensor contractions are mapped to sparse matrix-matrix multiplications. An in-between tall-and-skinny matrix layer reduces memory requirements for storage and reduces communication costs for multiplications by splitting the largest matrix dimension and running a simplified variant of the CARMA algorithm~\cite{6569817}. The tensor extension of the DBCSR library leads to significant improvements in terms of performance and usability compared to the initial implementation of the cubic scaling RPA~\cite{cubic-rpa}.

\begin{figure}
\includegraphics[width=0.5\textwidth]{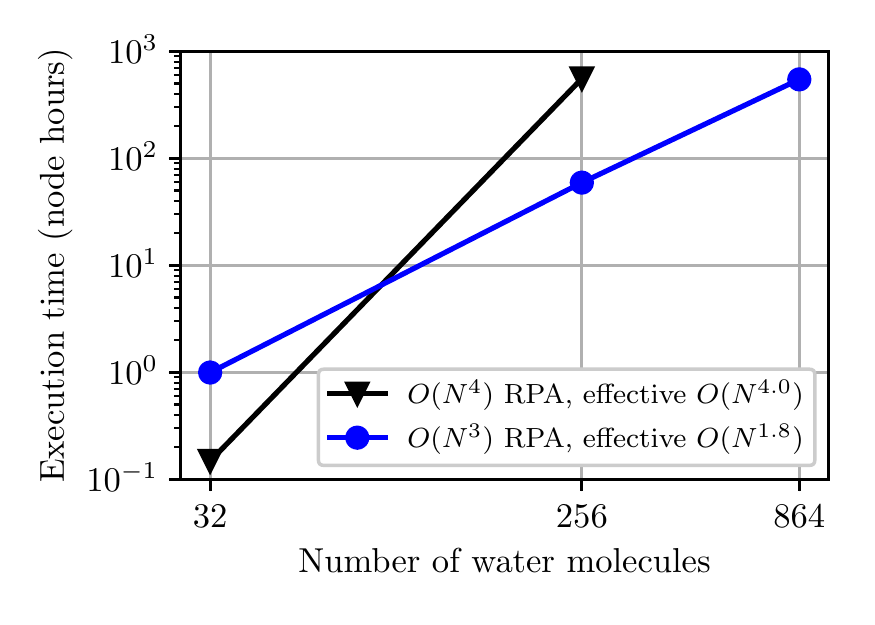}
\caption{Comparison of the timings for the calculation of the RPA correlation energy using the quartic- and the cubic-scaling implementation on a CRAY XC50 machine with 12 cores per nodes (flat MPI). The system sizes are $n\times n \times n$ supercells (with $n=1,2,3$) of a unit cell with 32 water molecules and a density of 1~g/cm$^3$. The intersection point where the cubic scaling methods becomes favorable is at approximately 90 water molecules. The largest system tested with the cubic scaling RPA contains 864 water molecules (6912 electrons, 49248 primary basis functions and 117 504 RI basis functions) and was calculated on 256 compute nodes (3072 cores). The largest tensor of size $117 504 \times 49248 \times 49248$ has an occupancy below $0.2\%$.}
\label{fig:rpa-scaling}
\end{figure}
In Fig.~\ref{fig:rpa-scaling}, we compare the computational costs of the quartic and the cubic scaling RPA energy evaluation for periodic water systems of different sizes. All calculations use the RI together with the overlap metric and a high-quality cc-TZVP primary basis with a matching RI basis. The neglect of small tensor elements in the $O(N^3)$ implementation is controlled by a filtering threshold parameter. This parameter has been chosen such that the relative error introduced in the RPA energy is below $0.01\%$. The favorable effective scaling of $O(N^{1.8})$ in the cubic scaling implementation leads to better absolute timings for systems of 100 or more water molecules. At 256 water molecules, the cubic scaling RPA outperforms the quartic scaling method by one order of magnitude.

The observed scaling is better than cubic in this example because the $O(N^3)$ scaling steps have a small prefactor and would start to dominate in systems larger than the ones presented here - they make up for around $20\%$ of the execution time for the largest system. The dominant sparse tensor contractions are quadratic scaling, closely matching the observed scaling of $O(N^{1.8})$. It is important to mention that the density matrices are not yet becoming sparse for these system sizes. Lower dimensional systems with large extend in one or two dimensions have an even more favorable scaling regime of $O(N)$ since the onset of sparse density matrices occurs at smaller system sizes. 

All aspects of the comparison discussed here also apply to SOS-MP2 because it shares the algorithm and implementation of the dominant computational steps with the cubic scaling RPA method. In general, the exact gain of the cubic scaling RPA/SOS-MP2 scheme depends on the specifics of the applied basis sets (locality and size). The effective scaling, however, is $O(N^3)$ or better for all systems, irrespective of whether the density matrix has a sparse representation, thus extending the applicability of the RPA to large systems containing thousands of atoms.

\subsection{Ionization potentials and electron affinities from \textit{GW}}
\label{subsection_gw}

\begin{figure} 
    \includegraphics[width=0.99\linewidth]{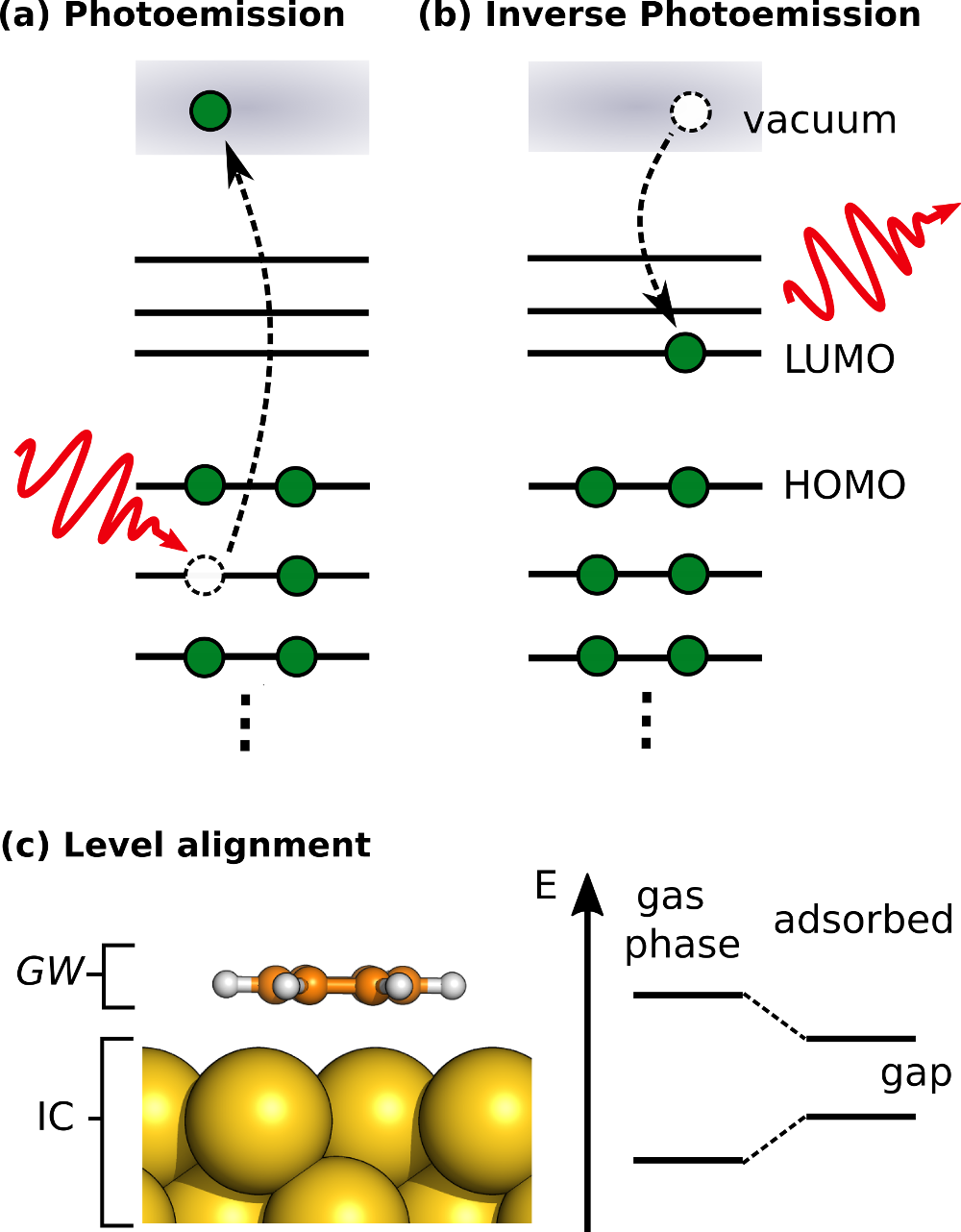}
	\caption{Properties predicted by the $GW$ method: Ionization potentials and electron affinities as measured by (a) photoemission and (b) inverse photoemission spectroscopy, respectively. (c) Level alignment of the HOMO and LUMO upon adsorption of a molecule at metallic surfaces, which are accounted for in CP2K by an image charge (IC) model avoiding the explicit inclusion of the metal in the $GW$ calculation. \label{fig:PES_IPES}
        }
\end{figure}
The $GW$ approach, 
which allows to approximately calculate the self-energy of a many-body system of electrons, has become a popular tool in theoretical spectroscopy to predict electron removal and addition energies as measured by direct and inverse photoelectron spectroscopy, respectively, see Figs.~\ref{fig:PES_IPES}(a) and (b)~\cite{HedinGW, PRBLouieBandstruc, GWreview}. In direct photoemission the electron is ejected from orbital $\psi_n$ to the vacuum level by irradiating the sample with visible, ultraviolet or X-rays, whereas in the inverse photoemission process an injected electron undergoes a radiative transition into an unoccupied state. 

The $GW$ approximation has been applied to a wide range of materials including two-dimensional systems, surfaces and molecules. A summary of applications and an introduction to the many-body theory behind $GW$ and practical aspects can be found in a  recent review article~\cite{GWreview}. The electron removal energies are  referred to as ionization potentials (IPs) and the negative of the electron addition energies as electron affinities (EAs); see Ref.~\onlinecite{GWreview} for details on the sign convention. The $GW$ method yields a set of energies $\{\varepsilon_n\}$ for all occupied and unoccupied orbitals $\{\psi_n\}$. For occupied states, $\varepsilon_n$ can be directly related to the IP and for the lowest unoccupied state (LUMO) to the EA. Hence, 
\begin{align}
    \mathrm{IP}_n = -\varepsilon_n, \hspace{0.1cm} n\in \mathrm{occ} \quad \& \quad
    \mathrm{EA}  = -\varepsilon_{\mathrm{LUMO}}.
    \label{eq:ea_ip}
\end{align}
 The difference between the IP of the highest occupied state (HOMO) and the EA is called the fundamental gap, a critical parameter for charge transport in, e.g, organic semiconductors~\cite{Kippelen2009}.
It should be noted that the fundamental HOMO-LUMO gap does not correspond to the first optical excitation  energy (also called optical gap) that is typically smaller than the fundamental gap due to the electron-hole binding energy~\cite{Baerends2013}.
For solids, the Bloch functions~$\psi_{n\mathbf{k}}$ carry an additional quantum number~$\mathbf{k}$ in the first Brillouin zone and give rise to a bandstructure~$\varepsilon_{n\mathbf{k}}$~\cite{PRBLouieBandstruc}. From the bandstructure we can determine the band gap, which is the solid-state equivalent to the HOMO-LUMO gap. Unlike for molecules, an angle-resolved photoemission experiment is required to resolve the $\mathbf{k}$-dependence.\par
For $GW$, mean absolute deviations of less than 0.2~eV from the higher-level CCSD(T) reference have been reported for IP$_{\mathrm{HOMO}}$ and EA~\cite{Gallandi2016,GWbenchmark}. The deviation from the experimental reference can be even reduced to $<0.1$~eV when including also vibrational effects~\cite{Gallandi2015}. On the contrary, DFT eigenvalues~$\varepsilon_n^\text{DFT}$ fail to reproduce spectroscopic properties. Relating $\varepsilon_n^\text{DFT}$ to the IPs and EA as in Eq.~\ref{eq:ea_ip}, is conceptually only valid for the HOMO level~\cite{Janak}.
Besides the conceptual issue, IPs from DFT eigenvalues are underestimated by several eV compared to experimental IPs due to the self-interaction error in GGA and LDA functionals yielding far too small HOMO-LUMO gaps~\cite{XavierBlase,Ke2011}. Hybrid XC functionals can improve the agreement with experiment, but the amount of exact exchange can strongly influence~$\varepsilon_n^\text{DFT}$ in an arbitrary way.\par
The most common $GW$ scheme is the $G_0W_0$ approximation, where the $GW$ calculation is performed non-self-consistently on top of an underlying DFT calculation. In $G_0W_0$, the MOs from DFT $\psi_n^\text{DFT}$ are employed and the DFT eigenvalues are corrected by replacing the incorrect XC contribution~$\braket{\psi_n^\text{DFT}|v^\text{xc}|\psi_n^\text{DFT}}$ by the more accurate energy-dependent XC self-energy~$\braket{\psi_n^\text{DFT}|\Sigma(\varepsilon)|\psi_n^\text{DFT}}$, i.e. 
\begin{align}
 \varepsilon_n = \varepsilon_n^\text{DFT} + \left<\left.\psi_n^\text{DFT}\right| \Sigma (\varepsilon_n) - v^\text{XC} \left|\psi^\text{DFT}_n\right.\right>.
\end{align}
The self-energy is computed from the Green's function $G$ and the screened Coulomb interaction $W$, $\Sigma=GW$, hence the name of the $GW$ approximation~\cite{GWreview}. \par
The $G_0W_0$ implementation in CP2K~\cite{GWCP2K} works routinely for molecules despite first attempts also have been made for periodic systems~\cite{periodicGWCP2K,Wilhelm2019,Yi2018}.
The standard $G_0W_0$ implementation is optimized for computing valence orbital energies~$\varepsilon_n$ (e.g. up to 10 eV below the HOMO and 10 eV above the LUMO)~\cite{GWCP2K}. The frequency integration of the self-energy is based on the analytic continuation using either the 2-pole model~\cite{Rojas1995}, or Pad\'{e} approximant as analytic form~\cite{Vidberg1977}. For core levels, more sophisticated implementations are necessary~\cite{gwaimsCD, Golze2020}. The standard implementation scales with $\mathcal{O}(N^4)$ with respect to system size $N$ and enables the calculation of molecules up to a few hundreds atoms on parallel supercomputers. Large molecules exceeding thousand atoms can be treated by the low-scaling $G_0W_0$ implementation within CP2K~\cite{lowscalingGW}, which effectively scales with $\mathcal{O}(N^2)$ to $\mathcal{O}(N^3)$. \par
The $GW$ equations should be in principle solved self-consistently. However, a fully self-consistent treatment is computationally very expensive. In CP2K, an approximate self-consistent scheme is available, where the wavefunctions $\psi_n^\text{DFT}$ from DFT are employed and only the eigenvalues in $G$ and $W$ are iterated, for details see Refs.~\onlinecite{XavierBlase,GWCP2K,GWreview}. This eigenvalue-self-consistent scheme (ev$GW$) includes more physics than $G_0W_0$, but is still computationally tractable. Depending on the underlying DFT functional, ev$GW$ improves the agreement of the HOMO-LUMO gaps with experiment by $0.1-0.3$~eV compared to $G_0W_0$~\cite{XavierBlase,GWCP2K}. For lower lying states the improvement over $G_0W_0$ might be not as consistent~\cite{Marom2012}.\par
Applications of the $GW$ implementation in CP2K have been focused on graphene-based nanomaterials on gold surfaces complementing scanning tunneling spectroscopy (STS) with $GW$ calculations to validate the molecular geometry and to obtain information about the spin configuration~\cite{applicationlowscalingGW,*urgel2019surface,*xusurface,*wangsurface,*digiovannsurface}.
The excitation process generates an additional charge on the molecule.
As a response, an image charge (IC) is formed inside the metallic surface which causes occupied states to move up in energy and unoccupied states to move down. The HOMO-LUMO gap of the molecule is thus significantly lowered on the surface compared to the gas phase, see Fig.~\ref{fig:PES_IPES}(c).
This effect has been accounted for by an IC  model~\cite{Neatonmodel}, which is implemented in CP2K. Adding the IC correction on-top of a gas phase ev$GW$ calculation of the isolated molecule, CP2K can efficiently compute HOMO-LUMO gaps of physisorbed molecules on metal surfaces~\cite{applicationlowscalingGW,*urgel2019surface,*xusurface,*wangsurface,*digiovannsurface}.
%

\section{Density Functional Perturbation Theory}
\label{section_DFPT}
Several experimental observables are measured by perturbing the system and then observing its response, hence they can be obtained  as derivatives of the energy or density with respect to some specific external perturbation. 
In the common perturbation approach, the perturbation is included in the Hamiltonian, i.e. as an external potential, then the electronic structure is obtained by applying the variational principle and the changes on the expectation values are evaluated. The perturbation Hamiltonian defines the specific problem. 
The perturbative approach applied in the framework of DFT turns out to perform better than the straightforward numerical methods, where the total energy is computed after actually perturbing the system. 
For all kinds of properties related to derivatives of the total energy, DFPT is derived from the following extended energy functional
\begin{equation}
\mathcal{E}[\{\phi_i\}] =  \mathcal{E}_{\text{KS}}[\{\phi_i\}] + \lambda \mathcal{E}_{\text{pert}} [\{\phi_i\}],
\end{equation}
where the external perturbation is added in the form of a functional and $\lambda$ is a small perturbative parameter representing the strength of the interaction with the static external field~\cite{Gonze1995,Putrino2000}. The  minimum of the functional is expanded perturbatively in powers of $\lambda$ as 
\begin{equation}
E = E^{(0)} + \lambda E^{(1)} + \lambda^2 E^{(2)} + \dots,
\end{equation}
whereas the corresponding minimising orbitals are
\begin{equation}
\phi_i = \phi_i^{(0)} + \lambda \phi_i^{(1)} + \lambda^2 \phi_i^{(2)} + \dots.
\end{equation}
If the expansion of the wavefunction up to an order $(n-1)$ is known, then the variational principle for the $2n$th-order derivative of the energy is given by 
\begin{equation}
E^{(2n)} = \min_{\phi_i^{(n)}} \left\{ \mathcal{E}\left[\sum_{k=0}^n \lambda^k \phi_i^{(k)}\right]\right\}
\end{equation}
under the constraint
\begin{equation}
\sum_{k=0}^n \langle \phi_i^{(n-k)}|\phi_j^{(k)}\rangle = 0.
\end{equation}
For zero-order, the solution is obtained from the ground state KS equations. The second-order energy is variational in the first-order wavefunction and is obtained as 
\begin{eqnarray}
E^{(2)}[\{\phi_i^{(0)}\},\{\phi_i^{(1)}\}] & = & \sum_{ij}\langle\phi_i^{(1)}| \mathcal{H}^{(0)}\delta_{ij} -\Lambda_{ij}|\phi_j^{(1)} \rangle \\
 & + & \sum_i \left[ \langle \phi_i^{(1)} | \frac{\delta \mathcal{E}_{\text{pert}}}{\delta \langle \phi_i^{(0)}|} +\frac{\delta \mathcal{E}_{\text{pert}}}{\delta |\phi_i^{(0)} \rangle} |\phi_i^{(1)}\rangle\right] \nonumber \\
 & + & \frac{1}{2}\int d{\bf r} \int d{\bf r}^\prime \mathcal{K}({\bf r},{\bf r}^\prime)n^{(1)}({\bf r})n^{(1)}({\bf r}^\prime). \nonumber
\end{eqnarray}
The electron density is also expanded in powers of $\lambda$ and the first-order term reads as
\begin{equation}
n^{(1)}({\bf r}) = \sum_i f_i [\phi^{(0)*}_i({\bf r})\phi^{(1)}_i({\bf r})+\phi^{(1)*}_i({\bf r})\phi^{(0)}_i({\bf r})].
\end{equation}
The Lagrange multipliers $\Lambda_{ij}$ are the matrix elements of the zeroth-order Hamiltonian, which is the KS Hamiltonian. Hence, 
\begin{equation}
\Lambda_{ij} = \langle \phi^{(0)}_i |\mathcal{H}^{(0)}|\phi^{(0)}_j\rangle,
\end{equation}
and the second-order energy kernel is
\begin{equation}
\mathcal{K}({\bf r},{\bf r}^\prime) = \left. \frac{\delta^2 \mathcal{E}_{\text{Hxc}}[n]}{\delta n({\bf r}) \delta n({\bf r}^\prime)}\right|_{n^{(0)}},
\end{equation}
where $\mathcal{E}_{\text{Hxc}}$ represents the sum of the Hartree and the XC energy functionals. Thus, the evaluation of the kernel requires the second-order functional derivative of the XC functionals. 
 
The second-order energy is variational with respect to the $\{\phi_i^{(1)}\}$, where the orthonormality condition of the total wavefunction gives at the first-order
\begin{equation}
\langle \phi^{(0)}_i|\phi^{(1)}_j \rangle + \langle \phi^{(1)}_i|\phi^{(0)}_j \rangle = 0. \qquad \forall i,j
\end{equation}
This also implies the conservation of the total charge.

The perturbation functional can often be written as the expectation value of a perturbation Hamiltonian 
\begin{equation}
\mathcal{H}^{(1)} = \frac{\delta \mathcal{E}_{\text{pert}}}{\delta \langle \phi_i^{0}|} + \int  d{\bf r}^\prime \mathcal{K}({\bf r},{\bf r}^\prime)n^{(1)}({\bf r}^\prime). 
\end{equation}
However, the formulation through an arbitrary functional also allows orbital specific perturbations. 
The stationary condition then yields the inhomogeneous, non-linear system for $\{\phi_i^{(1)} \}$, i.e.
\begin{widetext}
\begin{equation}
-\sum_j \left( \mathcal{H}^{(0)}\delta_{ij} -\Lambda_{ij} \right)|\phi_j^{(1)} \rangle = \mathcal{P} \bigg( \int d{\bf r}^\prime \mathcal{K}({\bf r},{\bf r}^\prime)n^{(1)}({\bf r}^\prime) |\phi^{(0)}_i\rangle + \frac{\delta \mathcal{E}_{\text{pert}}}{\delta \langle \phi_i^{0}|} \bigg),
\end{equation}
\end{widetext}
where $\mathcal{P} = 1 - \sum_j | \phi_j^{(0)} \rangle\langle \phi_j^{(0)} |$ is the projector upon the unoccupied states. Note that the right hand side still depends on the $\{\phi_i^{(1)} \}$ via the perturbation density $n^{(1)}$.
In our implementation the problem is solved directly using a  preconditioned conjugate-gradient (CG) minimisation algorithm. 

\subsection{Polarizability}
\label{subsection_polarizability}
One case where the perturbation cannot be expressed in a Hamiltonian form is the presence of an external electric-field, which couples with the electric polarisation ${\bf P}^{\text{el}}= e \langle {\bf r} \rangle $, where i$\langle {\bf r} \rangle $ is the expectation value of the position operator for the  system of $N$ electrons. In the case of periodic systems, the position operator is ill-defined, and we use the modern theory of polarization in the $\Gamma$-point-only to write the perturbation in terms of the Berry phase
\begin{equation}
\gamma_\alpha = \operatorname{Im}\log \det {\bf Q}^{(\alpha)}, 
\end{equation}
where the matrix is defined as 
\begin{equation}
 {\bf Q}^{(\alpha)}_{ij} = \langle \phi_i | \exp[\text{i} 2\pi {\bf h}_{\alpha}^{-1}\cdot {\bf r}]| \phi_j \rangle
\end{equation}
and ${\bf h}=[{\bf a},{\bf b},{\bf c}]$ is the $3 \times 3$ matrix defining the simulation cell and ${\bf h}_\alpha = (a_\alpha, b_{\alpha}, c_{\alpha})$~\cite{Resta1994,Resta1998}. The electric dipole is then given by 
\begin{equation}
P^{\text{el}}_\alpha = \frac{e}{2\pi}  h_\alpha \gamma_\alpha.
\end{equation}
Through the coupling to an external electric-field ${\bf E}^{\text{ext}}$, this induces a perturbation of the type
\begin{equation}
\lambda \mathcal{E}_{\text{pert}}[\{\phi_i\}] = -\sum_\alpha E^{\text{ext}}_\alpha   P^{\text{el}}_\alpha,
\end{equation}
where the perturbative parameter is the field component $E^{\text{ext}}_\alpha $. The functional derivative $\delta \mathcal{E}_{\text{pert}}/\delta \langle \phi_I^{(0)}|$ can be evaluated using the formula for the derivative of a matrix with respect to a generic variable~\cite{Putrino2000}, which gives the perturbative term as
\begin{equation}
\frac{\delta \log \det  {\bf Q}^{(\alpha)} }{\delta \langle \phi_I^{(0)}|} = \sum_j \left( {\bf Q}^{(\alpha)} \right)_{ij}^{-1} \exp[\text{i} 2\pi {\bf h}_{\alpha}^{-1}\cdot {\bf r}]| \phi_j \rangle.
\end{equation}
Once the first-order correction to the set of the KS orbitals has been calculated, the induced polarization is 
\begin{widetext}
\begin{equation}
\delta P^{\text{el}}_\alpha = -\sum_\beta \frac{e}{2\pi}  h_\beta  \operatorname{Im}\left[ \sum_{ij} \left( \langle \phi_i^{\beta(1)}|  \exp[\text{i} 2\pi {\bf h}_{\alpha}^{-1}\cdot {\bf r}] | \phi_j^{(0)}\rangle + \langle \phi_i^{(0)}|  \exp[\text{i} 2\pi {\bf h}_{\alpha}^{-1}\cdot {\bf r}] | \phi_j^{\beta(1)}\rangle \right) \left( {\bf Q}^{(\alpha)} \right)_{ij}^{-1}  \right]E_{\beta},
\end{equation}
\end{widetext}
while the elements of the polarizability tensor are obtained as $\alpha_{\alpha\beta} = \partial P^{\text{el}}_\alpha/\partial E_\beta$.

The polarizability can be looked as the deformability of the electron cloud of the molecule by the electric-field. In order for a molecular vibration to be Raman active, the vibration must be accompanied by a change in the polarizability.  In the usual Placzeck theory, ordinary Raman scattering intensities can be expressed in terms of the isotropic transition polarizability $\alpha^{\text{i}}= \frac{1}{3}\text{Tr}[\alpha]$, and the anisotropic transition polarizability $\alpha^{\text{a}} = \sum_{\alpha\beta} \frac{1}{2}(3\alpha_{\alpha\beta}\alpha_{\alpha\beta}-\alpha_{\alpha\alpha}\alpha_{\beta\beta})$~\cite{Putrino2002}. The Raman scattering cross section can be related to the dynamical autocorrelation function of the polarazability tensor. Along AIMD simulations, the polarizability can be calculated as a function of time~\cite{PartoviAzar2015a,PartoviAzar2015b}. As the vibrational spectra are obtained by the temporal Fourier transformation of the velocity autocorrelation function, and the IR spectra from that of the dipole autocorrelation function, the depolarized Raman intensity can be calculated from the autocorrelation of the polarizability components~\cite{Luber2014}.

\subsection{Nuclear magnetic resonance and electron paramagnetic resonance spectroscopy}
\label{subsection_nmr_epr}
The development of the DFPT within the GAPW formalism allows for an all-electron description, which is important when the induced current density generated by an external static magnetic perturbation is calculated. The so induced current density determines at any nucleus $A$ the nuclear magnetic resonance (NMR) chemical shift
\begin{equation}
\sigma_{\alpha\beta}^A = \frac{1}{c} \int \left[ \frac{{\bf r}-{\bf R}_A}{|{\bf r}-{\bf R}_A|^3} \times {\bf j}_\alpha \right]_\beta d{\bf r}
\end{equation}
and, for systems with net electronic spin $1/2$, the electron paramagnetic resonance (EPR) ${\bf g}$-tensor 
\begin{equation}
 g_{\alpha \beta} = g_e\delta_{\alpha \beta}+ \Delta g_{\alpha \beta}^{\text{ZKE}}+ \Delta g_{\alpha \beta}^{\text{SO}}+ \Delta g_{\alpha \beta}^{\text{SOO}}.
\end{equation}
In the above expressions, ${\bf R}_A$ is the position of the nucleus, ${\bf j}_\alpha$ is the current density induced by a constant external magnetic-field applied along the $\alpha$ axis, and $g_e$ is the free electron $g$-value. Among the different contributions to the ${\bf g}$-tensor, the current density dependent ones are the spin-orbit (SO) interaction 
\begin{equation}
\Delta g_{\alpha \beta}^{\text{SO}} = \frac{g_e -1}{c}\int [{\bf j}^{{\uparrow}}_\alpha({\bf r })\times {\boldsymbol \nabla}V_{\text{eff}}^{\uparrow}(\bf r) - {\bf j}^{\downarrow}_\alpha({\bf r })\times {\boldsymbol \nabla}V_{\text{eff}}^{\downarrow}(\bf r)]_\beta d{\bf r}
\end{equation}
and the spin-other-orbit (SOO) interaction 
\begin{equation}
\Delta g_{\alpha \beta}^{\text{SOO}} = 2\int B_{\alpha\beta}^{\text{corr}}(\bf r) n^{\text{spin}}({\bf r}) d{\bf r}, 
\end{equation}
where 
\begin{equation}
 \qquad B_{\alpha\beta}^{\text{corr}}(\bf r) = \frac{1}{c} \int\left[ \frac{{\bf r}^\prime-{\bf r}}{|{\bf r}^\prime-{\bf r}|^3} \times ({\bf j}_\alpha({\bf r}^\prime)-{\bf j}_\alpha^{\text{spin}}({\bf r}^\prime))\right]_\beta d{\bf r}^\prime,
\end{equation}
which also depends on the spin density $n^{\text{spin}}$ and the spin-current density ${\bf j}^{\text{spin}}$. Therein, $V_{\text{eff}}^{\uparrow}$ is an effective potential in which the spin up electrons are thought to move (similarly $V_{\text{eff}}^{\downarrow}$ for spin down electrons), whereas $ B_{\alpha\beta}^{\text{corr}}$ is the $\beta$ component of the magnetic-field originating from the induced current density along $\alpha$. The SO-coupling is the leading correction term in the  computation of the $g$-tensor. It is relativistic in origin and therefore becomes much more important for heavy elements.
In the current CP2K implementation the SO term is obtained by integrating the induced spin-dependent current densities and the gradient of the  effective potential over the simulation cell. The effective one-electron operator replaces the computationally demanding two-electrons integrals~\cite{Schreckenbach1997}.
A detailed discussion on the impact of the various relativistic and SO approximations, which are implemented in the various codes, is provided by Van Yperen-De Deyne et al.~\cite{Deyne2012}.

In the GAPW representation, the induced current density is decomposed with the same scheme applied for the electron density distinguishing among the soft contribution to the total current density, and the local hard and local soft contributions, i.e. 
\begin{equation}
{\bf j}({\bf r}) = \tilde{\bf j}({\bf r})  + \sum_{A} \left({\bf j}_A({\bf r})  + \tilde{\bf j}_A({\bf r})\right). 
\end{equation}
In the linear response approach, the perturbation  Hamiltonian at the first-order in the field strength is
\begin{equation}
\mathcal{H}^{(1)} = \frac{e}{m} {\bf p} \cdot {\bf A}({\bf r}),
\end{equation}
where ${\bf p}$ is the momentum operator and ${\bf A}$ is the vector potential representing the field ${\bf B}$. Thus, 
\begin{equation}
{\bf A}({\bf r}) = \frac{1}{2}({\bf r}-{\bf d}({\bf r})) \times {\bf B}, 
\end{equation}
with the cyclic variable ${\bf d}({\bf r})$ being the gauge origin. The induced current density is calculated as the sum of orbital contributions ${\bf j}_i$ and can be separated in a diamagnetic term ${\bf j}^{\text{d}}_i({\bf r}) = \frac{e^2}{m}{\bf A}({\bf r})|\phi^{(0)}_i({\bf r})|^2$, and a paramagnetic term ${\bf j}^{\text{p}}_i({\bf r}) = \frac{e^2}{m}\langle \phi_i^{(0)}| [{\bf p} | {\bf r}\rangle \langle  {\bf r}| + | {\bf r}\rangle \langle  {\bf r}| {\bf p} ] | \phi_i^{(1)} \rangle$. Both contributions individually are gauge dependent, whereas
the total current is gauge-independent. The position operator appears in the definition of the perturbation operators and of the current density. In order to be able to deal with periodic systems, where the multiplicative position operator is not a valid operator, first we perform a unitary transformation of the ground state orbitals to obtain their maximally localised Wannier functions (MLWFs) counterpart~\cite{Marzari2012}. Hence, we use the alternative definition of the position operator, which is unique for each localised orbital and showing a sawtooth-shaped profile centered at the orbital's Wannier center~\cite{Sebastiani2001}.

Since we work with real ground state MOs, in the unperturbed state, the current density vanishes. Moreover, the first-order perturbation wavefunction is purely imaginary, which results in a vanishing first-order change in the electronic density $n^{(1)}$.
This significantly simplifies the perturbation energy functional, since the second-order energy kernel can be skipped. The  system of linear equations to determine the matrix of the expansion coefficients of the linear response orbitals ${\bf C}^{(1)}$ reads as
\begin{equation}
-{\text{i}}\sum_{i \mu}\left( \mathcal{H}^{(0)}_{\nu\mu}\delta_{ij} -S_{\nu\mu}\langle \phi_i^{(0)}| \mathcal{H}^{(0)}| \phi_j^{(0)} \rangle \right) C_{\mu i} = \sum_\mu \mathcal{H}^{(1)}_{\nu\mu(j)} C^{(0)}_{\mu j}, 
\end{equation}
where  $i$ and $j$ are the orbital indexes, $\nu$ and $\mu$ are basis set function indexes, and $S_{\nu\mu}$ is an element of the overlap matrix. The optional subindex $(j)$, labelling the matrix element of the perturbation operator, indicates that the perturbation might be orbital dependent. 
In our CP2K implementation~\cite{Weber_JCP_2009}, the formalism proposed by Sebastiani et al. is employed~\cite{Sebastiani2001}, i.e. we split the perturbation in three different types of operators, which are ${\bf L} = ({\bf r}-{\bf d}_j)\times {\bf p}$, the orbital angular momentum operator ${\bf p}$, the momentum operator and $\boldsymbol \Delta_i = ({\bf d}_i-{\bf d}_j)\times {\bf p}$, the full correction operator. The vector $ {\bf d}_j$ is the Wannier center associated with the unperturbed $j$-th orbital, thus making ${\bf L}$ and $\boldsymbol \Delta_i $ dependent on the unperturbed orbital to which they are applied. By using the Wannier center as relative origin in the definition of ${\bf L}$, we introduce an individual reference system, which is then corrected by the $\boldsymbol \Delta_i $. As a consequence, the response orbitals are given by nine sets of expansion coefficients, as for each operator all three Cartesian components need to be individually considered. The evaluation of the orbital angular momentum contributions and of the momentum contributions can be done at the computational cost of just one total energy calculation. The full correction term, instead, requires one such calculation for each electronic state. This term does not vanish unless all ${\bf d}_i$ are equal. However, in most circumstances this correction is expected to be small, since it describes the reaction of the state $i$ to the perturbation of state $j$, which becomes negligible when the two states are far away. 
Once all contributions have been calculated, the $x$-component of the current density induced by the magnetic-field along $y$
 is
\begin{widetext}
\begin{eqnarray}
     j_{xy}({\bf r}) & = & -\frac{1}{2c} \sum_{i\nu\mu} \left[ C_{\nu i}^{(0)}  (C^{L_y}_{\mu i} + ({\bf d}({\bf r})-{\bf d}_i)_x C^{p_z}_{\mu i} - ({\bf d}({\bf r})-{\bf d}_i)_z C^{p_x}_{\mu i} -C^{\Delta i_y}_{\mu i}) \times \left(\nabla_x\varphi_\nu({\bf r})\varphi_\mu({\bf r}) - \varphi_\nu({\bf r})\nabla_x \varphi_\mu({\bf r})\right)  \right] \nonumber \\
     &+& ({\bf r} -{\bf d}({\bf r}))_z n^{(0)}({\bf r})),
\end{eqnarray}
\end{widetext}
where the first term is the paramagnetic contribution and the second the diamagnetic one.

The convergence of the magnetic properties  with respect to Gaussian basis set size is strongly dependent on the choice of the gauge. The available options in CP2K are the individual gauge for atoms in molecules (IGAIM) approach introduced by Keith and Bader~\cite{Keith1992}, and the continuous set of gauge transformation (CSGT) approach~\cite{Keith1993}. The diamagnetic part of the current density vanishes identically when the CSGT approach is used, i.e. ${\bf d}({\bf r} = {\bf r})$. Yet, this advantage is weakened by the rich basis set required  to obtain an accurate description of the current density close to the nuclei, which typically affects the accuracy within the NMR chemical shift. In the IGAIM approach, however, the gauge is taken at the closer nuclear center. Large-scale computations of NMR chemical shifts for extended paramagnetic solids are reported by Mondal et al.~\cite{Mondal_JCTC_2018}. They show that the contact, pseudocontact, and orbital-shift contributions to paramagnetic NMR can be calculated by combining hyperfine couplings obtained with hybrid functionals with g-tensors and orbital shieldings computed using gradient-corrected functionals.
%

\section{Time-Dependent Density Functional Theory}
\label{section_tddft}
The dynamics and properties of many-body systems in the presence of time-dependent potentials, such as electric or magnetic fields, can be investigated via TD-DFT.

\subsection{Linear-response time-dependent density functional theory}
\label{subsection_lr-tddft}
Linear-response TD-DFT (LR-TDDFT)~\cite{Casida1995}
is an inexpensive correlated method to compute vertical transition energies
and oscillator strengths between the ground and singly-excited electronic states. Optical properties are computed as a linear-response of the system to a perturbation caused by an applied weak electro-magnetic field.

The current implementation relies on Tamm-Dancoff and adiabatic approximations~\cite{Strand:2019}.
The Tamm-Dancoff approximation ignores electron de-excitation channels~\cite{Fetter1971,Hutter:2003:tda}, thus reducing the  LR-TDDFT equations to a standard Hermitian eigenproblem~\cite{Hirata:1999}, i.e. 
\begin{equation}\label{eq:TDDFPT_TDA}
   A X = \omega X,
\end{equation}
where $\omega$ is a transition energy, $X$ a response eigenvector, and $A$ a response operator. 
In addition, the adiabatic approximation postulates independence of the employed XC functional on time and leads to the following form for the matrix elements of the response operator~\cite{Dreuw:2003}:
\begin{align}
  A_{ia\sigma,jb\tau}
  &= \delta_{ij}\delta_{ab}\delta_{\sigma\tau}(\epsilon_{a\sigma} - \epsilon_{i\sigma}) +
     (i_\sigma a_\sigma \vert j_\tau b_\tau) \label{eq:Amatrix} \\
  &- c_\mathrm{HFX} \delta_{\sigma\tau} (i_\sigma j_\sigma \vert a_\tau b_\tau) + (i_\sigma a_\sigma \vert f_{\mathrm{xc};\sigma\tau} \vert j_\tau b_\tau). \nonumber
\end{align}
In the above equation, the indices $i$ and $j$ stand for occupied spin-orbitals, whereas $a$ and $b$ indicated virtual spin-orbitals,
and $\sigma$ and $\tau$ refers to specific spin components. The terms $(i_\sigma a_\sigma \vert j_\tau b_\tau)$ and $(i_\sigma a_\sigma \vert f_{\mathrm{xc};\sigma\tau} \vert j_\tau b_\tau)$ are standard electron repulsion and XC integrals over KS orbital functions $\{\phi(\mathbf{r})\}$ with corresponding KS orbital energies $\epsilon$, hence 
%
%
\begin{subequations}
\begin{equation}
  (i_\sigma a_\sigma \vert j_\tau b_\tau) = \int \varphi^{*}_{i\sigma}(\mathbf{r}) \varphi_{a\sigma}(\mathbf{r})
    \frac{1}{\vert \mathbf{r} - \mathbf{r'} \vert}
     \varphi^{*}_{j\tau}(\mathbf{r'}) \varphi_{b\tau}(\mathbf{r'}) \mathrm{d}\mathbf{r} \mathrm{d}\mathbf{r'}    
\end{equation}
and 
\begin{eqnarray}
  (i_\sigma a_\sigma \vert f_{\mathrm{xc};\sigma\tau} \vert j_\tau b_\tau) &=&  \int \varphi^{*}_{i\sigma}(\mathbf{r}) \varphi_{a\sigma}(\mathbf{r})
    f_{\mathrm{xc};\sigma\tau}(\mathbf{r}, \mathbf{r'}) \nonumber \\
 &\times&  \varphi^{*}_{j\tau}(\mathbf{r'}) \varphi_{b\tau}(\mathbf{r'}) \mathrm{d}\mathbf{r} \mathrm{d}\mathbf{r'}
\end{eqnarray}
\end{subequations}
Here, the XC kernel $f_{\mathrm{xc};\sigma\tau}(\mathbf{r}, \mathbf{r'})$ is simply the second functional derivative of the XC functional $E_\mathrm{xc}$ over the ground state electron density $n^\mathrm{(0)}(\mathbf{r})$~\cite{Chassaing:2005}, hence
\begin{equation}\label{eq:TDDFPT_kernel}
   f_{\mathrm{xc};\sigma\tau}(\mathbf{r}, \mathbf{r'}) = \left. \frac{\delta^2 E_\mathrm{xc}[n](\mathbf{r})}
         {\delta n_{\sigma}(\mathbf{r'}) \delta n_{\tau}(\mathbf{r'})} \right\vert_{n=n^\mathrm{(0)}}.
\end{equation}
To solve the Eq.~(\ref{eq:TDDFPT_TDA}), the current implementation uses a block Davidson iterative method~\cite{Crouzeix:1994}.
This scheme is flexible enough and allows to tune performance of the algorithm. In particular, it supports hybrid exchange functionals 
along with many acceleration techniques (see section~\ref{section_hfx}), such as integral screening~\cite{Guidon2008},
truncated Coulomb operator~\cite{Guidon2009}, and ADMM~\cite{Guidon2010}.
Whereas in most cases the same XC functional is used to compute both the ground state electron density and the XC kernel,
separate functionals are also supported.
This can be used, for instance, to apply a long-term correction to the truncated Coulomb operator during the LR-TDDFT stage~\cite{Guidon2009}, when such correction has been omitted during the reference ground state calculation.
The action of the response operator on the trial vector $X$ for a number of excited states may also be computed simultaneously.
This improves load-balancing and reduces communication costs allowing a larger number of CPU cores to be effectively utilized.

\subsubsection{Applications}
\label{subsubsection_lr-tddft_applications}
The favorable scaling and performance of the LR-TDDFPT code has been exploited to calculate the excitation energies of various systems with 1D, 2D and 3D periodicities, auch as cationic defects in aluminosilicate imogolite nanotubes~\cite{Poli:2019}, as well as surface and bulk canonical vacancy defects in MgO and HfO$_2$~\cite{Strand:2019}. Throughout, the dependence of results on the fraction of Hartree-Fock exchange was explored and the accuracy of the ADMM approximation verified. The performance was found to be comparable to ground state calculations for systems of $\approx1000$ atoms, which were shown to be sufficient to converge localized transitions from isolated defects, within these medium to wide band gap materials.

\subsection{Real-time time-dependent density functional theory and Ehrenfest dynamics}
\label{subsection_rt-tddft}
Alternatively to perturbation based methods, real-time propagation-based TDDFT is also available in CP2K.
The real-time TDDFT formalism allows to investigate non-linear effects and can be used to gain direct insights into the dynamics of processes driven by the electron dynamics.
For systems in which the coupling between of electronic and nuclear motion is of importance, CP2K provides the option to propagate cores and electrons simultaneously using the Ehrenfest scheme.
Both methods are implemented in a cubic- and linear scaling form.
The cubic scaling implementation is based on MO coefficients (MO-RTP), whereas the linear scaling version acts on the density matrix (P-RTP).\\
While the derivation of the required equations for origin independent basis functions is rather straightforward, additional terms arise for atom centred basis sets~\cite{KunertSchmidt}.
The time-evolution of the MO coefficients in a non-orthonormal Gaussian basis reads as 
\begin{equation}
\label{elProp}
 \dot a_\alpha^j = \sum_{\beta\gamma}\mathbf{S}_{\alpha\beta}^{-1} \left(i\mathbf{H}_{\beta\gamma}+\mathbf{B}_{\beta\gamma}\right)a_\alpha^j, 
\end{equation}
whereas the corresponding nuclear equations of motion is given by
\begin{equation}
\label{EOMRT}
   M_I \mathbf{\ddot{R}}_I = -\frac{\partial U(\mathbf{R}, t)}{\partial \mathbf{R}_I} + \sum_{j=1}^{N_e} \sum_{\alpha, \beta} {a_{\alpha}^j}^* \left(\mathbf{D}_{\alpha \beta}^{I} -\frac{\partial H_{\alpha \beta}}{\partial \mathbf{R}_I}  \right) a_{\beta}^j.
\end{equation}
Therein, $\mathbf{S}$ and $\mathbf{H}$ are the overlap and KS matrices, whereas $M_I$ and $R_I$ are the position and mass of ion I and U(R,t) is the potential energy of the ion–ion interaction.
The terms involving these variables represent the basis set independent part of the equations of motion.
The additional terms containing matrices $\mathbf{B}$ and $\mathbf{D}$ are arising as a consequence of the origin dependence and are defined as follows:
%
%
\begin{eqnarray}
  \mathbf{D}^{I} &=& \mathbf{B}^{I+} ( \mathbf{S}^{-1} \mathbf{H} - i \mathbf{S}^{-1} \mathbf{B}) + i {\mathbf{C}^{I+}} \nonumber \\
  &+& (\mathbf{HS}^{-1} + i \mathbf{B}^+ \mathbf{S}^{-1}) \mathbf{B}^{I} + i \mathbf{C}^{I}, 
\end{eqnarray}
with
%
%
\begin{eqnarray}
 \mathbf{B}_{\alpha\beta}=\Braket{\phi_\alpha |\frac{d}{d t} \phi_\beta } &;& \, \mathbf{B}^+_{\alpha\beta}=\Braket{\frac{d}{d t}\phi_\alpha | \phi_\beta } \nonumber \\
 \mathbf{B}^{I} = \Braket{\phi_\alpha |\frac{d}{d R^I} \phi_\beta } &;& \, \mathbf{B}^{I+} = \Braket{\frac{d}{d R^I}\phi_\alpha | \phi_\beta }\\
 \mathbf{C}^{I} = \Braket{\frac{d}{d t}\phi_\alpha |\frac{d}{d R^I} \phi_\beta } &;& \, \mathbf{C}^{I+} = \Braket{\frac{d}{d R^I}\phi_\alpha | \frac{d}{d t} \phi_\beta }. \nonumber
\end{eqnarray}
%
For simulations with fixed ionic positions, i.e. when only the electronic wavefunction is propagated, the $\mathbf{B}$ and $\mathbf{D}$ terms are vanishing and Ref.~\ref{elProp} simplifies to just the KS term.
The two most important propagators for the electronic wavefunction in CP2K are the exponential midpoint and the enforced time reversal symmetry (ETRS) propagators.
Both propagators are based on matrix exponentials. 
The explicit computation of it, however, can be easily avoided for both MO-RTP, as well as P-RTP techniques.
Within the MO-RTP scheme, the construction of a Krylov subspace $K_n(X,MO)$ together with Arnoldi's method allows for the direct computation of the action of the propagator on the MOs at a computational complexity of $O(N^2M)$.
For the P-RTP approach, the exponential is applied from both sides, which allows the expansion into a series similar to the Baker-Campbell-Hausdorff expansion.
This expansion is rapidly converging and in theory only requires sparse matrix-matrix multiplications. 
Unfortunately, pure linear scaling Ehrenfest dynamics seems to be impossible due to the non-exponential decay of the density matrix during such simulations~\cite{doi:10.1021/acs.jctc.6b00398}.
This leads to a densification of the involved matrices and eventually cubic scaling with system size. 
However, the linear scaling version can be coupled with subsystem DFT to achieve true linear scaling behaviour.\\
In subsystem DFT, as implemented in CP2K, follows the approach of Kim and Gordon~\cite{KimGordon1972,KimGordon1974}. 
In this approach, the different subsystems are minimised independently with the XC functional of choice.
The coupling between the different subsystem is added via a correction term using a kinetic energy functional: 
\begin{equation}
 E_{corr}=E_s[\rho]-\sum_i^{nsub}E_s[\rho_i], 
\end{equation}
where $\rho$ is the electron density of the full system and $\rho_i$ the one of subsystem $i$.
Using an orbital-free density functional to compute the interaction energy does not affect the structure of the overlap, which is block diagonal in this approach.
If P-RTP is applied to the Kim-Gordon density matrix, the block diagonal structure is preserved and linear scaling with the number of subsystems is achieved.
%

\section{Diagonalization-based and low-scaling Eigensolver}
\label{section_eigensolver}
After the initial Hamiltonian matrix for the selected method has been built by \cp,
like the KS matrix $\bK$ in case of a DFT-based method using the \qs\ module,
the calculation of the total (ground state) energy for the given atomic configuration 
is the next task. This requires an iterative self-consistent field (SCF) procedure, as
the Hamiltonian depends usually on the electronic density. In each SCF step, the eigenvalues
and at least the eigenvectors of the occupied MOs have to be calculated.
Various eigensolver schemes are provided by \cp\ for that task:
\begin{itemize}
 \item Traditional diagonalization (TD)
 \item Pseudo diagonalization (PD)~\cite{Stewart1982}
 \item Orbital Transformation (OT) method~\cite{VandeVondele2003}
 \item Purification methods~\cite{McWeeny_RMP_1960}
\end{itemize}
The latter method, OT, is the method of choice concerning computational efficiency and
scalability for growing system sizes. However, OT requires fixed integer MO occupations.
Therefore, it is not applicable for systems with a very small or no band gap, like metallic systems, 
which need fractional (``smeared'') MO occupations. Also, very large systems for which even the scaling
of the OT method becomes unfavorable, one has to resort to linear-scaling methods (see section~\ref{subsection_sign-method}).
By contrast, TD can also be applied with fractional MO occupations, but its cubic-scaling ($N^3$) limits
the accessible system sizes. In that respect, PD may provide significant speedups (factor of 2 or more) once a pre-converged
solution has been obtained with TD.\\
In the following, we will restrict the description of the implemented eigensolver methods to
spin-restricted systems, since the generalization to spin-unrestricted, i.e.\ spin-polarized systems
is straightforward and \cp\ can deal with both types of systems using each of these methods.

\subsection{Traditional diagonalization}
The TD scheme in \cp\ employs an eigensolver either from the parallel program library ScaLAPACK (Scalable Linear Algebra PACKage)~\cite{Choi1996},
or the ELPA library (Eigenvalue soLvers for Petascale Applications)~\cite{Auckenthaler2011,Marek2014}
to solve the general eigenvalue problem
\begin{equation}
 \bK\,\bC = \bS\,\bC\,\epsilon,
\end{equation}
where $\bK$ is the KS and $\bS$ is the overlap matrix. The eigenvectors $\bC$
represent the MO coefficients, and the $\epsilon$ are the corresponding MO eigenvalues.
Unlike to PW codes, the overlap matrix $\bS$ is not the unit matrix, since \cpqs\ employs atom-centered basis sets of non-orthogonal Gaussian-type functions
(see section~\ref{subsection_basis}).
Thus, the eigenvalue problem is transformed to its special form
\begin{subequations}
\begin{eqnarray}
 \bK\,\bC                                      &=& \bU^{\rm T}\bU\,\bC\,\epsilon\\
 (\bU^{\rm T})^{-1}\,\bK\,\bU^{-1}\,\bC^\prime &=& \bC^\prime\,\epsilon\\
 \bK^\prime\,\bC^\prime                        &=& \bC^\prime\,\epsilon \label{eq:kcce}
\end{eqnarray}
\end{subequations}
using a Cholesky decomposition of the overlap matrix
\begin{equation}
 \bS= \bU^{\rm T} \bU
\end{equation}
as the default method for that purpose.
Now, Ref.~\ref{eq:kcce} can simply be solved by a diagonalization of $\bK^\prime$.
The MO coefficient matrix $\bC$ in the non-orthogonal basis are finally obtained by the back-transformation
\begin{equation}
 \bC^\prime = \bU\,\bC.
\end{equation}
Alternatively, a symmetric L\"owdin orthogonalization instead of
a Cholesky decomposition can be applied~\cite{Loewdin1950}, i.e.
\begin{equation}
 \bU = \bS^{1/2}.
\end{equation}
On the one hand, the calculation of $\bS^{1/2}$ as implemented in \cp\ involves, however,
a diagonalization of $\bS$ which is computationally more expensive than a
Cholesky decomposition. On the other hand, however, linear dependencies in the basis
set introduced by small Gaussian function exponents can be detected
and eliminated when $\bS$ is diagonalized. Eigenvalues of $\bS$ smaller than $10^{-5}$ usually indicate significant linear dependencies in the basis set and a filtering of the corresponding eigenvectors can help to ameliorate numerical
difficulties during the SCF iteration procedure.
For small systems, the choice of the orthogonalisation has no crucial impact on the performance since it has to be performed only once for each configuration during the initialization of the SCF run.

Only the occupied MOs are required for the built-up of the density matrix $\bP$ in the AO basis
\begin{equation}
 \label{eq:P=CCT}
 \bP = 2\, \bC_{\rm occ}^{\phantom{\rm T}} \bC_{\rm occ}^{\rm T}.
\end{equation}
This saves not only memory, but also computational time since the orthonormalization of the eigenvectors is a time-consuming step. Usually only 10--20~\% of the orbitals are
occupied when standard Gaussian basis sets are employed with \cp.
\subsubsection{TD/DIIS}
The TD scheme is combined with methods to improve the convergence
of the SCF iteration procedure. The most efficient SCF convergence acceleration is achieved by the direct inversion in the iterative sub-space (DIIS) \cite{Pulay1982,Pulay1980}
exploiting the commutator relation
\begin{equation}
 \be = \bK \, \bP \, \bS - \bS \, \bP \, \bK
\end{equation}
between the KS and the density matrix, where the error matrix
$\be$ is zero for the converged density. The DIIS method can be very efficient
in the number of iterations required to reach convergence starting from a sufficiently
pre-converged density, which is significant if the cost of constructing the Hamiltonian
matrix is larger than the cost of diagonalization.

\subsubsection{TD/Broyden and Kerker mixing}
Yet, the DIIS method has frequently problems to converge for instance metallic systems because
of an imbalance between the short- and long-range charge redistribution of consecutive SCF steps.
Am effect that is also called ``charge sloshing''.

\subsection{Pseudo diagonalization}
Alternatively to TD, a pseudo diagonalization can be applied as soon as a sufficiently pre-converged wavefunction is obtained~\cite{Stewart1982}.
The KS matrix $\bK^{\rm AO}$ in the AO basis is transformed
into the MO basis in each SCF step via 
\begin{equation}
 \bK^{\rm MO} = \bC^{\rm T} \bK^{\rm AO} \bC
\end{equation}
using the MO coefficients $\bC$ from the preceding SCF step.
The converged $\bK^{\rm MO}$ matrix using TD is a diagonal matrix and
the eigenvalues are its diagonal elements. Already after a few SCF iteration
steps, the $\bK^{\rm MO}$ matrix becomes diagonally dominant.  Moreover, the $\bK^{\rm MO}$ matrix shows the following natural blocking
\begin{equation}
 \left(
  \begin{array}{c|c}
   oo & ou \\
   \hline
   uo & uu \\
 \end{array}
 \right)
\end{equation}
due to the two MO sub-sets of $\bC$, namely the occupied ($o$)
and the unoccupied ($u$) MOs.
The eigenvectors are used during the SCF iteration to calculate
the new density matrix (see Eq.~\ref{eq:P=CCT}), whereas the eigenvalues are not needed. The total energy only depends on the occupied MOs and thus a block diagonalization, which decouples the occupied and
unoccupied MOs allows to converge the wavefunctions. As a consequence, only all
elements of the block $ou$ or $uo$ have to become zero, since
$\bK^{\rm MO}$ is a symmetric matrix. Hence, the transformation
into the MO basis
\begin{equation}
  \bK_\text{ou}^\text{MO} = \bC_o^\text{T} \bK^\text{AO} \bC_u
\end{equation}
has only to be performed for that matrix block. Then the decoupling can be achieved iteratively by consecutive $2 \times 2$ Jacobi rotations, i.e.
\begin{subequations}
\begin{eqnarray}
\theta &=& \frac{\epsilon_q - \epsilon_p}{2\,K_{pq}^{\rm MO}}\\
 t &=& \frac{{\rm sgn}(\theta)}{|\theta| + \sqrt{1 + \theta^2}}\\
 c &=& \frac{1}{\sqrt{t^2 + 1}}\\
 s &=& tc\\
 {\tilde C}_p &=& c\,\bC_p - s\,\bC_q \\
 {\tilde C}_q &=& s\,\bC_p + c\,\bC_q,
\end{eqnarray}
\end{subequations}
%
where the angle of rotation $\theta$ is determined by the difference of
the eigenvalues of the MOs $p$ and $q$ divided by the corresponding
matrix element $K_{pq}^{\rm MO}$ in the $ou$ or $uo$ block.
The Jacobi rotations are computationally cheap as they can be performed
with BLAS level 1 routines ({\tt DSCAL} and {\tt DAXPY}).
The $oo$ block is significantly smaller than the $uu$ block, since only
10--20\% of the MOs are occupied using a standard basis set.
Consequently, the $ou$ or $uo$ block also includes only 10--20\% of the
$\bK^{\rm MO}$ matrix. Furthermore, an expensive
re-orthonormalization of the MO eigenvectors is not needed, since the
Jacobi rotations preserve the orthonormality of the MO eigenvectors.
Moreover, the number of non-zero blocks decreases rapidly when convergence
is approached which results in a decreasing compute time for the PD part.

\subsection{Orbital transformations}
\label{subsection_ot}
%

An alternative to the just described diagonalization-based schemes are algorithms that relies on a direct minimization of the electronic energy functional~\cite{McWeeny56,Fletcher70,Seeger1976,Stich89,Yang07,VandeVondele2003,Weber2008}. 
Convergence of this approach can in principle be guaranteed if the energy can be reduced in each step. The direct minimization approach is thus more robust. It also replaces the diagonalization step by having fewer matrix-matrix multiplications, significantly reducing the time-to-solution.
This is of great importance for many practical problems, in particular large systems that are difficult or sometimes even impossible to tackle with DIIS-like methods. However, preconditioners are often used in conjunction with direct energy minimization algorithms to facilitate faster and smoother convergence.

The calculation of the total energy within electronic structure theory can be formulated variationally in terms of an energy
functional of the occupied single-particle orbitals that are constrained with respect to an orthogonality condition.
With $M$ orbitals, $\mathbf{C}\in\mathbb{R}^{N \times M}$ is given in a nonorthogonal basis consisting of $N$ basis functions 
$\{\phi_i\}_{i=1}^N$ and its associated $N \times N$ overlap matrix $\mathbf{S}$, with element $S_{ij}=\langle\phi_i|\phi_j\rangle$.
The corresponding constrained minimization problem is expressed as
\begin{equation}\label{eq:minE_C}
  \mathbf{C}^* = \arg \min_{\mathbf{C}}\{E[\mathbf{C}]\;|\;\mathbf{C}^T\mathbf{SC} = \mathbf{1} \},
\end{equation}
where $E[\mathbf{C}]$ is an energy functional, $\mathbf{C}^*$ is the minimizer of $E[\mathbf{C}]$ that fulfills the condition of orthogonality $\mathbf{C}^T\mathbf{SC}=\mathbf{1}$, whereas 
$\arg\min$ stands for the argument of the minimum. The ground state energy is finally obtained as $E[\mathbf{C}^*]$.
The form of the energy functional $E[\mathbf{C}]$ is determined by the particular electronic structure theory used,
in the case of hybrid Hartree--Fock/DFT, the equation reads as
\begin{eqnarray}
  E_\text{HF/DFT}[\mathbf{C}] &=& \Tr[\mathbf{Ph}] + \frac{1}{2} \Tr[ \mathbf{P} ( \mathbf{J}[\mathbf{P}] + \alpha_\text{HF}\mathbf{K}[\mathbf{P}] )] \nonumber \\ 
  &+& E_\text{XC}[\mathbf{P}],
\end{eqnarray}
where $\mathbf{P}=\mathbf{CC}^T$ is the density matrix, whereas $\mathbf{h}$, $\mathbf{J}$ and $\mathbf{K}$ are the core Hamiltonian, the Coulomb, and Hartree--Fock exact exchange matrices, respectively, and $E_\text{XC}[\mathbf{P}]$ is the XC energy.

Enforcing the orthogonality constraints within an efficient scheme poses a major hurdle in the direct minimization of $E[\mathbf{C}]$. Hence, in the following, we describe two different approaches implemented in \cp~\cite{VandeVondele2003,Weber2008}. 

\subsubsection{Orthogonality constraints}
\label{subsubsection_orthogonality-constratins}
\paragraph{Orbital transformation functions: OT/Diag and OT/Taylor}
To impose the orthogonality constraints on the orbitals, VandeVondele and Hutter reformulated the non-linear constraint on $\mathbf{C}$ (see Eq.~\ref{eq:minE_C}) by a linear constraint on the auxiliary variable $\mathbf{X}$ via 
\begin{equation}
  \mathbf{X}^* = \arg\min_\mathbf{X}\{ E[\mathbf{C}(\mathbf{X})]\; |\; \mathbf{X}^T\mathbf{SC}_0 = \mathbf{0} \}
\end{equation}
and 
\begin{equation}
  \mathbf{C}^* = \mathbf{C}(\mathbf{X}^*),
\end{equation}
where $\mathbf{X}\in\mathbb{R}^{N \times M}$, and $\mathbf{C}_0$ is a set of initial orbitals that fulfill the orthogonality constraints $\mathbf{C}_0^T\mathbf{SC}_0=\mathbf{1}$~\cite{VandeVondele2003}. The OT is parametrized as follows:
\begin{equation}
  \mathbf{C}(\mathbf{X}) = \mathbf{C}_0 \cos \mathbf{U} + \mathbf{XU}^{-1} \sin \mathbf{U},
\end{equation}
where $\mathbf{U}=\mathbf{X}^T\mathbf{SX}^{1/2}$. This parametrization ensures that $\mathbf{C}^T\mathbf{XSCX}=\mathbf{1}$, for all $\mathbf{X}$ satisfying the constraints $\mathbf{X}^T\mathbf{SC}_0=\mathbf{0}$. The matrix functions $\cos \mathbf{U}$ and $\mathbf{U}^{-1} \sin \mathbf{U}$ are evaluated either directly by diagonalization, or by a truncated Taylor expansion in $\mathbf{X}^T\mathbf{SX}$~\cite{Higham08}.

\paragraph{Orbital transformation based on a refinement expansion: OT/IR}
In this method, Weber et al. replaced the constrained functional by an equivalent unconstrained functional ($\mathbf{C} \to f(\mathbf{Z})$)~\cite{Weber2008}.
The transformed minimization problem in Eq.~\ref{eq:minE_C} is then given by
\begin{equation}\label{eq:minE_Z}
  \mathbf{Z}^* = \arg\min_\mathbf{Z} E[f(\mathbf{Z})]
\end{equation}
and
\begin{equation}
  \mathbf{C}^* = f(\mathbf{Z}^*),
\end{equation}
where $\mathbf{Z} \in \mathbb{R}^{N \times M}$. The constraints have been mapped onto the matrix function $f(\mathbf{Z})$, which fulfills the orthogonality constraint $f^T(\mathbf{Z})\mathbf{S}f(\mathbf{Z})=\mathbf{1}$ for all matrices $\mathbf{Z}$.
The main idea of this approach is to approximate the OT in Eq.~\ref{eq:minE_Z} by $f_n(\mathbf{Z})\approx f(\mathbf{Z})$, where $f_n(\mathbf{Z})$ is an approximate constraint function, which is correct up to some order $n+1$ in $\delta \mathbf{Z}=\mathbf{Z}-\mathbf{Z}_0$, where $\mathbf{Z}_0^T\mathbf{SZ}_0=\mathbf{1}$. The functions derived by Niklasson for the iterative refinement of an approximate inverse matrix factorization are used to approximate $f(\mathbf{Z})$~\cite{Niklasson04}.
The first few refinement functions are given by
%
\begin{subequations}
\begin{eqnarray}
  f_1(\mathbf{Z}) &=& \frac{1}{2}\mathbf{Z}(3 - \mathbf{Y})\\
  f_2(\mathbf{Z}) &=& \frac{1}{8}\mathbf{Z}(15 - 10\mathbf{Y} + 3\mathbf{Y}^2)\\
  f_3(\mathbf{Z}) &=& \frac{1}{16}\mathbf{Z}(35 - 35\mathbf{Y} + 21\mathbf{Y}^2 - 5\mathbf{Y}^3),\\
  \cdots && \nonumber
\end{eqnarray}
\end{subequations}
where $\mathbf{Y} =\mathbf{Z}^T\mathbf{SZ}$ and $\mathbf{Z}=\mathbf{Z}_0+\delta \mathbf{Z}$. It can be shown that 
\begin{equation}
  f_n^T(\mathbf{Z})\mathbf{S}f_n(\mathbf{Z}) - \mathbf{1} = \mathcal{O}(\delta \mathbf{Z}^{n+1}).
\end{equation}
Using this general ansatz for $f_n(\mathbf{Z})$, it is possible to extend the accuracy to any finite order recursively
by an iterative refinement expansion $f_n(\cdots f_n(\mathbf{Z})\cdots)$.


\subsubsection{Minimizer}
\label{subsubsection_minimizer}
\paragraph{Direct inversion of the iterative subspace}
The DIIS method introduced by Pulay is an extrapolation technique based on minimizing the norm of a linear combination of gradient vectors~\cite{Pulay1980}. The problem is given by
\begin{equation}
  c^*=\arg\min_{c}\left\{ \left\|\sum_{i=1}^{m}c_i \mathbf{g}_i \right\| \; | \; \sum_{i=1}^m c_i = 1 \right\},  
\end{equation}
where $\mathbf{g}_i$ is an error vector and $c^*$ the optimal coefficients. 

The next orbital guess $\mathbf{x}_{m+1}$ is obtained by linear combination of the previous points using the optimal coefficients $c^*$, i.e.
\begin{equation}
  \mathbf{x}_{m+1} = \sum^m_{i=1} c^*_i (\mathbf{x}_i - \tau \mathbf{f}_i ),
\end{equation}
where $\tau$ is an arbitrary step size chosen for the DIIS method.
The method simplifies to a steepest descent (SD) for the initial step $m=1$.
While the DIIS method converges very fast in most of the cases, it is not particularly robust.
In \cp, the DIIS method is modified to switch to SD when a DIIS step brings the solution towards an ascent direction.
This safeguard makes DIIS more robust and is possible because the gradient of the energy functional is available.

\paragraph{Non-linear conjugate gradient minimization}
%


Non-linear CG leads to a robust, efficient and numerically stable energy minimization scheme.
In non-linear CG, the residual is set to the negation of the gradient $\mathbf{r}_i = -\mathbf{g}_i$ and
the search direction is computed by Gram-Schmidt conjugation of the residuals, i.e. 
\begin{equation}
  \mathbf{d}_{i+1} = \mathbf{r}_{i+1} + \beta_{i+1} \mathbf{d}_i.
\end{equation}
Several choices for updating $\beta_{i+1}$ are available and the Polak-Ribi{\`e}re variant with restart is used in \cp~\cite{Polak1969, Nocedal2006}: 
\begin{equation}
  \beta_{i+1} = \max\left(\frac{ \mathbf{r}^T_{i+1} ( \mathbf{r}_{i+1} - \mathbf{r}_i )}{ \mathbf{r}_i^T \mathbf{r}_i },\;0\right).
\end{equation}
Similar to linear CG, a step length $\alpha_i$ is found that minimizes the energy function
  $f( \mathbf{x}_i + \alpha_i \mathbf{d}_i )$
using an approximate line search. The updated position becomes
$\mathbf{x}_{i+1} = \mathbf{x}_i+\alpha_i \mathbf{d}_i$.
In \qs, a few different line searches are implemented. The most robust is the golden
section line search~\cite{Kiefer1953}, but the default quadratic
interpolation along the search direction suffices in most cases.
Regarding time-to-solution, the minimization performed with the latter quadratic interpolation is in general significantly faster than the golden section line search.

Non-linear CG are generally preconditioned by choosing an appropriate preconditioner $M$ that approximate $f''$ (see Section~\ref{sec:prec}).

\paragraph{Quasi-Newton method}
Newton's method can also be used to minimize the energy functional.
The method is scale invariant, and the zigzag behavior that can be seen in the SD method is not present. The iteration for Newton's method is given by
\begin{equation}
  \mathbf{x}_{k+1} = \mathbf{x}_k - \beta \mathbf{H}(\mathbf{x}_k)^{-1}\mathbf{f}_k,
\end{equation}
where $\mathbf{H}$ is the Hessian matrix.
On the one hand, the method exhibits super-linear convergence when the initial guess is close to the solution and $\beta = 1$, but, on the other hand, when the initial guess is further away from the solution, Newton's method may diverge.
This divergent behavior can be suppressed by the introduction of line search or backtracking.
As they require the inverse Hessian of the energy functional, the full Newton's method is in general too time-consuming or difficult to use.
Quasi-Newton methods~\cite{Fang2009}, however, are advantageous alternatives to Newton's method when the
Hessian is unavailable or is too expensive to compute. In those methods, an approximate Hessian is
updated by analyzing successive gradient vectors.
A quasi-Newton step is determined by
\begin{equation}
  \mathbf{x}_{k+1} = \mathbf{x}_k - \beta \mathbf{G}_k \mathbf{f}_k,
\end{equation}
where $\mathbf{G}_k$ is the approximate inverse Hessian at step $k$. Different update formulas exist to compute $\mathbf{G}_k$~\cite{Martinez00}.
In \qs, the Broyden’s type 2 update is implemented to construct the approximate
inverse Hessian with an adaptive scheme for estimating the curvature of the energy functional to increase the performance~\cite{Baarman11}.

\subsubsection{Preconditioners}\label{sec:prec}
\paragraph{Preconditioning the non-linear minimization}
 Gradient based OT methods are guaranteed to converge, but will exhibit slow convergence behavior if not appropriately preconditioned.
 A good reference for the optimization can be constructed from the generalized eigenvalue problem under the orthogonality constraint of Eq.~\ref{eq:minE_C}
%
and its approximate second derivative
\begin{equation}
 \frac{\partial^2 E}{\partial {X}_{\alpha i}\partial {X}_{\beta j}}=2{H}_{\alpha\beta}\delta_{ij}-2{S}_{\alpha\beta}\delta_{ij}\epsilon^0_i.
\end{equation}
Therefore, close to convergence, the best preconditioners would be of the form
\begin{equation}
  (\mathbf{H}-\mathbf{S}\epsilon_i^0)_{\alpha \beta}^{-1} \left(\frac{\delta E} {\delta {X}_{\beta i}}\right).
\end{equation}
As this form is impractical, requiring a different preconditioner for each orbital, a single positive definite preconditioning matrix $\mathbf{P}$ is constructed approximating
\begin{equation}
 \mathbf{P}(\mathbf{H}-\mathbf{S\epsilon})\mathbf{x}-\mathbf{x}\approx 0.
\end{equation}
In \cp, the closest approximation to this form is the \texttt{FULL\_ALL} preconditioner. 
%
It performs an orbital dependent eigenvalue shift of $\mathbf{H}$.
In this way, positive definiteness is ensured with minimal modifications.
The downside is that the eigenvalues of $\mathbf{H}$ have to be computed at least once using diagonalization and thus scales as $\mathcal{O}(N^3)$.\\
To overcome this bottleneck several more approximate, though lower scaling preconditioners have been implemented within \cp.
The simplest assume $\mathbf{\epsilon}=\mathbf{1}$ and $\mathbf{H}=\mathbf{T}$, with $\mathbf{T}$ being the kinetic energy matrix (\texttt{FULL\_KINETIC}), or even $\mathbf{H}=0$ (\texttt{FULL\_S\_INVERSE}) as viable approximations.
These preconditioners are obviously less sophisticated.
However, they are linear-scaling in their construction as $\mathbf{S}$ and $\mathbf{T}$ are sparse and still lead to accelerated convergence.
Hence, these preconditioners are suitable for large-scale simulations.\\
For many systems, the best trade off between quality and cost of the preconditioner is obtained with the \texttt{FULL\_SINLGE\_INVERSE} preconditioner. 
Instead of shifting all orbitals separately, only the occupied eigenvalues are inverted.
Thus, making the orbitals closest to the band gap most active in the optimization.
The inversion of the spectrum can be done without the need for diagonalization by  
\begin{equation}
 \mathbf{P}=\mathbf{H} - 2\mathbf{S}\mathbf{C}_0 (\mathbf{C}_0^T \mathbf{H} \mathbf{C}_0+ \delta) \mathbf{C}_0^T\mathbf{S}-\mathbf{\epsilon} \mathbf{S},
 \end{equation}
where $\delta$ represents an additional shift depending on the HOMO energy, which ensures positive definiteness of the preconditioner matrix.
It is important to note that the construction of this preconditioner matrix can be done with a complexity of $\mathcal{O}(NM^2)$ in the dense case and is therefore of the same complexity as the rest of the OT algorithm.

\paragraph{Reduced scaling and approximate preconditioning}
All of the above mentioned preconditioners still require the inversion of the preconditioning matrix $\mathbf{P}$.
In dense matrix algebra this leads to an $\mathcal{O}(N^3)$ scaling behavior independent of the chosen preconditioner.
For large systems, the inversion of $\mathbf{P}$ will become the dominant part of the calculation when low-scaling preconditioners are used.
As Schiffmann et al. has shown~\cite{Schiffmann15}, sparse matrix techniques are applicable for the inversion of the low-scaling preconditioners and can be activated using the \texttt{INVERSE\_UPDATE} option as preconditioner solver.
By construction, the preconditioner matrix is symmetric and positive definite.
This allows for the use of Hotteling's iterations to compute the inverse of $\mathbf{P}$ as
\begin{equation}
 \mathbf{P}^{-1}_{i+1}=\alpha \mathbf{P}^{-1}_{i} (2\mathbf{I}-\mathbf{P}\alpha_i\mathbf{P}^{-1}_{i}).
\end{equation}
Generally, the resulting approximants to the inverse become denser and denser~\cite{Lass2018}.
In \cp\ this is dealt with aggressive filtering on ${P}^{-1}_{i+1}$.
Unfortunately, there are limits to the filtering threshold.
While the efficiency of the preconditioner is not significantly affected by the loss of accuracy (see section~\ref{subsection_ac}), the Hotteling iterations may eventually become unstable~\cite{lass17-esl}.\\
Using a special way to determine the initial alpha based on the extremal eigenvalues of $\mathbf{P}\mathbf{P}^{-1}_0$ it can be shown that any positive definite matrix can be used as initial guess for the Hotteling iterations~\cite{Richters2019}.
For MD simulations or geometry optimizations this means the previous inverse can be used as initial guess as the changes in $\mathbf{P}$ are expected to be small~\cite{Richters2014}.
Therefore, only very few iterations are required until convergence after the initial approximation for the inverse is obtained.

\subsection{Purification methods}
\label{subsection_purification}
Linear-scaling DFT calculations can also be achieved by purifying the KS matrix $\mathbf{K}$ directly into the density matrix $\mathbf{P}$ without using the orbitals $\mathbf{C}$ explicitly~\cite{McWeeny_RMP_1960}. These density matrix-based methods exploit the fact that the KS matrix $\mathbf{K}$, as well as the density matrix $\mathbf{P}$, have by definition the same eigenvectors $\mathbf{C}$, and that a purification maps eigenvalues $\epsilon_i$ of $\mathbf{K}$ to eigenvalues $f_i$ of $\mathbf{P}$ via the Fermi-Dirac-function
\begin{equation}
    f_i=\frac{1}{\exp\left (\frac{\epsilon_i-\mu}{k_B T}\right)+1}, 
\end{equation}
with the chemical potential $\mu$, the Boltzmann constant $k_B$ and electron temperature $T$. In practice, the purification is computed by an iterative procedure that is constructed to yield a linear-scaling method for sparse KS matrices. 
By construction, such purifications are usually grand-canonical purifications so that additional measures such as modifications to the algorithms or additional iterations to find the proper value of the chemical potential are necessary to allow for canonical ensembles.
In \cp, the trace-resetting fourth-order method (TRS4~\cite{doi:10.1063/1.1559913}), the trace-conserving second order method (TC2~\cite{doi:10.1137/130911585}) and the purification via the sign function (SIGN~\cite{VandeVondele2012}) are available. 
Additionally, \cp\ implements an interface to the PEXSI (Pole EXpansion and Selected Inversion) library~\cite{Lin2009,Lin2013,Jacquelin2016,Jacquelin2018}, which allows to evaluate selected elements of the density matrix as the Fermi-Dirac-function of the KS matrix via a pole expansion.

\subsection{Sign-Method}
\label{subsection_sign-method}
The sign function of a matrix
\begin{equation}
    \sign(A)=A (A^2)^{-1/2} \label{eq:signdef}
\end{equation}
can be used as a starting point for the construction of various linear-scaling algorithms~\cite{Higham1997}. The relation
\begin{equation}
    \sign\begin{pmatrix}0 & A \\ I & 0\end{pmatrix}=\sign\begin{pmatrix}0 & A^{1/2} \\ A^{-1/2} & 0\end{pmatrix}
\end{equation}
together with iterative methods for the sign function, such as the Newton-Schulz iteration, are used by default in \cp\ for linear-scaling matrix inversions and (inverse) square roots of matrices~\cite{Ceriotti2008,Ceriotti2009}. Several orders of iterations are available: the second-order Newton-Schulz iteration, as well as third-order and fifth-order iterations based on higher-order Pad\'e-approximants~\cite{doi:10.1137/0612020,Higham1997,Richters2019}. 

\begin{figure}
\includegraphics*[width=8cm]{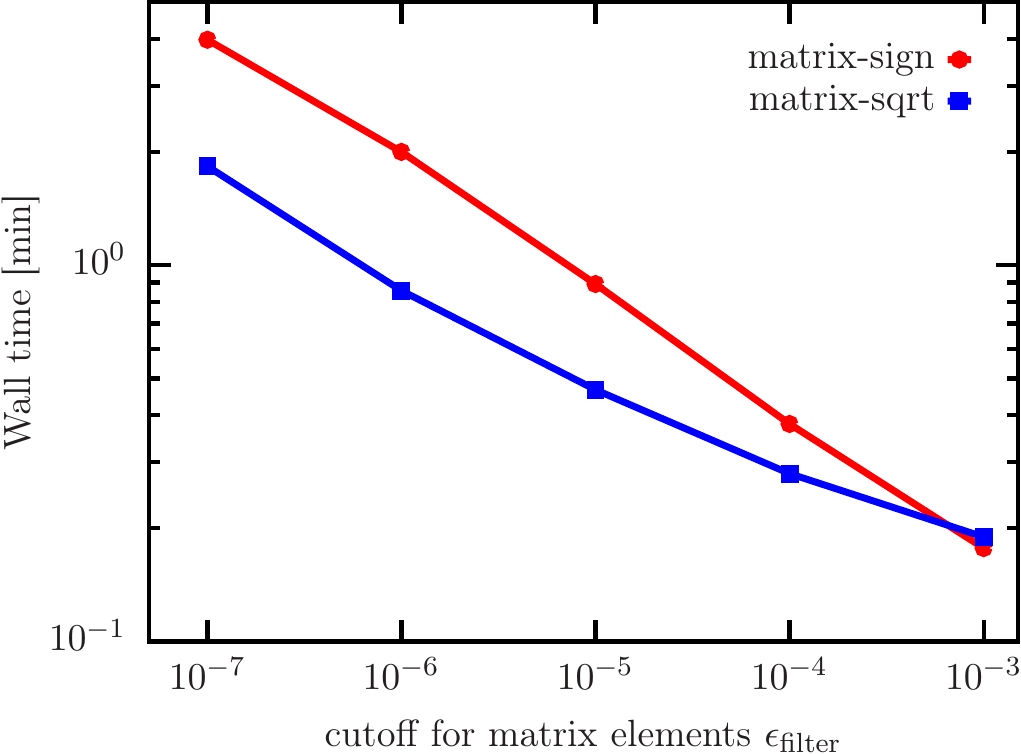}
\caption{\label{fig:NS_STMV_epsfilter} Wall time for the calculation of the computationally dominating matrix-sqrt (blue, boxes) and matrix-sign (red, circles) functions as a function of matrix truncation threshold $\epsilon_{\mathrm{filter}}$ for the STMV virus. The latter contains more than one million atoms and was simulated using the periodic implementation in \cp\ of the GFN2-xTB model~\cite{xtb}. The fifth-order sign-function iteration of Eq.~\ref{eq:fifth_order_sign} and the third-order sqrt-function iterations have been used. The calculations have been performed with 256 nodes (10240 cpu-cores) of the Noctua system at the Paderborn Center for Parallel Computing (PC$^2$).} 
\end{figure}

The sign function can also be used for the purification of the Kohn-Sham matrix $\mathbf{K}$ into the density matrix $\mathbf{P}$, i.e. via the relation
\begin{equation}
    \mathbf{P}=\frac{1}{2}\left(\mathbf{I}-\sign\left(\mathbf{S}^{-1} \mathbf{K}-\mu \mathbf{I}\right ) \right)\mathbf{S}^{-1}.
\end{equation}
Within \cp, the linear-scaling calculation of the sign-function is implemented up to seventh-order based on Pad\'e-approximants. For example, the fifth-order iteration has the form
%
\begin{subequations}
\begin{eqnarray}
    \label{eq:fifth_order_sign}
    \mathbf{X}_0&=&\mathbf{S}^{-1} \mathbf{K}-\mu \mathbf{I}\\
    \mathbf{X}_{k+1}&=&\frac{\mathbf{X}_k}{128} (35 \mathbf{X}_k^8 - 180 \mathbf{X}_k^6 + 378 \mathbf{X}_k^4 \nonumber\\&-& 420 \mathbf{X}_k^2 + 315)\\
    \lim_{k\rightarrow \infty}  \mathbf{X}_{k}&=&\sign\left(\mathbf{S}^{-1} \mathbf{K}-\mu \mathbf{I}\right )
\end{eqnarray}
\end{subequations}
and is implemented in \cp\ with just four matrix multiplications per iteration.
After each matrix multiplication, all matrix elements smaller than a threshold $\epsilon_{\mathrm{filter}}$ are flushed to zero to retain sparsity. The scaling of the wall-clock time for the computation of the sign- and sqrt-functions to simulate the STMV virus in water solution using the GFN2-xTB Hamiltonian is shown in Fig.~\ref{fig:NS_STMV_epsfilter}~\cite{doi:10.1021/acs.jctc.6b00398}.
The drastic speedup of the calculation, when increasing the threshold $\epsilon_{\mathrm{filter}}$, immediately suggests the combination of sign-matrix iteration based linear-scaling DFT algorithms with the ideas of approximate computing (AC), as discussed in section~\ref{subsection_ac}.

Recently, a new iterative scheme for the inverse p-th of a matrix has been developed which also allows to directly evaluate the density matrix via the sign function in Eq.~\ref{eq:signdef}~\cite{Richters2019}. An arbitrary-order implementation of this scheme is also available in \cp. 
\subsection{Submatrix Method}
\label{subsection_submatrix-method}
In addition to the sign method, the submatrix method presented in Ref.~\onlinecite{Lass2018} has been implemented in \cp\ as an alternative approach to calculate the density matrix $\mathbf{P}$ from the KS matrix $\mathbf{K}$. Instead of operating on the entire matrix, calculations are performed on principal submatrices thereof. Each of these submatrices covers a set of atoms that originates from the same block of the KS matrix in the DBCSR-format, as well as those atoms in their direct neighborhood whose basis functions are overlapping. As a result, the submatrices are much smaller than the KS matrix, but relatively dense. For large systems, the size of the submatrices is independent on the overall system size so that a linear-scaling method immediately results from this construction.
Purification of the submatrices can be performed either using iterative schemes to compute the sign function (see section~\ref{subsection_sign-method}), or via a direct eigendecomposition.
\begin{figure}
\includegraphics*[width=8cm]{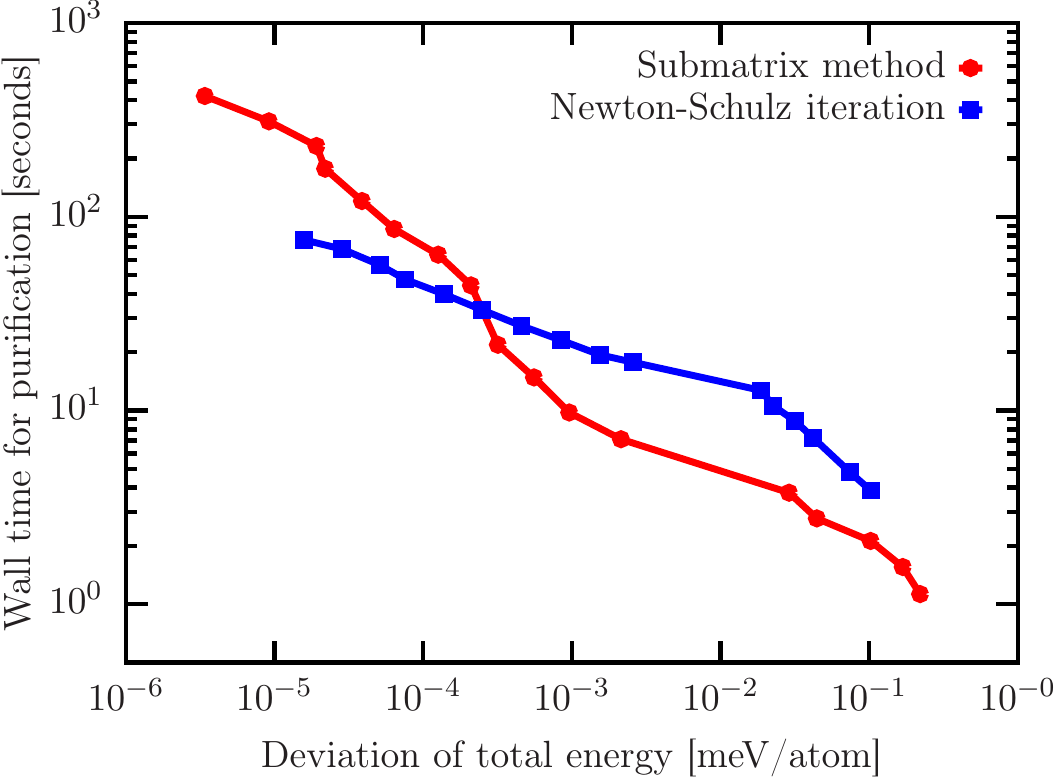}
\caption{\label{fig:submatrix_ns_err-time} Comparison of the wall time required for the purification using the submatrix method (red circles, with a direct eigendecomposition for the density matrix of the submatrices) and the 2nd order Newton-Schulz sign iteration (blue boxes) for different relative errors in total energy after one SCF iteration. The corresponding reference energy has been computed by setting $\epsilon_{\mathrm{filter}}=10^{-16}$. All calculations have been performed on a system consisting of 6192 H\textsubscript{2}O molecules, using KS-DFT together with a SZV basis set, utilizing two compute nodes (80 cpu-cores) of the ``Noctua'' system at PC$^2$.}
\end{figure}
The submatrix method provides an approximation of the density matrix $\mathbf{P}$, whose quality and computational cost depends on the truncation threshold $\epsilon_{\mathrm{filter}}$ used during the SCF iterations. Fig.~\ref{fig:submatrix_ns_err-time} compares the accuracy provided and wall time required by the submatrix method with accuracy and wall time of a Newton-Schulz sign iteration.

\section{Localized molecular orbitals}
\label{section_ALMO}
Spatially localized molecular orbitals (LMOs), also known as MLWFs in solid state physics and materials science, are widely used to visualize chemical bonding between atoms, help classify bonds and thus understand electronic structure origins of observed properties of atomistic systems (see section~\ref{sec:CLMO-EDA}). 
Furthermore, localized orbitals are a key igredient in multiple local electronic structure methods that dramatically reduce the computational cost of modeling electronic properties of large atomistic systems. 
LMOs are also important to many other electronic structure methods that require local states such as XAS spectra modeling or dispersion-corrected XC functionals based on atomic polarizabilities. 


\subsection{Localization of orthogonal and non-orthogonal molecular orbitals} 
\label{subsection_orbital_localization}
\CPXK\ offers a variety of localization methods, in which LMOs $\ket{j}$ are constructed by finding a unitary transformation $\mathbf{A}$ of canonical MOs $\ket{i_0}$, either occupied or virtual, i.e.
\begin{equation}
\begin{split}
\ket{j} = \sum_{i}\ket{i_0} A_{ij}, 
\end{split}
\end{equation}
which minimizes the spread of individual orbitals. \CPXK\ uses the localization functional proposed by Resta~\cite{Resta1998}, which was generalized by Berghold \emph{et al.} to periodic cells of any shape and symmetry: 
\begin{subequations}
\begin{eqnarray} \label{eq:fun-loc}
\Omega_L(\mathbf{A}) &= - \sum_K \sum_i \omega_K \vert z_{i}^{K} \vert^2 \\
z_{i}^{K} &= \sum_{mn} {A}_{mi} B^{K}_{mn} {A}_{ni} \\
B^{K}_{mn} &= \bra{m_0} \E^{\imi \mathbf{G}_K \cdot \mathbf{\op{r}}} \ket{n_0},
\end{eqnarray}
\end{subequations}
where $\mathbf{\op{r}}$ is the position operator in three dimensions, $\omega_K$ and $\mathbf{G}_K$ are suitable sets of weights and reciprocal lattice vectors, respectively~\cite{Berghold2000}. The functional in Eq.~(\ref{eq:fun-loc}) can be used for both gas-phase and periodic systems. In the former case, the functional is equivalent to the Boys-Foster localization~\cite{boys1960construction}. In the latter case, its applicability is limited to the electronic states described within the $\Gamma$-point approximation.

\CPXK\ also implements the Pipek-Mezey localization functional~\cite{pipek1989fast}, which can be written in the same form as the Berghold functional above with $K$ referring to atoms, $z_{i}^{K}$ measuring the contribution of orbital $i$ to the Mulliken charge of atom $K$ and $\mathbf{B}$ being defined as
\begin{equation} \label{eq:pipek}
\begin{split}
B^{K}_{mn} &= \frac{1}{2} \sum_{\mu \in K} \bra{m_0}  \left( \ketbra{\chi_{\mu}}{\chi^{\mu}} + \ketbra{\chi^{\mu}}{\chi_{\mu}} \right) \ket{n_0},
\end{split}
\end{equation}
where $\ket{\chi_\mu}$ and $\ket{\chi^\mu}$ are atom-centered covariant and contravariant basis set functions~\cite{Berghold2000}. The Pipek-Mezey functional has the advantage of preserving the separation of $\sigma$ and $\pi$ bonds and is commonly employed for molecular systems. 


\begin{figure}
\includegraphics*[width=8cm]{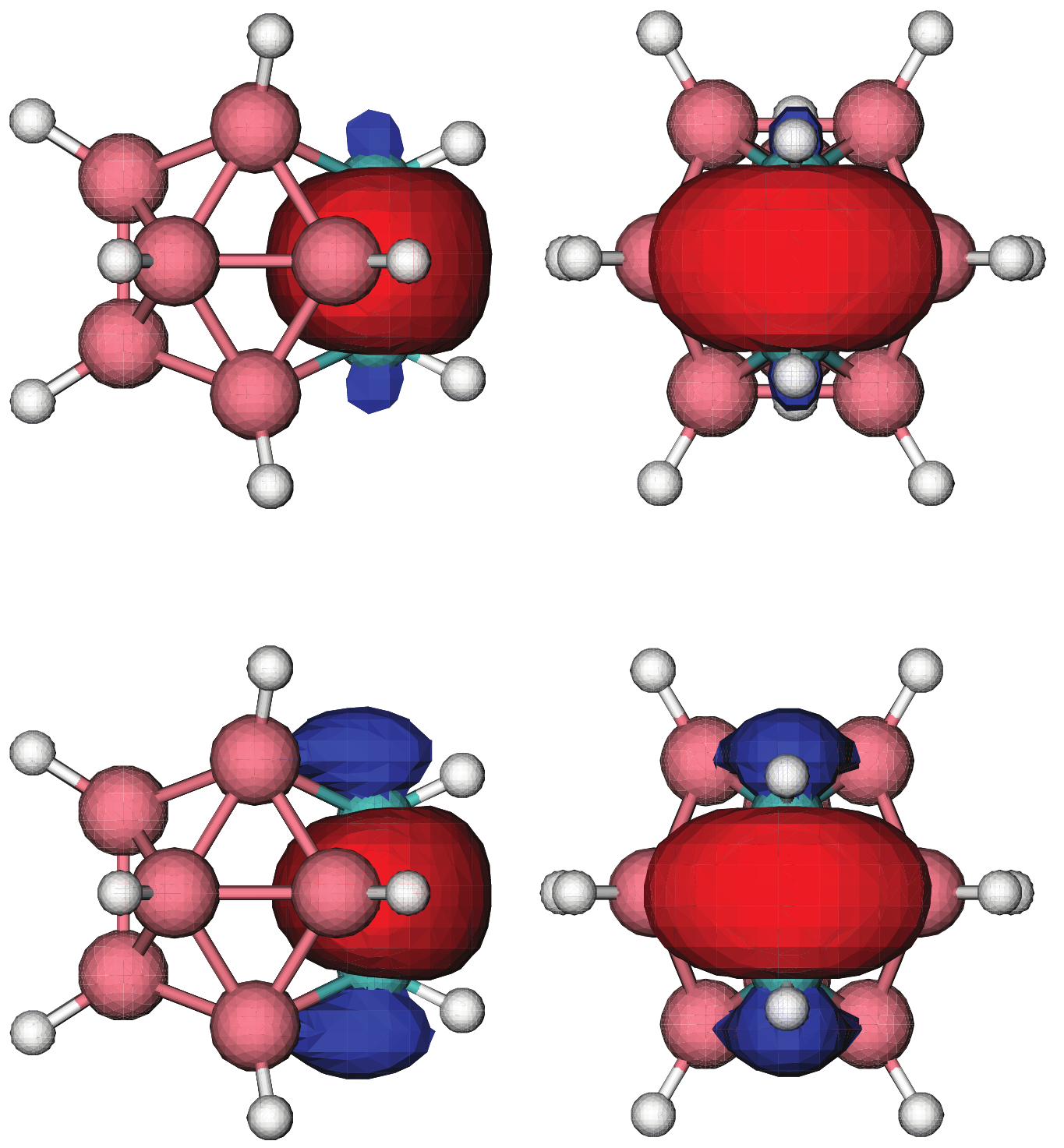}
\caption{\label{fig:nlmos} Orthogonal (bottom) and non-orthogonal (top) LMOs on the covalent bond of the adjacent carbon atoms in a carborane molecule C$_2$B$_{10}$H$_{12}$ (isosurface value is 0.04~a.u.). }
\end{figure}

In addition to the traditional localization techniques, \CPXK\ offers localization methods that produce non-orthogonal LMOs (NLMOs)~\cite{Cui2010}. In these methods, matrix $\mathbf{A}$ is not restricted to be unitary and the minimized objective function contains two terms: a localization functional $\Omega_L$ given by Eq.~(\ref{eq:fun-loc}) and a term that penalizes unphysical states with linearly dependent localize orbitals
\begin{equation} \label{eq:fun-pen}
\begin{split}
\Omega(\mathbf{A}) = \Omega_L(\mathbf{A}) - c_P \log \det \left[ \sigma_{\text{diag}}^{-1} (\mathbf{A}) \sigma (\mathbf{A}) \right], 
\end{split}
\end{equation}
where $c_P > 0$ is the penalty strength and $\sigma$ is the NLMO overlap matrix 
\begin{equation}
\begin{split}
\sigma_{kl} = \langle k | l \rangle = \sum_{ji} {A}_{jk} \langle {j_0}|{i_0} \rangle {A}_{il}.
\end{split}
\end{equation}
The penalty term varies from 0 for orthogonal LMOs to $+\infty$ for linearly dependent NLMOs, making the latter inaccessible in the localization procedure with finite penalty strength $c_P$. The inclusion of the penalty term converts the localization procedure into a straightforward unconstrained optimization problem and produces NLMOs that are noticeably more localized than their conventional orthogonal counterparts (Figure~\ref{fig:nlmos}). 

\subsection{Linear scaling methods based on localized one-electron orbitals} \label{sec:ALMO-ON}
Linear-scaling, or so-called \LS, methods described in section~\ref{subsection_sign-method} exploit the natural locality of the one-electron \DM. 
Unfortunately, the variational optimization of the \DM\ is inefficient if calculations require many basis functions per atom. 
From the computational point of view, the variation of localized one-electron states is preferable to the \DM\ optimization because the former requires only the occupied states, reducing the number of variational degrees of freedom significantly, especially in calculations with large basis sets. 
\CPXK\ contains a variety of orbital-based \LS\ DFT methods briefly reviewed here and in section~\ref{sec:CLMOAIMD}. 

Unlike density matrices, one-electron states tend to delocalize in the process of an unconstrained optimization and their locality must be explicitly enforced to achieve linear-scaling. 
To this end, each occupied orbital is assigned a \emph{localization center} -- an atom (or a molecule) -- and a \emph{localization radius} $R_{c}$. Then, each orbital is expanded strictly in terms of subsets of localized basis functions centered on the atoms (or molecules) lying within $R_c$ from the orbital's center. 
In \CPXK, contracted Gaussian functions are used as the localized basis set. 

Since their introduction~\cite{Matsuoka1977, Stoll1977, Mehler1977}, the orbitals with this strict \emph{a priori} localization have become known under different names including absolutely localized molecular orbitals~\cite{Stoll1980}, localized wavefunctions~\cite{Ordejon1995}, non-orthogonal generalized Wannier functions~\cite{Skylaris2005}, multi-site support functions~\cite{Nakata2015}, and non-orthogonal localized molecular orbitals~\cite{Burger2008}. 
Here, they are referred to as compact localized molecular orbitals (CLMOs) to emphasize that their expansion coefficients are set to zero for all basis functions centered outside orbitals' localization subsets. Unlike previous works~\cite{Khaliullin2013,Khaliullin2013a,scheiber2018compact}, the \mbox{ALMO} acronym is avoided~\cite{Stoll1980}, since it commonly refers to a special case of compact orbitals with $R_c=0$~\cite{Khaliullin2006, Khaliullin2007, Khaliullin2008, Kuehne2013, Horn2013, NatureComm2015}.

While the localization constraints are necessary to design orbital-based \LS\ methods, the reduced number of electronic degrees of freedom results in the variationally optimal CLMO energy being always higher than the reference energy of fully delocalized orbitals. 
From the physical point of view, enforcing orbital locality prohibits the flow of electron density between distant centers and thus switches off the stabilizing donor-acceptor (i.e. covalent) component of interactions between them. It is important to note that the other interactions such as long-range electrostatics, exchange, polarization, and dispersion are retained in the CLMO approximation. 
Thereby, the accuracy of the CLMO-based calculations depends critically on the chosen localization radii, which should be tuned for each system to obtain the best accuracy-performance compromise. 
In systems with non-vanishing band gaps, the neglected donor-acceptor interactions are typically short-ranged and CLMOs can accurately represent their electronic structure if $R_c$ encompasses the nearest and perhaps next nearest neighbors. On the other hand, the CLMO approach is not expected to be practical for metals and semi-metals because of their intrinsically delocalized electrons. 

The methods and algorithms in \CPXK\ are designed to circumvent the known problem of slow variational optimization of \mbox{CLMOs}~\cite{Mauri1993, Ordejon1995, Goedecker1999, Fattebert2004, Tsuchida2008, Peng2013}, the severity of which rendered early orbital-based \LS\ methods impractical.
Two solutions to the convergence problem are described here. The first approach is designed for systems without strong covalent interactions between localization centers such as ionic materials or ensembles of small weakly-interacting molecules~\cite{Tsuchida2007, Tsuchida2008, Khaliullin2013, Khaliullin2013a}. The second approach is proposed to deal with more challenging cases of strongly bonded atoms such as covalent crystals. 

The key idea of the first approach is to optimize CLMOs in a two-stage SCF procedure. 
In the first stage, $R_c$ is set to zero and the CLMOs -- they can be called ALMOs in this case -- are optimized on their centers.  In the second stage, the CLMOs are relaxed to allow delocalization onto the neighbor molecules within their localization radius $R_{c}$. 
To achieve a robust optimization in the problematic second stage, the delocalization component of the trial CLMOs must be kept orthogonal to the fixed ALMOs obtained in the first stage. 
If the delocalization is particularly weak, the CLMOs in the second stage can be obtained using the simplified Harris functional~\cite{Harris1985} -- orbital optimization without updating the Hamiltonian matrix -- or non-iterative perturbation theory. 
The mathematical details of the two-stage approach can be found in Ref.~\citenum{Khaliullin2013a}. A detailed description of the algorithms is presented in the supplementary material of Ref.~\cite{scheiber2018compact}. 

The two-stage SCF approach resolves the convergence problem only if the auxiliary ALMOs resembles the final variationally optimal CLMOs and, therefore, is not practical for systems with noticeable electron delocalization -- in other words, covalent bonds -- between atoms. The second approach, designed for systems of covalently bonded atoms, utilizes an approximate electronic Hessian that is inexpensive to compute and yet sufficiently accurate to identify and remove the problematic optimization modes. The accuracy and practical utility of this approach relies on the fact that the removed low-curvature modes are associated with the nearly-invariant occupied-occupied orbital mixing, which produce only an insignificant lowering in the total energy.

\begin{figure}
\includegraphics*[width=8cm]{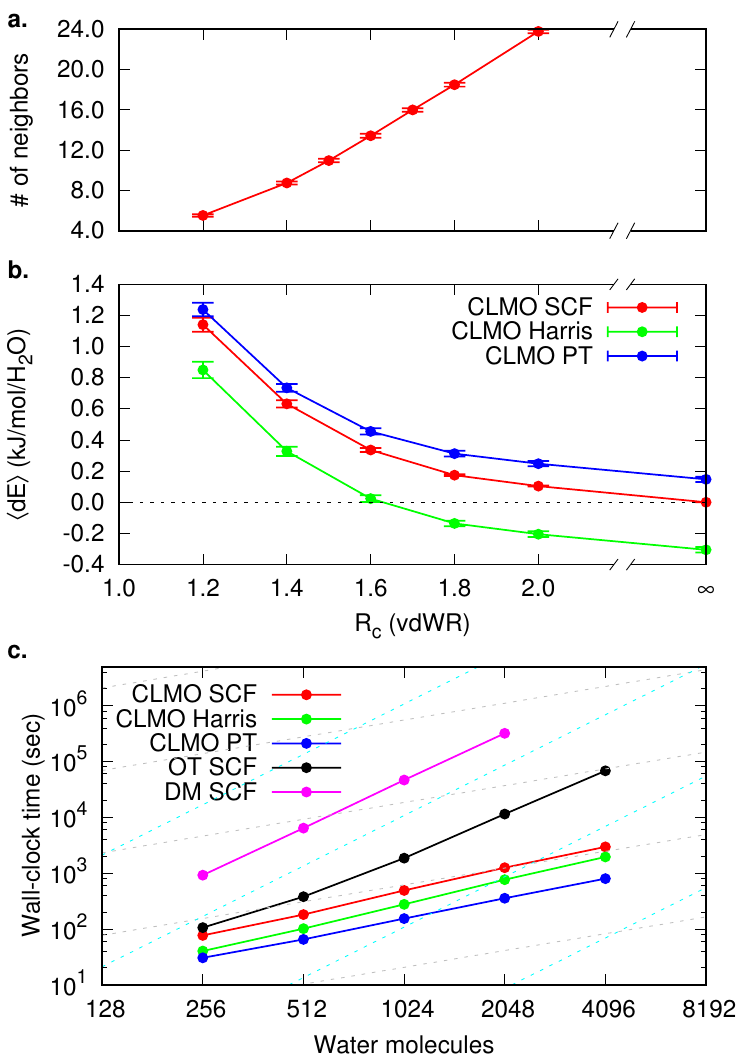}
\caption{\label{fig:clmos-mol} Accuracy and efficiency of \LS\ DFT for liquid water based on the two-stage CLMO optimization. 
Calculations were performed using the BLYP functional and a TZV2P basis set for 100 snapshots representing liquid water at constant temperature (300~K) and density (0.9966~g/cm$^3$). (a) Dependence of the average number of neighbors on the localization radius, expressed in units of the elements' van der Waals radii (vdWR). 
(b) Energy error per molecule relative to the energies of fully delocalized orbitals. For simulations, in which the coordination number of molecules does not change drastically, the mean error represents a constant shift of the potential energy surface and does not affect the quality of simulations. In such cases, the standard deviation (error bars) is better suited to measure the accuracy of the CLMO methods.
%
%
(c) Wall-time required for the variational minimization of the energy on 256 compute cores. The localization radius is set to 1.6~vdWR for the CLMOs methods. The CLMO methods are compared to the cubically-scaling optimization of delocalized orbitals (OT SCF)~\cite{VandeVondele2003,Weber2008}, as well as to the \LS\ optimization of \DM~\cite{VandeVondele2012}. Perfect linear- and cubic-scaling lines are shown in grey and cyan, respectively. See Ref.~\onlinecite{Khaliullin2013a} for details.}
\end{figure}

The robust variational optimization of CLMOs combined with \CPXK's fast \LS\ algorithms for the construction of the KS Hamiltonian enabled the implementation of a series of \LS\ orbital-based DFT methods with a low computational overhead. Fig.~\ref{fig:clmos-mol} shows that $R_c$ can be tuned to achieve substantial computational savings without compromising the accuracy of the calculations. It also demonstrates that CLMO-based DFT exhibits early-offset linear-scaling behavior even for challenging condensed-phase systems and works extremely well with large basis sets.

The current implementation of the CLMO-based methods is limited to localization centers with closed-shell electronic configurations. The nuclear gradients are available only for the methods that converge the CLMO variational optimization. This excludes CLMO methods based on the Harris functional and perturbation theory. 

Section~\ref{sec:CLMOAIMD} describes how the CLMO methods can be used in AIMD simulations by means of the second-generation Car-Parrinello MD (CPMD) method of K\"uhne and coworkers~\cite{Kuehne2007,Kuehne2014,Prodan2018}. Energy decomposition analysis (EDA) methods that exploit the locality of compact orbitals to understand the nature of intermolecular interactions in terms of physically meaningful components are described in section~\ref{sec:CLMO-EDA}.

\subsection{Polarized atomic orbitals from machine learning}
\label{subsection_ml-pao}
The computational cost of a DFT calculation depends critically on the size and condition number of the employed basis set.
Traditional basis sets are atom centered, static, and isotropic.
Since molecular systems are never isotropic, it is apparent that isotropic basis sets are sub-optimal.
The polarized atomic orbitals from machine learning (PAO-ML) scheme provides small adaptive basis sets, which adjust themselves to the local chemical environment~\cite{pao_ml}.
The scheme uses polarized atomic orbitals (PAOs), which are constructed from a larger primary basis function as introduced by Berghold~et~al.~\cite{Berghold}.
A PAO basis function $\tilde \varphi_\mu$ can be written as a weighted sum of primary basis functions $\varphi_\nu$, where $\mu$ and $\nu$ belong to the same atom:
\begin{equation}
\tilde \varphi_\mu = \sum_\nu B_{\mu\nu} \,\varphi_\nu.
\end{equation}
The aim of the PAO-ML method is to predict the transformation matrix $\mathbf{B}$ for a given chemical environment using machine learning (ML).
The analytic nature of ML models allows for the calculation of exact analytic forces, as they are required for AIMD simulations.
In order to train such an ML model, a set of relevant atomic motifs with their corresponding optimal PAO basis are needed.
This poses an intricate non-linear optimization problem as the total energy must be minimal with respect to the transformation matrix $\mathbf{B}$, and the electronic density, while still obeying the Pauli principle.
To this end, the PAO-ML scheme introduced an improved optimization algorithm based on the Li, Nunes, and Vanderbilt formulation for the generation of training data~\cite{Li1993}.

When constructing the ML model, great care must be taken to ensure the invariance of the predicted PAO basis sets with respect to rotations of the simulation cell, to prevent artificial torque forces.
The PAO-ML scheme achieve rotational invariance by employing potentials anchored on neighboring atoms.
The strength of individual potential terms is predicted by the ML model.
Collectively the potential terms form an \textit{auxiliary atomic Hamiltonian}, whose eigenvectors are then used to construct the transformation matrix $\mathbf{B}$.

The PAO-ML method has been demonstrated by means of AIMD simulations of liquid water.
A minimal basis set yielded structural properties in fair agreement with basis set converged results.
In the best case, the computational cost were reduced by a factor of 200 and the required flops by 4 orders of magnitude.
Already a very small training set gave satisfactory results as the variational nature of the method provides robustness.
%

\section{Ab-Initio molecular dynamics}
\label{section_AIMD}
The mathematical task of AIMD is to evaluate the expectation value $\langle \mathcal{O} \rangle$ of an arbitrary operator $\mathcal{O}(\bm{R}, \bm{P})$ with respect to the Boltzmann distribution
\begin{eqnarray}
  \langle \mathcal{O} \rangle = \frac{\int{d \bm{R} \, d \bm{P} \, \mathcal{O}(\bm{R}, \bm{P}) \, e^{- \beta E(\bm{R}, \bm{P})}}}{\int{d \bm{R} \, d \bm{P} \, e^{- \beta E(\bm{R}, \bm{P})}}}, 
\label{BoltzmannEquation}
\end{eqnarray}
where $\bm{R}$ and $\bm{P}$ are the nuclear positions and momenta, while $\beta = {1}/{k_{B}T}$ is the inverse temperature. The total energy function
\begin{eqnarray} 
  E(\bm{R}, \bm{P}) = \sum_{I=1}^{N}{ \frac{\bm{P}_{I}^{2}}{2 M_{I}} } + E[\{ \psi_{i} \}; \bm{R}], 
\end{eqnarray}
where the first term denotes the nuclear kinetic energy, $E[\{ \psi_{i} \}; \bm{R}]$ the potential energy function, $N$ the number of nuclei and $M_{I}$ the corresponding masses. 


However, assuming the ergodicity hypothesis, the thermal average $\langle \mathcal{O} \rangle$ can not only be determined as the ensemble average, but also as a temporal average 
\begin{eqnarray}
  \langle \mathcal{O} \rangle = \lim_{\mathcal{T} \rightarrow \infty} \frac{1}{\mathcal{T}} \int{d t \, \mathcal{O}(\bm{R}(t), \bm{P}(t))}
  \label{TemporalAverage}
\end{eqnarray}
by means of AIMD.

In the following, we will assume that the potential energy function is calculated on-the-fly using KS-DFT, so that $E[\{ \psi_{i} \}; \bm{R}] = E^{\text{KS}} \left[ \{ \psi_{i}[\rho(\bm{r})] \}; \bm{R} \right] + E_{II}(\bm{R})$. In \cp, AIMD comes in two distinct flavors, which are both outlined in this section.

\subsection{Born-Oppenheimer molecular dynamics}
\label{subsection_BOMD}
In Born-Oppenheimer MD (BOMD) the potential energy $E \left[ \{ \psi_{i} \}; \bm{R} \right]$ is minimized at every AIMD step with respect to $\{ \psi_{i}({\bm{r}}) \}$ under the holonomic orthonormality constraint $\langle {\psi_{i}(\bm{r})}|{\psi_{j}(\bm{r})} \rangle = \delta_{ij}$. 
This leads to the following Lagrangian:
\begin{eqnarray}
  \mathcal{L}_{\text{BO}} \left( \{\psi_{i}\}; \bm{R}, \dot{\bm{R}} \right) &=& \frac{1}{2} \sum_{I=1}^{N}{M_{I}\dot{\bm{R}}_{I}^{2}} - \min_{ \{\psi_{i}\} } E \bigl[ \{ \psi_{i} \}; \bm{R} \bigr] \nonumber \\
  &+& \sum_{i,j} { \Lambda_{ij} \left( \langle {\psi_{i}}|{\psi_{j}} \rangle - \delta_{ij} \right) }, \label{BO_Lagrangian}
\end{eqnarray}
where $\bm{\Lambda}$ is a Hermitian Lagrangian multiplier matrix. By solving the corresponding Euler-Lagrange equations
\begin{subequations}
\begin{eqnarray}
  \frac{d}{dt} \frac{\partial \mathcal{L}}{\partial \dot{\bm{R}}_{I}} &=& \frac{\partial \mathcal{L}}{\partial \bm{R}_{I}} \label{ElectronEulerLagrangian} \\
  \frac{d}{dt} \frac{\partial \mathcal{L}}{\partial \bra{\dot{\psi}_{i}}} &=& \frac{\partial \mathcal{L}}{\partial \bra{\psi_{i}}} \label{IonEulerLagrangian}
\end{eqnarray}
\end{subequations}
one obtains the associated equations of motion 
\begin{subequations}
\begin{eqnarray}
  M_{I} \Ddot{\bm{R}}_{I} &=& -\nabla_{\bm{R}_{I}} \left[ \min_{ \{\psi_{i}\} } E \bigl[ \{ \psi_{i} \}; \bm{R} \bigr] \biggm|_{ \{ \langle {\psi_{i}}|{\psi_{j}} \rangle = \delta_{ij} \} } \right] \nonumber \\
  &=& - \frac{\partial E}{\partial \bm{R}_{I}} + \sum_{i,j} \Lambda_{ij} \frac{\partial}{\partial \bm{R}_{I}} \langle {\psi_{i}}|{\psi_{j}} \rangle \nonumber \\ 
  &-& 2 \sum_{i} \frac{\partial \bra{\psi_{i}}}{\partial \bm{R}_{I}} \left[ \frac{\delta E}{\delta \bra{\psi_{i}}} - \sum_{j} \Lambda_{ij} \ket{\psi_{j}} \right] \label{IonBOMD_EOM} \\
  0 &\lesssim& - \frac{\delta E}{\delta \bra{\psi_{i}}} + \sum_{j} \Lambda_{ij} \ket{\psi_{j}} \nonumber \\
  &=& -\hat{H}_{e} \bra{\psi_{i}} + \sum_{j} \Lambda_{ij} \ket{\psi_{j}} \label{ElectronBOMD_EOM}
\end{eqnarray}
\end{subequations}

The first term on the right-hand side (RHS) of Eq.~(\ref{IonBOMD_EOM}) is the so-called Hellmann-Feynman force~\cite{Hellmann1937, Feynman1939}. The second term that is denoted as Pulay~\cite{PulayForces}, or wavefunction force $\mathbf{F}_{\text{WF}}$, is a constraint force due to the holonomic orthonormality constraint, and is nonvanishing if, and only if, the basis functions $\phi_{j}$ explicitly depend on $\bm{R}$. The final term stems from the fact that, independently of the particular basis set used, there is always an implicit dependence on the atomic positions. 
%
%
The factor 2 in Eq.~(\ref{IonBOMD_EOM}) stems from the assumption that the KS orbitals are real, an inessential simplification. Nevertheless, the whole term vanishes whenever $\psi_{i}(\bm{R})$ is an eigenfunction of the Hamiltonian within the subspace spanned by the not necessarily complete basis set~\cite{Almloef1981, Scheffler1985}. Note that this is a much weaker condition than the original Hellmann-Feynman theorem, of which we hence have not availed ourselves throughout the derivation, except as an eponym for the first RHS term of Eq.~(\ref{IonBOMD_EOM}). However, as the KS functional is nonlinear, eigenfunctions of its Hamiltonian $\hat{H}_{e}$ are only obtained at exact self-consistency, which is why the last term of Eq.~(\ref{IonBOMD_EOM}) is also referred to as non-self-consistent force $\mathbf{F}_{\text{NSC}}$~\cite{Bendt1983}. Unfortunately, this can not be assumed in any numerical calculation and results in immanent inconsistent forces as well as the inequality of Eq.~(\ref{ElectronBOMD_EOM}). Neglecting either $\mathbf{F}_{\text{WF}}$ or $\mathbf{F}_{\text{NSC}}$, i.e. applying the Hellmann-Feynman theorem to a non-eigenfunction, leads merely to a perturbative estimate of the generalized forces 
\begin{eqnarray}
  \mathbf{F} = \mathbf{F}_{\text{HF}} + \mathbf{F}_{\text{WF}} + \mathbf{F}_{\text{NSC}},
  \label{GeneralizedForce}
\end{eqnarray}
which, contrary to the energies, depends just linearly on the error in the electronic charge density. That is why it is much more exacting to calculate accurate forces rather than total energies.


\subsection{Second-generation Car-Parrinello molecular dynamics}
\label{subsection_CPMD}
Until recently, AIMD has mostly relied on two general methods: The original CPMD and the direct BOMD approach, each with its advantages and shortcomings. In BOMD, the total energy of a system, as determined by an arbitrary electronic structure method, is fully minimized in each MD time step, which renders this scheme computationally very demanding~\cite{Payne1992}. By contrast, the original CPMD technique obviates the rather time-consuming iterative energy minimization by considering the electronic degrees of freedom as classical time-dependent fields with a fictitious mass $\mu$ that are propagated by the use of Newtonian dynamics~\cite{Car1985}. In order to keep the electronic and nuclear subsystems adiabatically separated, which causes the electrons to follow the nuclei very close to their instantaneous electronic ground state, $\mu$ has to be chosen sufficiently small~\cite{Pastore1991}. However, in CPMD the maximum permissible integration time step scales like $\sim \sqrt{\mu}$ and therefore has to be significantly smaller than that of BOMD, hence limiting the attainable simulation timescales~\cite{Bornemann1998}. 

The so-called second-generation CPMD method combines the best of both approaches by retaining the large integration time steps of BOMD, while, at the same time, preserving the efficiency of CPMD~\cite{Kuehne2007,Kuehne2014,Prodan2018}. In this Car-Parrinello-like approach to BOMD, the original fictitious Newtonian dynamics of CPMD for the electronic degrees of freedom is substituted by an improved coupled electron-ion dynamics that keeps the electrons very close to the instantaneous ground-state without the necessity of an additional fictitious mass parameter. The superior efficiency of this method originates from the fact that only one preconditioned gradient computation is necessary per AIMD step. In fact, a gain in efficiency between one to two orders of magnitude has been observed for a large variety of different systems ranging from molecules and solids~\cite{Camellone2009, Cucinotta2009, Richters2013, Roehrig2013, Spura2015, Sanna2016, Richters2018, Oschatz2019}, 
including phase-change materials~\cite{Caravati2007, Caravati2009a, Caravati2009b, Caravati2011, Gabardi2013a, Gabardi2013b, Gabardi2016}, 
over aqueous systems~\cite{Kuehne2009, Pascal2012,  Zhang2013, Zhang2015, John2016, Ojha2018, Elgabarty2019, Clark2019, Ojha2019}, 
to complex interfaces~\cite{Kuehne2011, Luduena2011, Torun2014, Kessler2015, PartoviAzar2016, Nagata2016, Lan2018}. 

Within mean-field electronic structure methods, such as Hartree-Fock and KS-DFT, $E \left[ \{ \psi_i \}; \bm{R} \right]$ is either a functional of the electronic wavefunction that is described by a set of occupied MOs $\ket{\psi_i}$ or, equivalently, of the corresponding one-particle density operator $\rho = \sum_i {\ket{\psi_i} \bra{\psi_i}}$. The improved coupled electron-ion dynamics of second-generation CPMD obeys the following equations of motion for the nuclear and electronic degrees of freedom: 
\begin{subequations}
\begin{eqnarray}
M_{I} \ddot{\bm{R}}_I &=& -\nabla_{{\bm{R}}_{I}} \left[ \left. \min_{ \{ \psi_i \}} \, E \left[ \{ \psi_i \}; \bm{R}_{I} \right] \right|_{\{ \langle {\psi_i}|{\psi_j} \rangle = \delta_{ij} \}} \right] \nonumber \\
&=& - \frac{\partial E}{\partial \bm{R}_{I}} + \sum_{i,j} \Lambda_{ij} \frac{\partial}{\partial \bm{R}_{I}} \langle {\psi_i}|{\psi_j} \rangle \\
&-& 2 \sum_{i} \frac{\partial \bra{\psi_i}}{\partial \bm{R}_I} \left[ \frac{\partial E \left[ \{ \psi_i \}; \bm{R}_{I} \right]}{\partial \bra{\psi_i}} - \sum_j \Lambda_{ij} \ket{\psi_j} \right] \nonumber \label{NuclEOM} \\
\frac{d^2}{d \tau^2} \ket{\psi_i (\bm{r}, \tau)} &=& - \frac{\partial E \left[ \{ \psi_i \}; \bm{R}_{I} \right]} {\partial \bra{\psi_i (\bm{r}, \tau)}} - \gamma \omega \frac{d}{d\tau} \ket{\psi_i (\bm{r}, \tau)} \nonumber \\ 
&+& \sum_j \Lambda_{ij} \ket{\psi_j(\bm{r}, \tau)}. \label{ElecEOM}
\end{eqnarray}
\end{subequations}
The former is the conventional nuclear equation of motion of BOMD consisting of Hellmann-Feynman, Pulay and non-self-consistent force contributions~\cite{Hellmann1937,Feynman1939,PulayForces,Bendt1983}, whereas the latter constitutes an universal oscillator equation 
as obtained by a nondimensionalization. The first term on the RHS in Eq.~\ref{ElecEOM} can be sought of as an ``electronic force'' to propagate $\ket{\psi_i}$ in dimensionless time $\tau$. The second term is an additional damping term to relax more quickly to the instantaneous electronic ground-state, where $\gamma$ is an appropriate damping coefficient and $\omega$ the undamped angular frequency of the universal oscillator. The final term derives from the constraint to fulfill the holonomic orthonormality condition $\langle {\psi_i}|{\psi_j} \rangle = \delta_{ij}$, by using the Hermitian Lagrangian multiplier matrix $\bm{\Lambda}$. As can be seen in Eq.~\ref{ElecEOM}, not even a single diagonalization step, but just one ``electronic force'' calculation is required. In other words, the second-generation CPMD method not only entirely bypasses the necessity of a SCF cycle, but also the alternative iterative wavefunction optimization. 

However, contrary to the evolution of the nuclei, for the short-term integration of the electronic degrees of freedom accuracy is crucial, which is why a highly accurate yet efficient propagation scheme is essential. As a consequence, the evolution of the MOs is conducted by extending the always-stable predictor-corrector integrator of Kolafa to the electronic structure problem~\cite{Kolafa2004}. But, since this scheme was originally devised to deal with classical polarization, special attention must be paid to the fact that the holonomic orthonormality constraint, which is due to the fermionic nature of electrons that forces the wavefunction to be antisymmetric in order to comply with the Pauli exclusion principle, is always satisfied during the dynamics. 
For that purpose, first the predicted MOs at time $t_n$ are computed based on the electronic degrees of freedom from the K previous AIMD steps: 
\begin{equation}
\ket{\psi_i^p(t_n)} = \sum_{m}^{K} (-1)^{m+1} m \frac{\binom{2K}{K-m}}{\binom{2K-2}{K-1}} \rho(t_{n-m}) \ket{\psi_i(t_{n-1})}.
\end{equation}
This is to say that the predicted one-electron density operator $\rho^p(t_{n})$ is used as a projector onto to occupied subspace $\ket{\psi_i(t_{n-1})}$ of the previous AIMD step. In this way, we take advantage of the fact that $\rho^p(t_n)$ evolves much more smoothly than $\ket{\psi_i^p(t_{n})}$ and is therefore easier to predict. This is particularly true for metallic systems, where many states crowd the Fermi level. Yet, to minimize the error and to further reduce the deviation from the instantaneous electronic ground-state, $\ket{\psi_i^p(t_n)}$ is corrected by performing a single minimization step $\ket{\delta \psi_i^p(t_n)}$ along the preconditioned electronic gradient direction, as computed by the orthonormality conserving orbital OT method described in section~\ref{subsection_ot}~\cite{VandeVondele2003}. Therefore, the modified corrector can be written as: 
\begin{eqnarray}
\ket{\psi_i(t_n)} &=& \ket{\psi_i^p(t_n)} + \omega \left( \ket{\delta \psi_i^p(t_n)} - \ket{\psi_i^p(t_n)} \right), \nonumber \\
\text{with}~\omega &=& \frac{K}{2K-1}~\text{for}~K \ge 2.
\end{eqnarray}
The eventual predictor-corrector scheme leads to an electron dynamics that is rather accurate and time-reversible up to $\mathcal{O}(\Delta t^{2K-2})$, where $\Delta t$ is the discretized integration time step, while $\omega$ is chosen so as to guarantee a stable relaxation towards the minimum. In other words, the efficiency of the present second-generation CPMD method stems from the fact that the instantaneous electronic ground state is very closely approached within just one electronic gradient calculation. We thus totally avoid the SCF cycle and any expensive diagonalization steps, while remaining very close to the BO surface and, at the same time, $\Delta t$ can be chosen to be as large as in BOMD. 

However, in spite of the close proximity of the propagated MOs to the instantaneous electronic ground state, the nuclear dynamics is slightly dissipative, most likely because the employed predictor-corrector scheme is not symplectic. The validity of this assumption has been independently verified by various numerical studies of others~\cite{Dai2009, Hutter2011, Luo2014, Musso2018}.
Nevertheless, presuming that the energy is exponentially decaying, which had been shown to be an excellent assumption~\cite{Kuehne2007,Kuehne2014,Prodan2018}, it is possible to rigorously correct for the dissipation by modeling the nuclear forces of second-generation CPMD $\bm{F}^{CP}_I = - \nabla_{\bm{R}_I} E \left[ \{ \psi_i \}; \bm{R}_{I} \right]$ by 
\begin{equation}
\bm{F}_I^{CP} = \bm{F}_I^{BO}-\gamma_D M_I \dot{\bm{R}}_I, 
\end{equation}
where $\bm{F}_I^{BO}$ are the exact, but inherently unknown BO forces and $\gamma_D$ is an intrinsic, yet to be determined friction coefficient to mimic the dissipation. The presence of damping immediately suggests a canonical sampling of the Boltzmann distribution by means of the following modified Langevin-type equation: 
\begin{subequations}\label{Eq-Langevin}
\begin{eqnarray}
M_I \ddot{\bm{R}}_I &=& \underbrace{\bm{F}_I^{BO} - \gamma_D M_I \dot{\bm{R}}_I} + \bm{\Xi}_I^{D} \\
&=& \qquad \; \; \bm{F}_I^{CP}+ \bm{\Xi}_I^{D}, 
\end{eqnarray}
\end{subequations}
where $\bm{\Xi}_I^{D}$ is an additive white noise term, which must obey the fluctuation-dissipation theorem $\langle \bm{\Xi}_I^D(0) \bm{\Xi}_I^D(t) \rangle = 2 \gamma_D M_I k_B T \delta(t)$ in order to guarantee an accurate sampling of the Boltzmann distribution.  

This is to say that if one knew the unknown value of $\gamma_D$ it would nevertheless be possible to ensure an exact canonical sampling of the Boltzmann distribution in spite of the dissipation. Fortunately, the inherent value of $\gamma_D$ does not need to be known \textit{a priori}, but can be bootstrapped so as to generate the correct average temperature~\cite{Kuehne2007,Kuehne2014,Prodan2018}, as measured by the equipartition theorem 
\begin{equation} \label{EPT}
  \left< \frac{1}{2} M_I \dot{\bm{R}}_I^2 \right> = \frac{3}{2} k_B T. 
\end{equation}
More precisely, in order to determine the hitherto unknown value of $\gamma_D$, we perform a preliminary simulation in which we vary $\gamma_D$ on-the-fly using a Berendsen-like algorithm until Eq.~\ref{EPT} is eventually satisfied~\cite{Kuehne2009}. Alternatively, $\gamma_D$ can also be continously adjusted automatically using the adaptive Langevin technique of Leimkuhler and coworkers~\cite{JonesLeimkuhler2011, Mones2015, LeimkuhlerStoltz2019}. In this method, the friction coefficient $\gamma_D$ of Eq.~\ref{Eq-Langevin} is reinterpreted as a dynamical variable, defined by a negative feedback loop control law as in the Nos\'e-Hoover scheme~\cite{Nose,Hoover}. The corresponding dynamical equation for $\gamma_D$ reads as %
\begin{equation}
  \dot{\gamma}_D= (2K-n k_B T)/\mathcal{Q}, 
\end{equation}
where $K$ is the kinetic energy, $n$ is the number of degrees of freedom and $\mathcal{Q}=k_B T \tau^2_{NH}$ is the Nose-Hoover fictitious mass with time constant $\tau_{NH}$.

\subsection{Low-cost linear-scaling \emph{ab-initio} molecular dynamics based on compact localized molecular orbitals}
\label{sec:CLMOAIMD}
%

The computational complexity of CLMO DFT described in section~\ref{sec:ALMO-ON} grows linearly with the number of molecules, while its overhead cost remains very low because of the small number of electronic descriptors and efficient optimization algorithms (see Fig.~\ref{fig:clmos-mol}). These features make CLMO DFT a promising method for accurate AIMD simulations of large molecular systems.

The difficulty of adopting CLMO DFT for dynamical simulations arises from the nonvariational character of compact orbitals. While CLMOs can be reliably optimized using the two-stage SCF procedure described in section~\ref{sec:ALMO-ON}, the occupied subspace defined in the first stage must remain fixed during the second stage to achieve convergence. In addition, electron delocalization effects can suddenly become inactive in the course of a dynamical simulation when a neighboring molecule crosses the localization threshold $R_{c}$. Furthermore, the variational optimization of orbitals is in practice inherently not perfect and terminated once the norm of the electronic gradient drops below a small, but nevertheless finite convergence threshold $\epsilon_{\text{SCF}}$. These errors accumulate in long AIMD trajectories leading to non-physical sampling and/or numerical problems. 

Traditional strategies to cope with these problems are computationally expensive and include increasing $R_c$, decreasing $\epsilon_{\text{SCF}}$ and computing the nonvariational contribution to the forces via a coupled-perturbed procedure~\cite{Kussmann2013,Benoit2001}. \CPXK\ implements another approach that uses the CLMO state obtained after the two-stage CLMO SCF procedure to compute only the inexpensive Hellmann-Feynman and Pulay components and neglects the computationally intense nonvariational component of the nuclear forces. To compensate for the missing component, \CPXK\ uses a modified Langevin equation of motion that is fine-tuned to retain stable dynamics even with imperfect forces. This approach is known as the second-generation CPMD methodology of K\"uhne and workers~\cite{Kuehne2007, Kuehne2014, Prodan2018}, which is discussed in detail in section~\ref{subsection_CPMD}. Its combination with CLMO DFT is described in Ref.~\onlinecite{scheiber2018compact}. 

\begin{figure}
\includegraphics*[width=8cm]{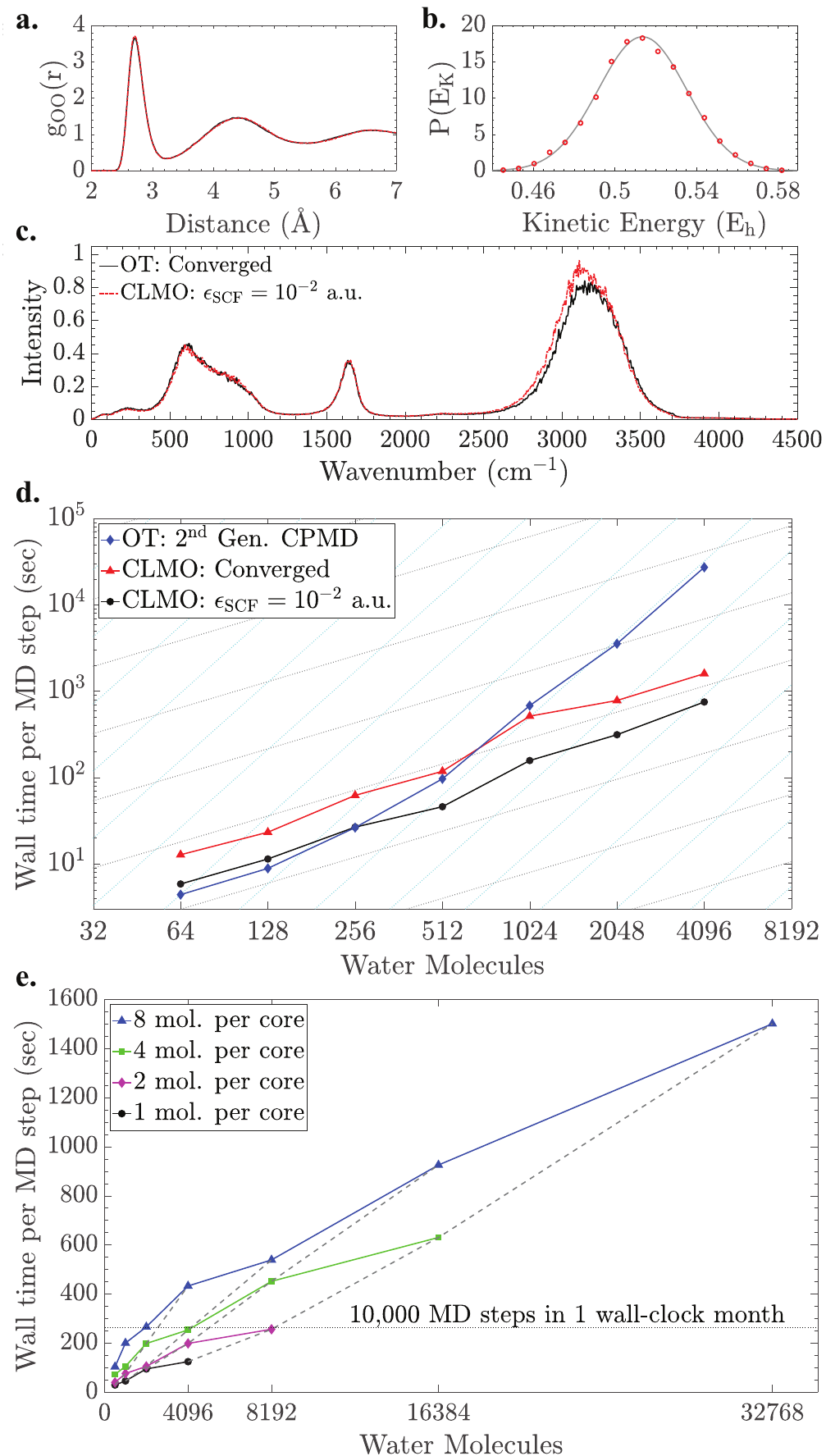}
\caption{\label{fig:clmos-aimd} Accuracy and efficiency of \LS\ AIMD based on the two-stage CLMO SCF for liquid water. 
All simulations were performed using the PBE XC functional and a TZV2P basis set at constant temperature (300~K) and density (0.9966~g/cm$^3$). For the CLMO methods, $R_{c} = 1.6$ vdWR.
(a) Radial distribution function, (b) kinetic energy distribution (the gray curve shows the theoretical Maxwell-Boltzmann distribution) and (c) IR spectrum were calculated using the fully converged OT method for delocalized orbitals (black line) and CLMO AIMD with $\epsilon_{\text{SCF}} = 10^{-2}$~a.u. 
(d) Timing benchmarks for liquid water on 256 compute cores. Cyan lines represent perfect cubic-scaling, whereas gray lines represent perfect linear-scaling. 
(e) Weak scalability benchmarks are performed with $\epsilon_{\text{SCF}} = 10^{-2}$~a.u. 
Dashed gray lines connect systems simulated on the same number of cores to confirm \LS\ behavior. 
See Ref.~\onlinecite{scheiber2018compact} for details.}
\end{figure}
An application of CLMO AIMD to liquid water demonstrates that the compensating terms in the modified Langevin equation can be tuned to maintain a stable dynamics and reproduce accurately multiple structural and dynamical properties of water with tight orbital localization ($R_c = 1.6$~vdWR) and $\epsilon_{\text{SCF}}$ as high as $10^{-2}$~a.u.~\cite{scheiber2018compact}. The corresponding results are shown in Fig.~\ref{fig:clmos-aimd}.
The low computational overhead of CLMO AIMD, afforded by these settings, makes it ideal for routine medium-scale AIMD simulations, while its linear-scaling complexity allows to extend first-principle studies of molecular systems to previously inaccessible length scales. 

It is important to note that AIMD in \CPXK\ cannot be combined with the CLMO methods based on the Harris functional and perturbative theory (see Section~\ref{sec:ALMO-ON}). A generalization of the CLMO AIMD methodology to systems of strongly interacting atoms (e.g. covalent crystals) is underway.

\subsection{Multiple-time-step integrator}
\label{subsection_mts}
The AIMD-based MTS integrator presented here is based on the reversible reference system propagator algorithm (r-RESPA), which was developed for classical molecular dynamics (MD) simulations~\cite{Tuckerman1992}.
Using a carefully constructed integration scheme, the time evolution is reversible, and the MD simulation remains accurate and energy conserving.
In AIMD-MTS, the difference in computational cost between a hybrid and a local XC functional is exploited, by
performing a hybrid calculation only after several conventional DFT integration time-steps.
r-RESPA is derived from the Liouville operator representation of Hamilton mechanics:
\begin{equation}
iL= \sum_{j=1}^{f}\left[\frac{\partial H}{\partial {p_j}} \frac{\partial{}}{\partial {x_j}}+\frac{\partial H}{\partial {x_j}} \frac{\partial{}}{\partial {p_j}} \right ],
\end{equation}
where $L$ is the Liouville operator for the system containing $f$ degrees of freedom. The corresponding positions and momenta are denoted as $x_j$ and $p_j$, respectively.
This operator is then used to create the classical propagator $U(t)$ for the system, which reads as
\begin{equation}
U(t)=e^{iLt}.
\end{equation}
Decomposing the Liouville operator into two parts
\begin{equation}
iL=iL_1 + iL_2,
\end{equation}
and applying a 2nd-order Trotter-decomposition to the corresponding propagator yields
\begin{eqnarray}
e^{i(L_1+L_2)\Delta t}&=&\left[e^{i(L_1 + L_2)\Delta t/n}\right]^n \\ &=&\left[e^{iL_1(\delta t / 2)} e^{iL_2\delta t}e^{iL_1(\delta t / 2)} \right]^n +O(\delta t^3), \nonumber
\end{eqnarray}
with $\delta t = \Delta t/n$.
For this propagator, several integrator schemes can be derived~\cite{Martyna1996}.
The extension for AIMD-MTS is obtained from a decomposition of the force
in the Liouville operator into two or more separate forces, i.e.
\begin{equation}
iL= \sum_{j=1}^{f}\left[\dot{x}_j \frac{\partial{}}{\partial {x_j}}+F_j^1 \frac{\partial{}}{\partial {p_j}}+F_j^2 \frac{\partial{}}{\partial {p_j}} \right ].
\end{equation}
For that specific case, the propagator can be written as
\begin{eqnarray}
e^{iL\Delta t}&=&e^{(\Delta t / 2)\mathbf{F}^2 \frac{\partial{}}{\partial \mathbf{p}}}\left[e^{(\delta t / 2)\mathbf{F}^1 \frac{\partial{}}{\partial \mathbf{p}}} e^{\delta t\dot{x}_j \frac{\partial{}}{\partial {x_j}}}e^{(\delta t / 2)\mathbf{F}^1 \frac{\partial{}}{\partial \mathbf{p}}} \right]^n \nonumber \\
&\times& e^{(\Delta t / 2)\mathbf{F}^2 \frac{\partial{}}{\partial \mathbf{p}}}.
\end{eqnarray}
This allows to treat $\mathbf{F}^1$ and $\mathbf{F}^2$ with different time-steps, while the whole propagator still remains time-reversible.
The procedure for $\mathbf{F}^1$ and $\mathbf{F}^2$ will be referred to as the inner and the outer loop, respectively.
In the AIMD-MTS approach, the forces are split in the following way
\begin{subequations}
\begin{eqnarray}
\mathbf{F}^1 &=& \mathbf{F}^{\textrm{approx}} \\
\mathbf{F}^2 &=& \mathbf{F}^{\textrm{accur}}-\mathbf{F}^{\textrm{approx}}, 
\end{eqnarray}
\end{subequations}
where $\mathbf{F}^{\textrm{accur}}$ are the forces computed by the more accurate method, e.g. hybrid DFT, whereas  $\mathbf{F}^{approx}$ are forces as obtained from a less accurate method, e.g. by GGA XC functionals.
Obviously, the corresponding Liouville operator equals the purely accurate one.
The advantage of this splitting is that the magnitude of $\mathbf{F}^2$ is usually much smaller than that of $\mathbf{F}^1$.
To appreciate that, it has to be considered how closely geometries and frequencies obtained by hybrid DFT normally match the ones obtained by a local XC functional, in particular for stiff degrees of freedom.
The difference of the corresponding Hessians is therefore small and low-frequent.
However, high-frequency parts are not removed analytically, thus the theoretical upper limit for the time-step of the outer loop remains half the period of the fastest vibration~\cite{Batcho2001}.
The gain originates from an increased accuracy and stability for larger time-steps in the outer loop integration.
Even using an outer loop time-step close to the theoretical limit, a stable and accurate AIMD is obtained.
Additionally, there is no shift to higher frequencies as the (outer loop) time-step is increased, contrary to the single time-step case.
%

\section{Energy decomposition and spectroscopic analysis methods}
\label{section_analysis}
Within \cp, the nature of intermolecular bonding and vibrational spectra can be rationalized by EDA, normal mode analysis (NMA) and mode selective vibrational analysis (MSVA) methods, respectively. 

\subsection{Energy decomposition analysis based on compact localized molecular orbitals} \label{sec:CLMO-EDA}
Intermolecular bonding is a result of the interplay of electrostatic interactions between permanent charges and multipole moments on molecules, polarization effects, Pauli repulsion, donor-acceptor orbital interactions (also known as covalent, charge-transfer or delocalization interactions) and weak dispersive forces. The goal of EDA is to measure the contribution of these components within the total binding energy, thus gaining deeper insight into physical origins of intermolecular bonds.

To that extent, \CPXK\ contains an EDA method based on ALMOs -- molecular orbitals localized entirely on single molecules or ions within a larger system~\cite{Khaliullin2007, Khaliullin2013}. 
The ALMO EDA separates the total interaction energy ($\Delta E_{\text{TOT}}$) into a frozen density (FRZ), polarization (POL) and charge-transfer (CT) terms, i.e.
\begin{equation}
	\Delta E_{\text{TOT}} = \Delta E_{\text{FRZ}} + \Delta E_{\text{POL}} + \Delta E_{\text{CT}}, 
\end{equation}
which is conceptually similar to the Kitaura-Morokuma EDA~\cite{Kitaura1976} -- one of the first EDA methods. The frozen interactions term is defined as the energy required to bring isolated molecules into the system without any relaxation of their MOs, apart from modifications associated with satisfying the Pauli exclusion principle: 
\begin{equation}
\Delta E_{\text{FRZ}} \equiv E(\mathbf{R}_{\text{FRZ}}) - \sum_{x} E(\mathbf{R}_x), 
\end{equation}
where $E(\mathbf{R}_x)$ is the energy of isolated molecule $x$ and $\mathbf{R}_x$ the corresponding density matrix, whereas $\mathbf{R}_{\text{FRZ}}$ is the density matrix of the system constructed from the unrelaxed MOs of the isolated molecules. 
The ALMO EDA is also closely related to the block-localized wavefunction EDA~\cite{Mo2000}, because both approaches use the same variational definition of the polarisation term as the energy lowering due to the relaxation of each molecule's ALMOs in the field of all other molecules in the system:
\begin{equation}
\Delta E_{\text{POL}} \equiv E(\mathbf{R}_{\text{ALMO}}) - E(\mathbf{R}_{\text{FRZ}}).
\end{equation}
The strict locality of ALMOs is utilized to ensure that the relaxation is constrained to include only intramolecular variations. This approach, whose mathematical and algorithmic details have been described by many authors~\cite{Stoll1980, Gianinetti1996,Nagata2001, Khaliullin2006, Khaliullin2013}, gives an upper limit to the true polarisation energy~\cite{azar2013useful}.
The remaining portion of the total interaction energy, the CT term, is calculated as the difference in the energy of the relaxed ALMO state and the state of fully delocalized optimized orbitals ($\mathbf{R}_{\text{SCF}}$): 
\begin{equation}
\Delta E_{CT} \equiv E(\mathbf{R}_{\text{SCF}}) - E(\mathbf{R}_{\text{ALMO}}).
\end{equation}
%
A distinctive feature of the ALMO EDA is that the charge-transfer contribution can be separated into contributions associated with forward and back-donation for each pair of molecules, as well as a many-body higher-order (induction) contribution (HO), which is very small for typical intermolecular interactions. Both, the amount of the electron density transferred between a pair of molecules and the corresponding energy lowering can be computed:
\begin{subequations}
\begin{eqnarray}
\label{eq:qedeldec}
\Delta Q_{\text{CT}} &=& \sum_{x,y > x} \{ \Delta Q_{x\rightarrow y} + \Delta Q_{y\rightarrow x} \} + \Delta Q_{\text{HO}} \\
\Delta E_{\text{CT}} &=& \sum_{x,y > x} \{ \Delta E_{x\rightarrow y} + \Delta E_{y\rightarrow x} \} + \Delta E_{\text{HO}}.
\end{eqnarray}
\end{subequations}

The ALMO EDA implementation in \CPXK\ is currently restricted to closed-shell fragments. The efficient \LS\ optimization of ALMOs serves as its underlying computational engine~\cite{Khaliullin2013a}. The ALMO EDA in \CPXK\ can be applied to both gas-phase and condensed matter systems. It has been recently extended to fractionally occupied ALMOs~\cite{staub2019energy}, thus enabling the investigation of interactions between metal surfaces and molecular adsorbates. Another unique feature of the implementation in \CPXK\ is the ability to control the spatial range of charge-transfer between molecules using the cutoff radius $R_c$ of CLMOs (see Section~\ref{sec:ALMO-ON}). Additionally, the ALMO EDA in combination with \CPXK's efficient AIMD engine allows to switch off the CT term in AIMD simulations, thus measuring the CT contribution to dynamical properties of molecular systems~\cite{shi2018contribution}.

\begin{figure}
\includegraphics*[width=8cm]{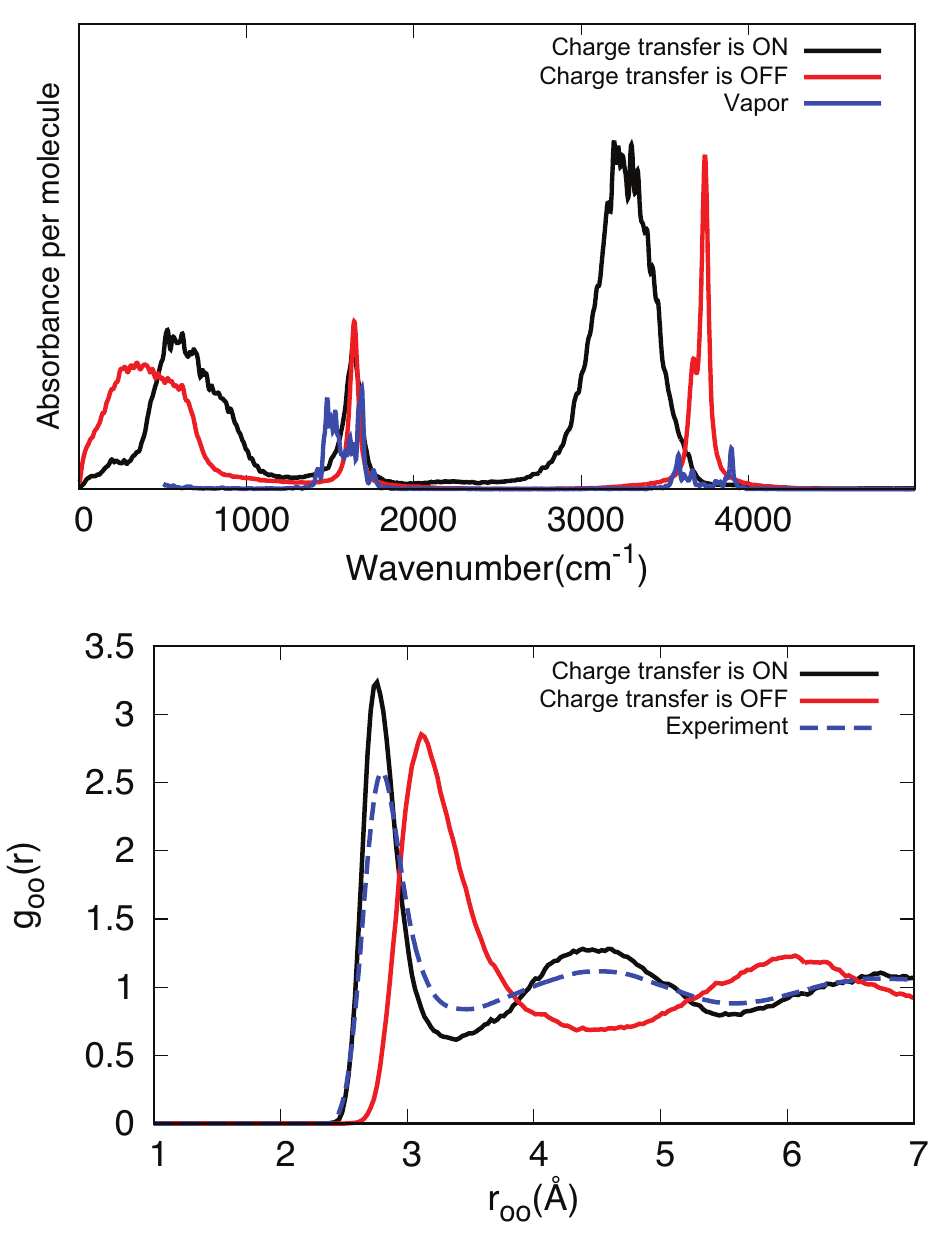}
\caption{\label{fig:covout} Comparison of the calculated radial distribution functions and IR spectra computed for liquid water based on AIMD simulations at the the BLYP+D3/TZV2P level of theory with and without the CT terms. See Ref.~\onlinecite{shi2018contribution} for details.}
\end{figure}
The ALMO EDA has been applied to study intermolecular interactions in a variety of gas- and condensed-phase molecular systems~\cite{luo2018why, staub2019energy}. 
The \CPXK\ implementation of ALMO EDA has been particularly advantageous to understand the nature of hydrogen bonding in bulk liquid water~\cite{Kuehne2013, Khaliullin2013, Zhang2013, NatureComm2015, shi2018contribution, Ojha2018, Elgabarty2019, Ojha2019}, ice~\cite{Kuehne2014a} and water phases confined to low dimensions~\cite{yun2019low}. Multiple studies have pointed out a deep connection between the donor-acceptor interactions and features of the X-ray absorption~\cite{Fransson2016, Kuehne2013}, infrared (IR)~\cite{Zhang2013, Ramos-Cordoba2011, Lenz2006, Zhang2015, shi2018contribution, Ojha2018, Ojha2019} (Fig.~\ref{fig:covout}), and nuclear magnetic resonance spectra of liquid water~\cite{NatureComm2015}. 
Extension of ALMO EDA to AIMD simulations have shown that the insignificant covalent component of HB determines many unique properties of liquid water including its structure (Fig.~\ref{fig:covout}), dielectric constant, hydrogen bond lifetime, rate of self-diffusion and viscosity~\cite{scheiber2018compact,shi2018contribution}.

\subsection{Normal mode analysis of infrared spectroscopy}
\label{subsection_NMA}
IR vibrational spectra can be obtained with CP2K by carrying out a NMA, within the Born-Oppenheimer approximation. The expansion of the potential energy in a Taylor series around the equilibrium geometry gives
\begin{eqnarray}
E_{\text{pot}}(\{{\bf R}_A\}) &=& E_{\text{pot}}^{(0)} + \sum_A\left( \frac{\partial E_{\text{pot}}}{\partial {\bf R}_A}\right) \delta{\bf R}_A \\ 
&+& \frac{1}{2}\sum_{AB} \left( \frac{\partial^2E_{\text{pot}}}{\partial {\bf R}_A\partial {\bf R}_B}\right) \delta{\bf R}_A \delta{\bf R}_B + \dots. \nonumber
\end{eqnarray}
At the equilibrium geometry, the first derivative is zero by definition, hence in the harmonic approximation only the second derivative matrix (Hessian) has to be calculated. 
In our implementation, the Hessian is obtained with the three-point central difference approximation, which for a system of $M$ particles corresponds to 6$M$ force evaluations. 
The Hessian matrix is diagonalised to determine the $3M$ $\mathcal{Q}$ vectors representing a set of decoupled coordinates, which are the normal modes. In mass weighted coordinates, the angular frequencies of the corresponding vibrations are then obtained by the eigenvalues of the second derivative matrix.

\subsection{Mode selective vibrational analysis}
\label{subsection_MSVIB}
For applications in which only few vibrations are of interest, the MSVA presents a possibility to reduce the computational cost significantly~\cite{Reiher2003,Reiher2004}.
Instead of computing the full second derivative matrix, only a subspace is evaluated iteratively using the Davidson algorithm.
As in NMA, the starting point of this method is the eigenvalue equation 
\begin{equation}
{\bf H}^{(m)} \mathcal{Q}_k = \lambda_k \mathcal{Q}_k,
\end{equation}
where ${\bf H}^{(m)}$ is the mass weighted Hessian in Cartesian coordinates. 
Instead of numerically computing the Hessian elementwise, the action of the Hessian on a predefined or arbitrary collective displacement vector
\begin{equation}
{\bf d}=\sum_i d_i e^m_i
\end{equation} 
is chosen, where $e^{(m)}_i$ are the 3$M$ nuclear basis vectors. 
The action of the Hessian on ${\bf d}$ is given in first approximation as
\begin{equation}
\sigma_k = ({\bf H}^{(m)} {\bf d})_k = \sum_l \frac{\partial^2 E_{\text{pot}}}{\partial R^{(m)}_l\partial R^{(m)}_k} d_l = \frac{\partial^2 E_{\text{pot}}}{\partial R^{(m)}_k \partial {\bf d}}. 
\end{equation}
The first derivatives with respect to the nuclear positions is simply the force which can be computed analytically.
The derivative of the force along components $i$ with respect to $d$ can then be evaluated as a numerical derivative using the three-point central difference approximation to yield the vector ${\boldsymbol \sigma}$.
%
The subspace approximation to the Hessian at the $i$-th iteration of the procedure is a $i \times i$ matrix
\begin{equation}
{\bf H}^{(m), i}_\text{approx} = {\bf D}^T{{\bf H}^{(m)}} {\bf D} = {\bf D}^T{\boldsymbol \Sigma},
\end{equation}
 where ${\bf B}$ and ${\Sigma}$ are the collection of the ${\bf d}$ and ${\boldsymbol \Sigma}$ vectors up to
the actual iteration step. Solving the eigenvalue problem for the small Davidson matrix 
\begin{equation}
{\bf H}^{(m), i}_\text{approx} {\bf u}^i = \tilde{\lambda}^{i}, {\bf u}^i 
\end{equation}
we obtain approximate eigenvalues $\tilde{\lambda}_k^{i}$. From the resulting eigenvectors.
The residuum vector 
\begin{equation}
{\bf r}_m^i = \sum_{k=1}^{i} u^i_{mk} [{\boldsymbol \sigma}^k - \tilde{\lambda}_m^{i} {\bf d}^k],
\end{equation}
where the sum is over all the basis vectors ${\bf d}^k$, the number of which increases at each new iteration $i$. The residuum vector is used as displacement vector for the next iteration. The approximation to the the exact eigenvector $m$ at the $i$-th iteration is
\begin{equation}
 \mathcal{Q}_m \approx \sum_{k=1}^{i} u^i_{mk} {\bf d}^k.
\end{equation}
This method avoids the evaluation of the complete Hessian, and therefore requires fewer force evaluations.
Yet, in the limit of $3M$ iterations, the exact Hessian is obtained and thus the exact frequencies and normal modes in the limit of the numerical difference approximation.
 
As this is an iterative procedure (Davidson subspace algorithm), the initial guess is important for convergence. Moreover, there is no guaranty that the created subspace will contain the desired modes in case of a bad initial guess. In CP2K the choice of the target mode can be a single mode, a range of frequencies, or modes localised on preselected atoms. If little is known about the modes of interest (e.g. a frequency range and contributing atoms) an initial guess can be build by a random displacement of atoms. In case, where a single mode is tracked, one can use normal modes obtained from lower quality methods. The residuum vector is calculated with respect to the mode with the eigenvalue closest to the input frequency. The method will only converge to the mode of interest if the initial guess is suitable. With the implemented algorithm always the mode closest to the input frequency is improved. Using an arbitrary initial guess, the mode of interest might not be present in the subspace at the beginning. It is important to note that the Davidson algorithm might converge before the desired mode becomes part of the subspace. Therefore, there is no warranty that the algorithm would always converges to the desired mode. By giving a range of frequencies as initial guess, it might happen that either none or more than one mode is already present in the spectrum. In the first case, the mode closest to the desired range will be tracked. In the second case, always the least converged mode will be improved. If the mode is selected by a list of contributing atoms, at each step the approximations to the eigenvectors of the full Hessian are calculated, and the vector with the largest contributions of the selected atoms is tracked for the next iteration step.
 
The MSVA scheme can be efficiently parallelized by either distributing the force calculations or using the block Davidson algorithm. In the latter approach, the parallel environment is split into $n$ sub-environments, each consisting out of $m$ processors performing the force calculations. The initial vectors are constructed for all $n $ environments, such that $n$ orthonormal vectors are generated. After each iteration step, the new ${\bf d}$ and ${\boldsymbol \sigma}$ vectors are combined into a single ${\bf D}$ and ${\boldsymbol \Sigma}$ matrix. These matrices are then used to construct the approximate Hessian, and from this the $n$ modes closest to the selected vectors are again distributed.
 
The IR intensities are defined as the derivatives of the dipole vector with respect to the normal mode. Therefore, it is necessary to activate the computation of the dipoles together with the NMA. By applying the mode selective approach large condensed matter systems can be addressed. An illustrative example is the study by Schiffmann et al. \cite{Schiffmann2010}, where the interaction of the N3, N719, and N712 dyes with anatase(101) have been modelled. The vibrational spectra for all low-energy conformations have been computed and used to assist the assignment of the experimental IR spectra, revealing a protonation dependent binding mode and the role of self-assembly in reaching high coverage.
%

\section{Embedding Methods}
\label{section_embedding}
%
%
\CPXK\ aims to provide a wide range of potential energy methods, ranging from empirical approaches such as classical force-fields over DFT-based techniques to quantum chemical methods. In addition multiple descriptions can be arbitrarily combined at the input level, so that many combinations of methods are directly available. Examples are schemes that combine two or more potential energy surfaces via 
\begin{equation}
    E[\mathbf{R}] = E^{MM}[\mathbf{R}_{I+II}] -  E^{MM}[\mathbf{R}_{I}] + E^{QM}[\mathbf{R}_{I}],
\end{equation}
linear combinations of potentials as necessary in alchemical free-energy calculations 
\begin{equation}
    E_{\lambda}[\mathbf{R}] = \lambda E_{I}[\mathbf{R}] + (1-\lambda) E_{II}[\mathbf{R}],
\end{equation}
or propagation of the lowest potential energy surfaces  
\begin{equation}
    E[\mathbf{R}] = \min \left[ E_{I}[\mathbf{R}], E_{II}[\mathbf{R}], ... \right].
\end{equation}
However, beside such rather simplistic techniques~\cite{Schusteritsch}, more sophisticated embedding methods described in the following are also available within \CPXK.

\subsection{QM/MM methods}
\label{subsection_qmmm}
The QM/MM multi-grid implementation in CP2K is based on the use of an additive QM/MM scheme~\cite{Sherwood2000,Field1990,Singh1986}. The total energy of the molecular system can be partitioned into three
disjointed terms: 
\begin{eqnarray}
   E_{TOT} (\Rqm, \Rmm) &=& E^{QM} (\Rqm) + E^{MM}(\Rmm) \nonumber \\ 
   &+& E^{QM/MM}(\Rqm, \Rmm). 
\end{eqnarray}
These energy terms depend parametrically on the coordinates of the nuclei in the quantum region ($\Rqm$) and on classical atoms ($\Rmm$). Hence, $E^{QM}$ is the pure quantum energy, computed using the \quickstep code~\cite{VandeVondele2005}, whereas $E^{MM}$ is the classical energy, described through the use of the internal classical molecular-mechanics (MM) driver called \textsc{FIST}. The latter allows the use of the most common force-fields employed in MM simulations~\cite{Brooks1983,Case2005}.
The interaction energy term $E^{QM/MM}$ contains all non-bonded
contributions between the QM and the MM subsystems, and in a DFT
framework we express it as: 
\begin{eqnarray}
\label{qmmm_base:qmmm_hamiltonian}
   E^{QM/MM} (\Rqm,\Rmm) &=& \sum_{a \in MM} q_a \int \rho(\rr, \Rqm) v_a(|\rr-\Rmm|)
d \rr \nonumber \\
&+& \sum_{a \in MM, \alpha \in QM} v_{\text{NB}}(\Rqm, \Rmm), 
\end{eqnarray}
where $\Rmm$ is the position of the MM atom $a$ with charge $q_a$, $\rho(\rr,\Rqm)$ is the total (electronic plus nuclear) charge density of the quantum system, and $v_{\text{NB}}(\Rqm, \Rmm)$ is the non--bonded interaction between classical atom $a$ and quantum atom $\alpha$. The electrostatic potential of the MM atoms $v_a(|\rr-\Rmm|)$ is described using for each atomic charge a Gaussian charge distribution
\begin{equation}
\rho(|\rr-\Rmm|)=\left(\frac{1}{\sqrt{\pi} r_{c,a}}\right)^3
\exp(-(|\rr-\Rmm|/r_{c,a})^2), 
\end{equation}
with width $r_{c,a}$, eventually resulting in:
\begin{equation}
 v_a(|\rr-\Rmm|) = \frac{\erf(|\rr-\Rmm|/r_{c,a})}{|\rr-\Rmm|}. 
\end{equation}
This renormalised potential has the desired property of tending to $1/r$ at
large distances and going smoothly to a constant
for small $r$ (see Ref.~\onlinecite{TLFMALMP:2005} for renormalization details).
Due to the Coulomb long-range behavior, the  computational cost of the 
integral in Eq.~\ref{qmmm_base:qmmm_hamiltonian} can be very large.
In CP2K, we designed a decomposition of the
electrostatic potential in terms of Gaussian functions with different
cutoffs combined with a real-space multi-grid framework to accelerate the calculation 
of the electrostatic interaction. We named this method Gaussian Expansion of the Electrostatic Potential or GEEP~\cite{TLFMALMP:2005}.
The advantage of this methodology is that grids of different
spacing can be used to represent the different contributions of
$v_{a}(\rr,\Rmm)$, instead of using only the same grid employed for the mapping of the electronic wavefunction. 
In fact, by writing a function as a sum of terms with compact
support and with different cutoffs, the mapping 
of the function can be efficiently achieved using different grid levels, in
principle as many levels as contributing terms, each optimal to
describe the corresponding term. 

\subsubsection{QM/MM for isolated systems}
\label{subsubsection_isolated_qmmm}
For isolated systems, each MM atom is represented as a continuous Gaussian charge
distribution and each GEEP term is mapped on one of the available grid
levels, chosen to be the first grid whose cutoff is equal to or bigger
than the cutoff of that particular GEEP contribution. 
However, all atoms contribute to the coarsest grid level through the long-range $R_{low}$ part, which is
the smooth GEEP component~\cite{TLFMALMP:2005}.
The result of this collocation procedure is a multi-grid
representation of the QM/MM electrostatic potential
$V_i^{QM/MM}(\rr,\Rmm)$, where $i$ labels the grid level,
represented by a sum of single atomic contributions
$V_i^{QM/MM}(\rr,\Rmm) = \sum_{a \in MM} v_a^i(\rr, \Rmm)$, on that
particular grid level.
In a realistic system, the collocation represents most of the
computational time spent in the evaluation of 
the QM/MM electrostatic potential that is around $60-80 \%$.
Afterwards, the multi-grid expansion $V_i^{QM/MM}(\rr,\Rmm)$ is
sequentially interpolated starting from the coarsest grid level up to the
finest level, using real-space interpolator and restrictor operators.

Using the real-space multi-grid operators together with the GEEP
expansion, the prefactor in the evaluation of the QM/MM electrostatic
potential has been lowered from $N_f*N_f*N_f$ to
$N_c*N_c*N_c$, where $N_f$ is the number of grid points on the
finest grid and $N_c$ is the number of grid points on the coarsest
grid. The computational cost of the other operations for evaluating
the electrostatic potential, such as the mapping of the Gaussians and
the interpolations, becomes negligible in the limit of a large MM
system, usually more than 600-800 MM atoms.

Using the fact that grids are commensurate ($N_f/N_c =
2^{3(N_{grid}-1)}$), and employing for every calculation 4 grid
levels, the speed-up factor is around 512 ($2^9$). This means
that the present implementation is 2 orders of magnitude
faster than the direct analytical evaluation of the potential on the
grid.

\subsubsection{QM/MM for periodic systems}
\label{subsubsection_periodic_qmmm}
The effect of the periodic replicas of the MM subsystem is only in the
long-range term, and comes
entirely from the residual function $R_{low}(\rr,\Rmm)$:
\begin{eqnarray}
V^{QM/MM}_{recip}(\rr, \Rmm) &=& \sum_{\nn}^{\infty}
\sum_a q_a v_a^{recip} \\
&=& \sum_{\nn}^{\infty} \sum_a q_a R_{low}(|\rr-\Rmm+\nn|), \nonumber
\end{eqnarray}
where $L$ labels the infinite sums over the period replicas. Performing the same manipulation used in  Ewald summation~\cite{Ewald1921}, the previous equation can be computed more efficiently in reciprocal space, i.e.
\begin{eqnarray}
\label{qmmm_periodic:pot_rec}
V^{QM/MM}_{recip}(\rri, \Rmm) &=& L^{-3} \sum^{k_{cut}}_{\kk} \sum_a
q_a \tilde{R}_{low}(\kk) \nonumber \\
&\times& \cos{[2 \pi \kk \cdot (\rri-\Rmm)]}.
\end{eqnarray}
The term $\tilde{R}_{low}(\kk)$, representing the Fourier transform of
the smooth electrostatic potential, can be evaluated analytically via:
\begin{eqnarray}
\label{qmmm_periodic:potential_gspace}
\tilde{R}_{low}(\kk) &=& \left[ \frac{4 \pi}{|\kk|^2} \right] \exp 
\left( - \frac{|\kk|^2 r_{c,a}^2}{4} \right) \nonumber \\ 
&-& \sum_{N_g} A_{g}
(\pi)^{\frac{3}{2}} G_g^3 \exp \left(-\frac{G_g^2 |\kk|^2}{4}\right).
\end{eqnarray}
The potential in Eq.~\ref{qmmm_periodic:pot_rec} can be mapped on the coarsest
available grid. Once the electrostatic potential of a single MM charge within periodic
boundary conditions is derived, the evaluation of the electrostatic potential due to the MM subsystem
is easily computed employing the same multi-grid operators (interpolation and restriction) used for isolated systems.

The description of the long-range QM/MM interaction with periodic
boundary conditions requires the description of the QM/QM periodic
interactions, which plays a significant role if the QM subsystem
has a net charge different from zero, or a significant dipole moment.
Here, we exploit a technique  proposed few years ago by Bl\"{o}chl 
\cite{PEB:1995}, for decoupling the periodic images and restoring the
correct periodicity also for the QM part. A full and comprehensive description of the methods is reported in~\cite{TLFMALMP:2006}.
%
%

\subsubsection{Image charge augmented QM/MM}
\label{subsubsection_ic-QMMM}
\begin{figure} 
    \includegraphics[width=0.99\linewidth]{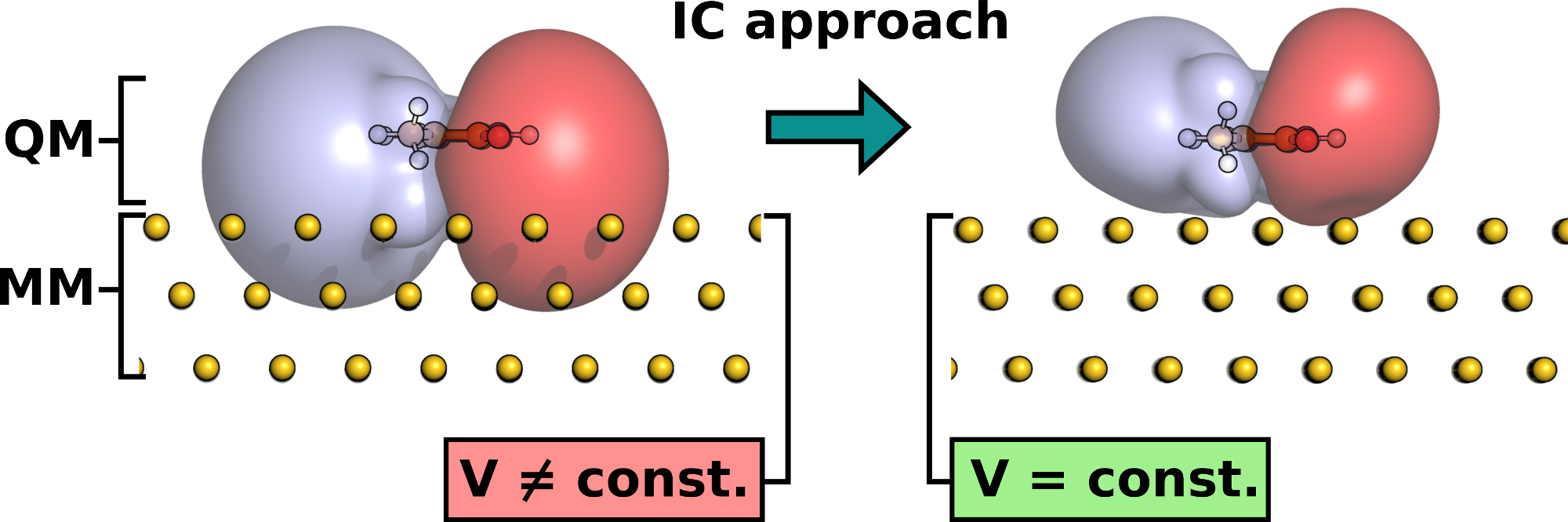}
    \caption{\label{fig:icqmmm} Simulation of molecules at metallic surfaces using the IC-QM/MM approach. Isosurface of the electrostatic potential at 0.0065~a.u. (red) and $-$0.0065~a.u. (blue) for a single thymine molecule at Au(111). Left: standard QM/MM approach, where the electrostatic potential of the molecule extends beyond the surface into the metal. Right: IC-QM/MM approach reproducing the correct electrostatics expected for a metallic conductor.
    }
\end{figure}
The image charge (IC) augmented QM/MM model in CP2K has been developed for the simulation of molecules adsorbed at metallic interfaces~\cite{Golze2013}. In the IC-QM/MM scheme, the adsorbates are described by KS-DFT and the metallic substrate is treated at the MM level of theory. The interactions between the QM and MM subsystems are modeled by empirical potentials of, e.g. the Lennard-Jones-type to reproduce dispersion and Pauli repulsion. Polarization effects due to electrostatic screening in metallic conductors are explicitly accounted for by applying the IC formulation.\par
The charge distribution of the adsorbed molecules generates an electrostatic potential $V_e(\mathbf{r})$, which extends into the substrate. If the electrostatic response of the metal is not taken into account in our QM/MM setup, the electrostatic potential has different values at different points $\mathbf{r}$ inside the metal slab, as illustrated in Fig.~\ref{fig:icqmmm}. However, the correct physical behavior is that $V_e(\mathbf{r})$ induces an electrostatic response such that the potential in the metal is zero or at least constant. Following Ref.~\onlinecite{Siepmann1995}, we describe the electrostatic response of the metallic conductor by introducing an image charge distribution $\rho_m(\mathbf{r})$, which is modeled by a set of Gaussian charges $\{g_a\}$ centered at the metal atoms, so that
\begin{equation}
    \rho_m(\mathbf{r}) = \sum_a c_a g_a(\mathbf{r,\mathbf{R}_a}),
\end{equation}
where $\mathbf{R}_a$ are the positions of the metal atoms. The unknown expansion coefficients $c_a$ are determined self-consistently imposing the constant-potential condition, i.e. the potential $V_m(\mathbf{r})$ generated by $\rho_m(\mathbf{r})$ screens $V_e(\mathbf{r})$ within the metal, so that $V_e(\mathbf{r})+V_m(\mathbf{r}) = V_0$, where $V_0$ is a constant potential that can be different from zero if an external potential is applied. The modification of the electrostatic potential upon application of the IC correction is shown in Fig.~\ref{fig:icqmmm}. Details on the underlying theory and implementation can be found in Ref.~\onlinecite{Golze2013}.\par
The IC-QM/MM scheme provides a computationally efficient way to describe the MM-based electrostatic interactions between adsorbate and metal in a fully self-consistent fashion allowing the charge densities of the QM and MM parts to mutually modify each other. Except for the positions of the metal atoms no input parameters are required. The computational overhead compared to a conventional QM/MM scheme is in the current implementation negligible. The IC augmentation adds an attractive interaction because $\rho_m(\mathbf{r})$ mirrors the charge distribution of the molecule, but has the opposite sign (see Ref~\onlinecite{Golze2013} for the energy expression). Therefore, the IC scheme strengthens the interactions between molecules and metallic substrates, in particular if adsorbates are polar or even charged. It also partially accounts for the rearrangement of the electronic structure of the molecules when they approach the surface~\cite{Golze2013}. \par
The IC-QM/MM approach is a good model for systems, where the accurate description of the molecule-molecule interactions is of primary importance and the surface acts merely as a template. For such systems, it is sufficient when the augmented QM/MM model predicts the structure of the first adsorption layer correctly. The IC-QM/MM approach has been applied, for example, to liquid water/Pt(111) interfaces~\cite{Golze2013,Steinmann2018}, organic thin-film growth on Au(100)~\cite{Evers2019}, as well as aqueous solutions of DNA molecules at gold interfaces~\cite{Dou2019,Maier2019}. Instructions how to set up an IC-QM/MM calculation with CP2K can be found in Ref.~\onlinecite{icqmmm_tutorial}.

\subsubsection{Partial atomic charges from electrostatic potential fitting}
\label{subsubsection_RESP}
Electrostatic potential (ESP) fitting is a popular approach to determine a set of partial atomic point charges $\{q_a\}$, which can then be used in, e.g. classical MD simulations to model electrostatic contributions in the force-field. The fitting procedure is applied such that the charges $\{q_a\}$ optimally reproduce a given potential $V_{\textnormal{QM}}$ obtained from a quantum chemical calculation. To avoid unphysical values for the fitted charges, restraints (and constraints) are often set for $q_a$, which are then called RESP charges~\cite{Bayly1993}.\par
The QM potential is typically obtained from a preceding DFT or Hartree-Fock calculation. The difference between $V_{\mathrm{QM}}(\mathbf{r})$ and the potential $V_{\mathrm{RESP}}(\mathbf{r})$ generated by $\{q_a\}$ is minimized in a least squares fitting procedure at defined grid points $\mathbf{r}_k$. The residual $R_{\mathrm{esp}}$ that is minimized is
\begin{equation}
R_{\mathrm{esp}}=\frac{1}{N}\sum_k^N{(V_{\mathrm{QM}}(\mathbf{r}_k)-V_{\mathrm{RESP}}(\mathbf{r}_k))^2}, 
\label{eq:residual_esp}
\end{equation}
where $N$ is the total number of grid points included in the fit.
The choice of $\{\mathbf{r}_k\}$ is system dependent and choosing $\{\mathbf{r}_k\}$ carefully is important to obtain meaningful charges. For isolated molecules or porous periodic structures (e.g. metal-organic frameworks) $\{\mathbf{r}_k\}$ are sampled within a given spherical shell around the atoms defined by a minimal and maximal radius. The minimal radius is usually set to values larger than the van der Waals radius to avoid fitting in  spatial regions, where the potential varies rapidly, which would result in a destabilization of the fit. As a general guideline, the charges should be fitted in the spatial regions relevant for interatomic interactions. CP2K offers also the possibility to fit RESP charges for slab-like systems. For these systems, it is important to reproduce the potential accurately above the surface where, e.g. adsorption processes take place. The sampling technique is flexible enough to follow a corrugation of the surface, see Ref~\onlinecite{Golze2015}.\par
When calculating RESP charges for periodic systems, periodic boundary conditions have to be employed for the calculation of $V_{\mathrm{RESP}}$. In CP2K, the RESP point charges $q_a$ are represented as Gaussian functions. The resulting charge distribution is presented on a regular real-space grid and the GPW formalism is employed to obtain a periodic RESP potential. Details of the implementation can be found in Ref.~\onlinecite{Golze2015}.\par
CP2K features also a GPW implementation of the REPEAT method~\cite{Campana2009}, which is a modification of the RESP fitting for periodic systems. The residual in  Eq.~\ref{eq:residual_esp} is modified such that the variance of the potentials instead of the absolute difference is fitted:
\begin{equation}
R_{\mathrm{repeat}}=\frac{1}{N}\sum_k^N{(V_{\mathrm{QM}}(\mathbf{r}_k)-V_{\mathrm{RESP}}(\mathbf{r}_k) -\delta)^2}, \label{eq:residual_repeat}
\end{equation}
where
\begin{equation}
          \delta = \frac{1}{N}\sum_k^N{(V_{\mathrm{QM}}(\mathbf{r}_k)-V_{\mathrm{RESP}}(\mathbf{r}_k) )}.
\end{equation}
The REPEAT method was originally introduced to account for the arbitrary offset of the electrostatic potential in infinite systems, which depends on the numerical method used.  In \cp, $V_{\mathrm{QM}}$ and $V_{\mathrm{RESP}}$ are both evaluated with the same method, the GPW approach, and have thus the same offset. However, fitting the variance is an easier task than fitting the absolute difference in the potentials and stabilizes significantly the periodic fit. Using the REPEAT method is thus recommended for the periodic case, in particular for systems that are not slab-like. For the latter, the potential above the surface changes very smoothly and we find that $\delta\approx 0$~\cite{Golze2015}.\par
The periodic RESP and REPEAT implementation in CP2K has been used to obtain atomic charges for surface systems, such as corrugated hexagonal boron nitride (hBN) monolayers on Rh(111)~\cite{Golze2015}. These charges were then used in MD QM/MM simulations of liquid water films (QM) on the hBN@Rh(111) substrate (MM). Other applications comprise two-dimensional periodic supramolecular structures~\cite{Korolkov2017},  metal-organic frameworks~\cite{Witman2016,Witman2017a,Witman2017b}, as well as graphane~\cite{Ciammaruchi2019}. Detailed instructions how to set up a RESP or REPEAT calculation with \cp\ can be found under Ref.~\onlinecite{resp_tutorial}.
%

\subsection{Density functional embedding theory}
\label{subsection_DFET}
%
\subsubsection{Theory}
\label{subsubsection_DFET_Theory}
%
Quantum embedding theories are multi-level approaches applying different electronic structure methods to subsystems, interacting with each other quantum-mechanically~\cite{Weselowski_CR_2015}. Density functional embedding theory (DFET) introduces high-order correlation to a chemically relevant subsystem (cluster), whereas the environment and the interaction between the cluster and the environment is described with DFT via the unique local embedding potential $v_{emb}(\mathbf{r})$~\cite{Huang_JCP_2011}. The simplest method to calculate the total energy is in the first-order perturbation theory fashion:
\begin{equation}
    E^{DFET}_{total} = E^{DFT}_{total} + (E^{CW}_{cluster, emb} - E^{DFT}_{cluster, emb} ),
\end{equation}
where $E^{DFT}_{cluster, emb}$ and $E^{DFT}_{env, emb}$ 
are DFT energies of the embedded subsystems, whereas $E^{CW}_{cluster, emb}$ is the energy of the embedded cluster at the correlated wavefunction (CW) level of theory. All these entities are computed with an additional one-electron embedding term in the Hamiltonian. 

The embedding potential is obtained from the condition that the sum of embedded subsystem densities should reconstruct the DFT density of the total system. This can be achieved by maximizing the Wu-Yang functional with respect to $v_{emb}$~\cite{Wu_JCP_2003}:
\begin{eqnarray}
    W[V_{emb}] &=& E_{cluster}[\rho_{cluster}] + E_{env}[\rho_{env}] \nonumber \\
    &+& \int V_{emb}(\rho_{total} - \rho_{cluster} - \rho_{env}) d\mathbf{r},
\end{eqnarray}
with the functional derivative to be identical to the density difference $ \frac{\delta W}{\delta V_{emb}} = \rho_{total} - \rho_{cluster} - \rho_{env}$.

\subsubsection{Implementation}
\label{subsubsection_DFET_Implementation}
The DFET workflow consists of computing the total system with DFT and obtaining $v_{emb}$ by repeating DFT calculations on the subsystems with updated embedding potential until the total DFT density is reconstructed. When the condition is fulfilled, the higher-level embedded calculation is performed on the cluster. The main computational cost comes from the DFT calculations. The cost is reduced by a few factors such as employing an wavefunction extrapolation from the previous iterations via the ASPC integrator~\cite{Kolafa2004}. 

The DFET implementation is available for closed and open electronic shells in unrestricted and restricted open-shell variants for the latter. It is restricted to GPW calculations with pseudopotentials describing the core electrons (i.e. full-electron methods are currently not available). Any method implemented within \qs\ is available as a higher-level method, including hybrid DFT, MP2 and RPA. It is possible to perform property calculations on the embedded clusters using an externally provided $v_{emb}$. The subsystems can employ different basis sets, although they must share the same PW grid.

In our implementation, $v_{emb}$ may be represented on the real-space grid, as well as using a finite Gaussian basis set, the first option being preferable, as it allows a much more accurate total density reconstruction and is computationally cheaper. 

\subsection{Implicit solvent techniques}
\label{subsection_SCCS}
AIMD simulations of solutions, biological systems or surfaces in presence of a solvent are often computationally
dominated by the calculation of the explicit solvent portion,  which may easily amount to 70\% of the total number
of atoms in a model system. Although the first and second sphere or layer of the solvent molecules around a solute
or above a surface might have a direct impact via chemical bonding, for instance via a hydrogen bonding network,
the bulk solvent mostly interacts electrostatically as a continuous dielectric medium. This triggered the development of methods treating the bulk solvent implicitly. Such implicit solvent methods have also the further advantage that they
provide a statistically averaged effect of the bulk solvent. That is beneficial for AIMD simulations that consider only a relatively small amount of bulk solvent and quite short sampling times.
The self-consistent reaction field (SCRF) in a sphere, which goes back to the early work by Onsager~\cite{Onsager1936},
is possibly the most simple implicit solvent method implemented in \cp~\cite{Barker1973,Watts1974}. 

More recent continuum solvation models like the conductor-like screening model (COSMO)~\cite{Klamt1993}, as well as the
polarizable continuum model (PCM)~\cite{Tomasi1994,Tomasi2005}, take also into account the explicit shape of the
solute. The solute-solvent interface is defined as the surface of a cavity around the solute. This cavity is
constructed by interlocking spheres centered on the atoms or atomic groups composing the solute~\cite{Connolly1983}.
This introduces a discontinuity of the dielectric function at the solute-solvent interface and thus causes
non-continuous atomic forces, which may impact the convergence of structural relaxations, as well as the energy
conservation within AIMD runs. To overcome these problems, Fatteberg and Gygi proposed a smoothed
self-consistent dielectric model function of the electronic density $\rho^\text{elec}(\br)$~\cite{Fattebert2002,Fattebert2003}:
\begin{equation}
 \epsilon\left[\rho^\text{elec}(\br)\right] = 1 + \frac{\epsilon_0 - 1}{2}
  \left(1 + \frac{1 - \left(\rho^\text{elec}(\br)/\rho_0\right)^{2\beta}}
                 {1 + \left(\rho^\text{elec}(\br)/\rho_0\right)^{2\beta}}
  \right),
\end{equation}
with the model parameters $\beta$ and $\rho_0$, which fulfills the asymptotic behaviour
\begin{equation}
 \label{epsFattebert}
 \epsilon(\br) \equiv \epsilon\left[\rho^\text{elec}(\br)\right] =
  \begin{cases}
   1 & \text{large }\rho^\text{elec}(\br) \\
   \epsilon_0 & \rho^\text{elec}(\br) \rightarrow 0.
  \end{cases}
\end{equation}
The method requires a self-consistent iteration of the polarization charge density spreading across
the solute-solvent interface in each SCF iteration step, since the dielectric function depends on
the actual electronic density $\rho^\text{elec}(\br)$. This may cause a non-negligible computational
overhead depending on the convergence behaviour.

More recently, Andreussi et al. proposed in the framework of a revised
self-consistent continuum solvation (SCCS) model an improved piecewise defined dielectric model function~\cite{Andreussi2012}:
\begin{equation}
 \label{epsAndreussi}
 \epsilon\left[\rho^\text{elec}(\br)\right] =
  \begin{cases}
   1 & \rho^\text{elec}(\br) > \rho_\text{max} \\
   \exp(t) & \rho_\text{min} \le \rho^\text{elec}(\br) \le \rho_\text{max} \\
   \epsilon_o & \rho^\text{elec}(\br) < \rho_\text{min},
  \end{cases}
\end{equation}
with $t = t\left[\ln\left(\rho^\text{elec}(\br)\right)\right]$ that employs a more elaborated switching function for the transition from the solute to the solvent region using the smooth function
\begin{eqnarray}
  t(x) &=& \frac{\ln\epsilon_0}{2\pi}\left[2\pi\frac{(\ln\rho_\text{max} - x)}{(\ln\rho_\text{max} - \ln\rho_\text{min})}\right. \nonumber \\
       &-& \left.\sin\left(2\pi\frac{(\ln\rho_\text{max} - x)}{(\ln\rho_\text{max} - \ln\rho_\text{min})}\right)\right].
\end{eqnarray}
Both models are implemented in \cp~\cite{Yin2015,Yin2017}.

The solvation free energy $\Delta G^\text{sol}$ can be computed by
\begin{equation}
 \Delta G^\text{sol} = \Delta G^\text{el} + G^\text{rep} + G^\text{dis}+ G^\text{cav} + G^\text{tm} + P\Delta V
\end{equation}
as the sum of the electrostatic contribution $\Delta G^\text{el} = G^\text{el} - G^0$,
where $G^0$ is the energy of the solute in vacuum, the repulsion energy $G^\text{rep} = \alpha\, S$, 
the dispersion energy $G^\text{dis} = \beta\, V$, 
and the cavitation energy $G^\text{cav} = \gamma\, S$, 
with adjustable solvent specific parameters $\alpha$, $\beta$ and $\gamma$~\cite{Scherlis2006}. Therein, 
$S$ and $V$ are the (quantum) surface and volume of the solute cavity, respectively, which are
evaluated in \cp~based on the quantum surface
\begin{eqnarray}
  S &=& \int \left[\vartheta\left(\rho^\text{elec}(\br) - \frac{\Delta}{2}\right) - \left(\rho^\text{elec}(\br) + \frac{\Delta}{2}\right)\right] \nonumber \\
      &\times&\frac{\vert\nabla\rho^\text{elec}(\br)\vert}{\Delta}\,d\br
\end{eqnarray}
and quantum volume
\begin{equation}
 V = \int\vartheta\left[\rho^\text{elec}(\br)\right]\,d\br, 
\end{equation}
a definition that was introduced by Cococcioni et al. with
\begin{equation}
 \vartheta\left[\rho^\text{elec}(\br)\right] = \frac{\epsilon_0 - \epsilon\left[\rho^\text{elec}(\br)\right]}{\epsilon_0 - 1}
\end{equation}
using the smoothed dielectric function either from Eq.~\ref{epsFattebert} or Eq.~\ref{epsAndreussi}, respectively~\cite{Cococcioni2005,Andreussi2012}.
The thermal motion term $G^\text{tm}$ and the volume change term $P\Delta V$ are often ignored and are not yet available in \cp.

\subsection{Poisson solvers}
\label{sec:Poisson_Solver}
The Poisson equation describes how the electrostatic potential $V(\bm{r})$, as a contribution to the Hamiltonian, relates to the charge distribution within the system with electron density $n(\bm{r})$. In a general form, which incorporates a non-homogeneous position-dependent dielectric constant $\varepsilon(\bm{r})$, the equation is formulated as
\begin{equation}
\label{eq:gpoisson}
	- \nabla \cdot \left( \varepsilon(\bm{r}) \nabla V (\bm{r}) \right)= 4 \pi n(\bm{r}),
\end{equation}
endowed with some suitable boundary conditions determined by the physical characteristics or setup of the considered system. 

Under the assumption that everywhere inside the simulation domain the dielectric constant is equal to that of free space, i.e. 1, and for periodically repeated systems or boundary conditions for which the analytical Green's function of the standard Poisson operator is known, like free, wire or surface boundary conditions, a variety of methods has been proposed to solve Eq.~\ref{eq:gpoisson}. Among these approaches, CP2K implements the conventional PW scheme, Ewald summation based techniques~\cite{Ewald1921, ps_Ewald_1, ps_Ewald_2, ps_Ewald_3, ps_Ewald_4}, the Martyna-Tuckerman method~\cite{Martyna1999}, and a wavelet based Poisson solver~\cite{Genovese_JCP_2006,Genovese_JCP_2007}. Within the context of continuum solvation models~\cite{Tomasi1994}, where solvent effects are included implicitly in the form of a dielectric medium~\cite{Fattebert2002}, the code also provides the solver proposed by Andreussi et al. that can solve Eq.~\eqref{eq:gpoisson} subject to periodic boundary conditions and with $\varepsilon$ defined as a function of the electronic density~\cite{Andreussi2012}.

Despite the success of these methods in numerous scenarios (for some recent applications see, e.g.,~\cite{Yin2015,Yazdani2019}), the growing interest in simulating atomistic systems with more complex boundary configurations, e.g. nanoelectronic devices, at an \textit{ab-inito} level of theory, demands for Poisson solvers with capabilities that exceed those of the existing solvers. To this end, a generalized Poisson solver has been developed and implemented in CP2K with the following key features~\cite{ps_BaniHashemian}: 
\begin{itemize}
\item The solver converges exponentially with respect to the density cutoff.
\item Periodic or homogeneous von Neumann (zero normal electric field) boundary conditions can be imposed on the boundaries of the simulation cell. The latter models insulating interfaces such as e.g. air-semiconductor and oxide-semiconductor interfaces in a transistor.
\item Fixed electrostatic potentials (Dirichlet-type boundary conditions) can be enforced at arbitrarily-shaped regions within the domain. These correspond e.g. to source, drain and gate contacts of a nanoscale device.
\item The dielectric constant can be expressed as any sufficiently smooth function.
\end{itemize}
Therefore, the solver offers advantages associated with the two categories of Poisson solvers, i.e. PW and real-space-based methods. 

The imposition of the above-mentioned boundary setups is accomplished by solving an equivalent constrained variational problem that reads as:
\begin{equation}
\label{eq:minimax_1}
	 \text{Find} \,\, \left( V,  \lambda \right) \,\,\text{s.t.}\,\, J(V,\lambda) = \underset{u}{\min} \,\, \underset{\mu}{\max} \,\, J(u,  \mu), 
\end{equation}
where
\begin{eqnarray}
\label{eq:minimax_2}
	J(u, \mu) &=& \, \frac{1}{2} \int_{\Omega} \varepsilon \lvert \nabla u \rvert^{2} \, d\bm{r} - \int_{\Omega} 4 \pi n u \, d\bm{r} \nonumber \\
	&+& \int_{\Omega_{D}} \mu \left( u - V_{D} \right) \, d\bm{r}.
\end{eqnarray}
Here, $V_{D}$ is the potential applied at the predefined subdomain $\Omega_{D}$ that may have an arbitrary geometry, whereas $u(V)$ satisfies the desired boundary conditions at the boundaries of the cell $\Omega$ and $\mu(\lambda)$ are Lagrange multipliers introduced to enforce the constant potential. Note that, the formulation of the problem within Eqs.~\ref{eq:minimax_1}-\ref{eq:minimax_2} is independent of how $\varepsilon$ is defined. 

Furthermore, consistent ionic forces have been derived and implemented which make the solver applicable to a wider range of applications like, energy-conserving BOMD~\cite{Praveen2017}, as well as EMD simulations~\cite{MicrocanonicalRTTDDFT}.


\section{\textbf{\DBCSR} library}
\label{sec:DBCSR}
\DBCSR has been specifically designed to efficiently perform block-sparse and dense matrix operations on distributed multicore CPUs and GPUs systems, covering a range of occupancy between $0.01\%$ up to dense~\cite{Borstnik2014, Schuett2016, Lazzaro:2017:IES:3093172.3093228}.
The library is written in Fortran and is freely available under the GPL license from \url{https://github.com/cp2k/dbcsr}. Operations include sum, dot product, and multiplication of matrices, and the most important operations on single matrices, such as transpose and trace. 

The \DBCSR library was developed to unify the previous disparate implementations of distributed dense and sparse matrix data structures and the operations among them. The chief performance optimization target is parallel matrix multiplication. Its performance objectives have been to be comparable to ScaLAPACK's PDGEMM for dense or nearly-dense matrices~\cite{Choi1996}, while achieving high performance when multiplying sparse matrices~\cite{Borstnik2014}.

\DBCSR matrices are stored in a blocked compressed sparse row (CSR) format distributed over a two-dimensional grid of $P$ message passing interface (MPI) processes.
Although the library accepts single and double precision complex and real numbers, it is only optimized for the double precision real type. The sizes of the blocks in the data structure are not arbitrary, but are determined by the properties of the data stored in the matrix, such as the basis sets of the atoms in the molecular system. While this property keeps the blocks in the matrices dense, it often results in block sizes that are suboptimal for computer processing, e.g. blocks of $5\times 5$, $5\times 13$, and $13\times 13$ for the H and O atoms described by a DZVP basis set.

\begin{figure}
\centerline{\includegraphics[scale=0.8]{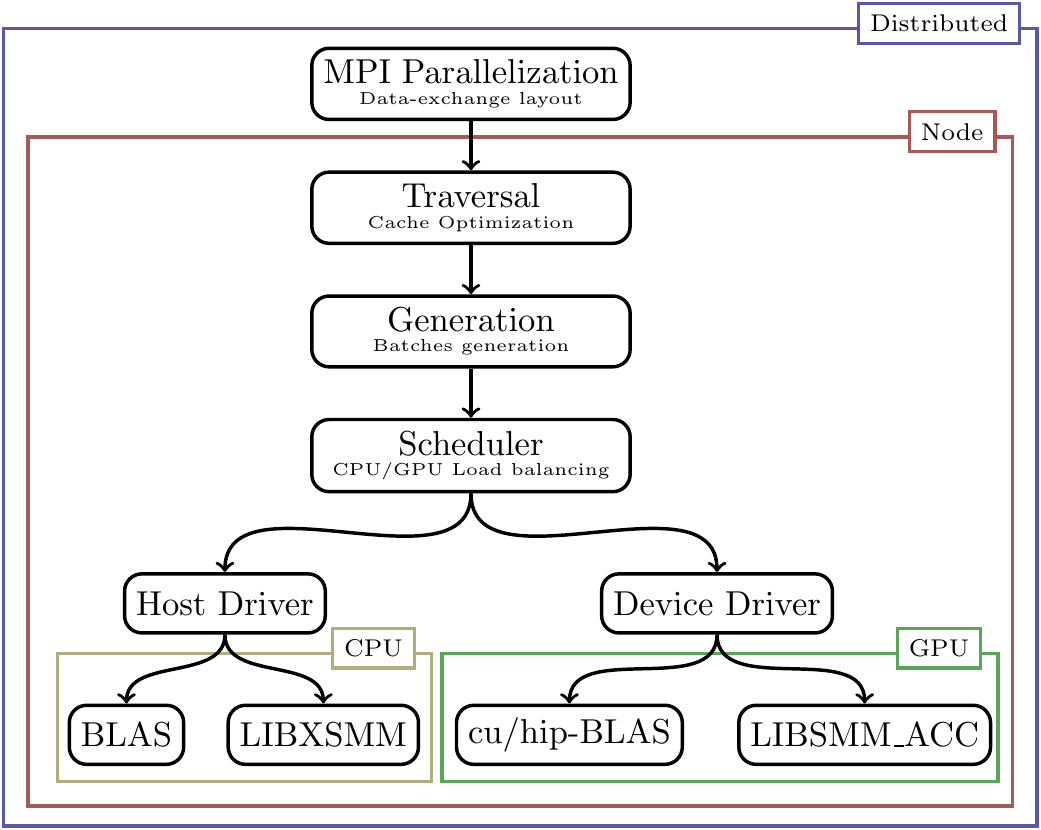}}
\caption{Schema of the \DBCSR library for the matrix-matrix multiplication (see text for description).} 
\label{fig:diagram}
\end{figure}

To achieve high performance for the distributed multiplication of two possibly sparse matrices, several techniques have been used. One is to focus on reducing communication costs among MPI ranks. Another is to optimize the performance of multiplying communicated data on a node by introducing a separation of concerns: \emph{what} will be multiplied vs.\ \emph{performing} the multiplications. The CPUs on a node determine what needs to be multiplied, creating batches of work. These batches are handled by hardware-specific drivers, such as CPUs, GPUs, or a combination of both. \DBCSR's GPU backend, \LIBSMMACC, supports both NVIDIA and AMD GPUs via CUDA and HIP, respectively. A schema of the library is shown in Fig.~\ref{fig:diagram}.

\subsection{Message passing interface parallelization}
\label{subsection_DBCSR_MPI}
At the top level, we have the MPI parallelization. The data-layout exchange is implemented with two different algorithms, depending on the sizes of the involved matrices in the multiplications:
\begin{itemize}
    \item for general matrices (any size) we use the Cannon algorithm, where the amount of communicated data by each process scales as $\mathcal{O} (1/\sqrt{P})$~\cite{Cannon1969, Borstnik2014};
    \item only for ``tall-and-skinny'' matrices (one large dimension) we use an optimized algorithm, where the amount of communicated data by each process scales as $\mathcal{O} (1)$~\cite{Sivkov2019}.
\end{itemize}
The communications are implemented with asynchronous point-to-point MPI calls. The local multiplication will start as soon as all the data has arrived at the destination process, and it is possible to overlap the local computation with the communication if the network allows that. 

\subsection{Local multiplication}
\label{subsection_local_multiplication}
The local computation consists of pairwise multiplications of small dense matrix blocks, with dimensions $(m \times k)$ for $A$ blocks and $(k \times n)$ for $B$ blocks. It employs a cache-oblivious matrix traversal to fix the order in which matrix blocks need to be computed, to improve memory locality ({\tt Traversal} phase in Fig.~\ref{fig:diagram}). First, the algorithm loops over $A$ matrix row-blocks and then, for each row-block, over $B$ matrix column-blocks. Then, the corresponding multiplications are organized in batches ({\tt Generation} phase in Fig.~\ref{fig:diagram}), where each batch consists of maximum $30{,}000$ multiplications. 
During the {\tt Scheduler} phase, a static assignment of batches with a given $A$ matrix row-block to OpenMP threads is employed to avoid data-race conditions. Finally, batches assigned to each thread can be computed in parallel on the CPU and/or executed on a GPU. For the GPU execution, batches are organized in such a way that the transfers between the host and the GPU are minimized. The multiplication kernels take full advantage of the opportunities for coalesced memory operations and asynchronous operations. Moreover, a double-buffering technique, based on CUDA streams and events, is used to maximize the occupancy of the GPU and to hide the data transfer latency~\cite{Schuett2016}.
When the GPU is fully loaded, the computation may be simultaneously done on the CPU. Multi-GPU execution on the same node is made possible by distributing the cards to multiple MPI ranks via a round-robin assignment.

\subsection{Batched execution}
\label{subsection_batched_execution}
Processing batches of small-matrix-multiplications (SMMs) has to be highly efficient. For this reason specific libraries were developed that outperform vendor BLAS libraries, namely \LIBSMMACC (previously called \LIBCUSMM, which is part of \DBCSR) for GPUs~\cite{Sivkov2019}, as well as \LIBXSMM for CPUs~\cite{Heinecke2016, parco_knl}.

\begin{figure}
\centerline{\includegraphics[scale=0.55]{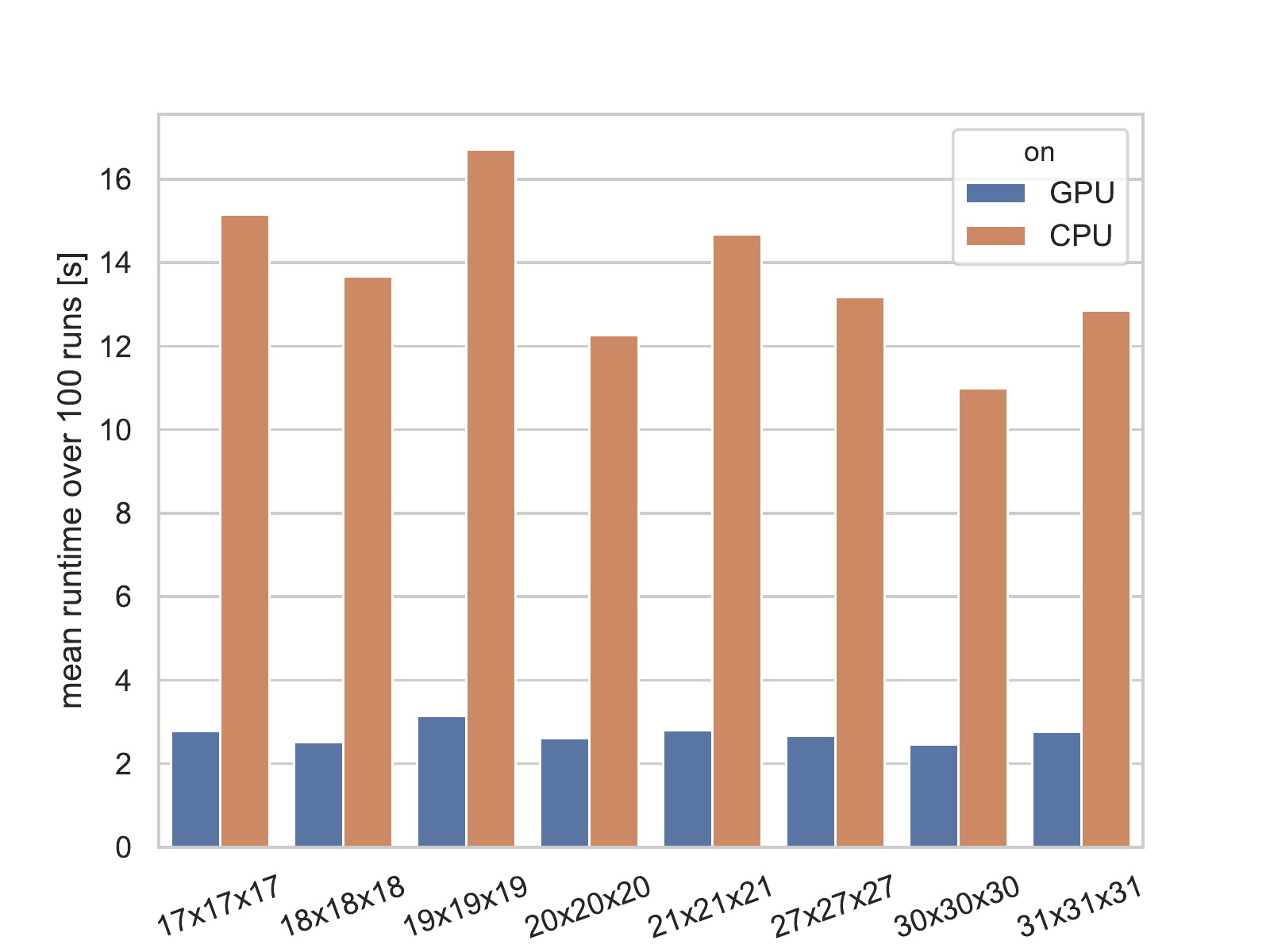}}
\caption{Comparison of \DBCSR dense multiplication of square matrices of size 10{,}000, dominated by different sub-matrix block sizes. With the JIT framework in place, these blocks are batch-multiplied on a GPU using \LIBSMMACC instead of on the CPU using \LIBXSMM. Multiplications were run on a heterogenous Piz Daint CRAY XC50 node containing a 12 core Intel Haswell CPU and on a NVIDIA V100 GPU.} 
\label{fig:dbcsr_performance}
\end{figure}

In \LIBSMMACC, GPU kernels are just-in-time (JIT) compiled at runtime. This allows to reduce \DBCSR's compile time by more than half and its library's size by a factor of 6 compared to generating and compiling kernels ahead-of-time. Fig.~\ref{fig:dbcsr_performance} illustrates the performance gain that can be observed since the introduction of the JIT framework: because including a new (m,n,k)-kernel to the library incurs no additional compile time, nor does it bloat the library size, all available (m,n,k) batched-multiplications can be run on GPUs, leading to important speedups.

\LIBSMMACC's GPU kernels are parametrized over 7 parameters, affecting the memory usages and patterns of the multiplication algorithms, the amount of work and number of threads per CUDA block, the number of matrix elements computed by each CUDA thread and the tiling sizes. An autotuning framework finds the optimal parameter set for each (m,n,k)-kernel, exploring about 100{,}000 possible parameter combinations. These parameter combinations result in vastly different performances: for example, the batched multiplication of kernel $5 \times 6 \times 6$ can vary between 10 Gflops and 276 Gflops in performance, with only 2\% of the possible parameter combinations yielding performances within 10\% of the optimum. 
Kernels are autotuned for NVIDIA's P100 (1{,}649 (m,n,k)s) and V100 (1{,}235 (m,n,k)s), as well as for AMD's Mi50 (10 kernels). This autotuning data is then used to train a machine learning performance prediction model, which finds near-optimal parameter sets for the rest of the 75{,}000 available (m,n,k)-kernels.
In this way, the library can achieve a speedup in the range of 2--4x with respect to batched DGEMM in cuBLAS for $\{m,n,k\}< 32$, while the effect becomes less prominent for larger sizes~\cite{parco_knl}. Performance of \LIBSMMACC and \LIBXSMM saturate for $\{m,n,k\}> 80$, for which \DBCSR directly calls cuBLAS or BLAS respectively.

\LIBXSMM generates kernels using a machine model, \eg considering the size of the \mbox{(vector-)}register file. Runtime code generation in \LIBXSMM does not depend on the compiler or flags used and can naturally exploit instruction set extensions presented by CPUID features. Raw machine code is emitted JIT with no compilation phase and one-time cost (10k--50k kernels per second depending on system and kernel). Generated code is kept for the runtime of an application (like normal functions), and ready-for-call at a rate of 20M--40M dispatches per second. Both timings include the cost of thread-safety or potentially concurrent requests. Any such overhead is even lower when amortized with homogeneous batches (same kernel on a per batch basis).

\subsection{Outlook}

An arithmetic intensity of AI\,=\,0.3--3~FLOPS\,/\,Byte (double-precision) is incurred by, \eg \CPTWOK's needs. This low AI is bound by Stream Triad given that $C=C+A*B$ is based on SMMs rather than scalar values. Since memory bandwidth is precious, lower or mixed precision (including single-precision) can be of interest given emerging support in CPUs and GPUs (see section~\ref{subsection_ac}).
%

\section{Interfaces to other programs}
\label{section_interfaces}
Beside containing native f77 and f90 interfaces, \cp\ also provides a library, which can be used in your own executable~\cite{Los2013,Los2016}, and comes with a small helper program called \cp-shell that includes a simple interactive command line interface with a well defined, parseable syntax. In addition, \cp\ can be interfaced with external programs such as i-PI~\cite{i-Pi}, PLUMED~\cite{Plumed}, or PyRETIS~\cite{Riccardi2019}, and its AIMD trajectories analyzed using TAMkin~\cite{TAMkin}, MD-Tracks~\cite{Verstraelen2008}, and TRAVIS~\cite{Travis}, to name just a few.

\subsection{Non-equilibrium Green's function formalism}
\label{sec:DFT-NEGF}
The non-equilibrium Green's function (NEGF) formalism provides a comprehensive framework for studying the transport mechanism across open nanoscale systems in the ballistic, as well as scattering regimes~\cite{Kadanoff1962, Keldysh1964, Datta1995}. Within the NEGF formalism, all observables of a system can be obtained from the single-particle Green's functions. In the ballistic regime, the main quantity of interest is the retarded Green's function $G$, which solves the following equation: 
\begin{equation}
\label{eq:negf_formalism}
	 \left( E \cdot S - H - \Sigma \right) \cdot G(E) = I, 
\end{equation}
where $E$ is the energy level of the injected electron, $S$ and $H$ are the overlap and Hamiltonian matrices, respectively, $I$ is the identity matrix and $\Sigma$ the boundary (retarded) self-energy that incorporates the coupling between the central active region of the device and the semi-infinite leads. A major advantage of the NEGF formalism is that no restrictions are made on the choice of the Hamiltonian that means, empirical, semi-empirical, or \textit{ab initio} Hamiltonians can be used in Eq.~\eqref{eq:negf_formalism}. Utilizing a, typically Kohn-Sham, DFT Hamiltonian, the DFT+NEGF method has established itself as a powerful technique due to its capability of modeling charge transport through complex nanostructures without any need for system-specific parameterizations~\cite{Taylor2001,Brandbyge2002}. With the goal of designing a DFT+NEGF simulator that can handle systems of unprecedented size, e.g. composed of tens of thousands of atoms, the quantum transport solver OMEN has been integrated within \cp~\cite{Luisier2006, Luisier2014, BaniHashemian_PhDThesis}. For a given device configuration, \cp\ constructs the DFT Hamiltonian and overlap matrices. The matrices are passed to OMEN where the Hamiltonian is modified for open boundaries and charge and current densities are calculated in the NEGF or, the equivalently, wave function/quantum transmitting boundary formalisms~\cite{Lent1990}. Due to the robust and highly efficient algorithms implemented in OMEN and exploiting hybrid computational resources effectively, devices with realistic sizes and structural configurations can now be routinely simulated~\cite{Brueck2014,Calderara2015,Brueck2017}.


\subsection{SIRIUS: Plane wave density functional theory support}
\label{subsection_sirius}
\cp\ supports computations of the electronic ground state, including forces and stresses~\cite{PhysRevLett.50.697, PhysRevB.32.3780}, in a PW basis. The implementation relies on the quantum engine
{\sc SIRIUS}~\cite{Kozhevnikov2019}.

The {\sc SIRIUS} library has full support for GPUs and CPUs and brings additional
functionalities to \cp. Collinear and non-collinear magnetic systems with or
without spin-orbit coupling can be studied with both pseudopotential
PW~\cite{PhysRevB.71.115106}, as well as full-potential linearized augmented PW methods~\cite{PhysRevB.12.3060, PhysRevB.64.195134}. All GGA XC functionals implemented in
{\it libxc} are available~\cite{Lehtola2018}. Hubbard corrections are based on Refs.~\onlinecite{PhysRevB.44.943} and \onlinecite{PhysRevB.52.R5467}.

To demonstrate a consistent implementation of the PW energies, forcesand stresses in \cp\ + {\sc SIRIUS}, an AIMD of a Si$_7$Ge supercell has been performed in the isothermal–isobaric NPT ensemble. To establish the accuracy of the
calculations, the ground state energy of the obtained atomic configurations at
each time step has been recomputed with {\sc Quantum ESPRESSO} (QE) using the same cutoff parameters, XC functionals and
pseudopotentials~\cite{Giannozzi2017,Giannozzi2009}.
\begin{figure}
  \includegraphics[width=8cm]{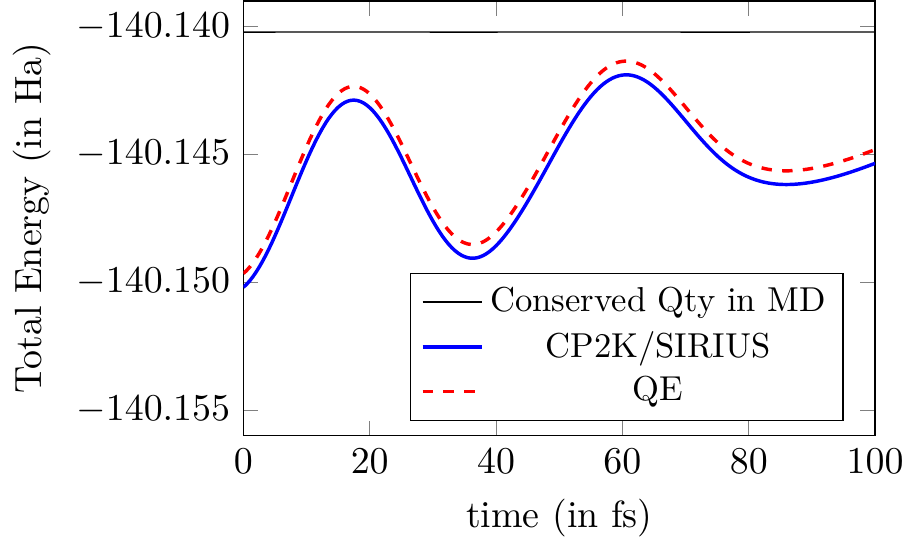}\\
  \includegraphics[width=8cm]{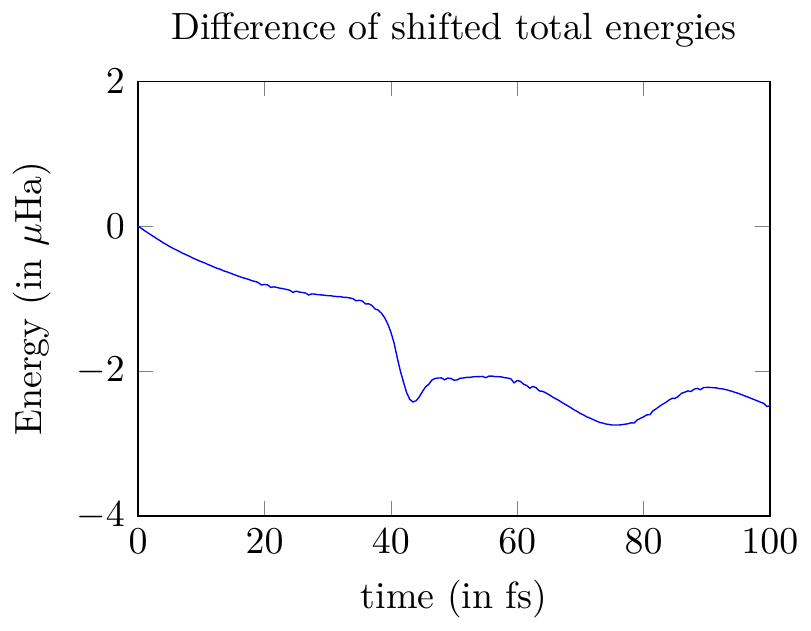}
\caption{\label{cp2k-sirius-fig} Comparison of potential energies between \cp\ + {\sc SIRIUS} and QE during an AIMD simulation. The total energy (black curve) remains constant over time (it is a conserved quantity in the NPT ensemble), while the potential energy calculated with \cp\ + {\sc SIRIUS} (blue curve) and QE (red dashed curve) have the same time evolution up to a constant shift in energy (see lower panel).}
\end{figure}
In Fig.~\ref{cp2k-sirius-fig}, it is shown that the total energy is conserved up to $10^{-6}$~Ha. It also shows the evolution of the potential energies
calculated with \cp\ + {\sc SIRIUS} (blue line) and QE (red dashed line). The two curves
are shifted by a constant offset, which can be attributed to differences in
various numerical schemes employed. After removal of this constant difference, the agreement between the two codes is of the order of $10^{-6}$~Ha.
%

\section{Technical and Community Aspects}
\label{section_technical}
%
%
With an average growth of 200 lines of code per day, the \cp\ code comprises currently more than a million lines of code. 
Most of \cp\ is written in Fortran95, with elements from Fortran03 and extensions such as OpenMP, OpenCL and CUDA C. It also employs various external libraries in order to incorporate new features and to decrease the complexity of \cp, thereby increasing the efficiency and robustness of the code. The libraries range from basic functionality such as MPI~\cite{Gropp2014}, fast Fourier transforms (FFTW)~\cite{FFTW}, dense linear algebra (BLAS, LAPACK, ScaLAPACK, ELPA)~\cite{Choi1996,Auckenthaler2011,Marek2014}, to more specialized chemical libraries to evaluate electron repulsion integrals (libint) and XC functionals (libxc)~\cite{libint,Lehtola2018}. \cp\ itself can be built as a library, allowing for easy access to some part of the functionality by external programs. Having lean, library-like interfaces within \cp\ has facilitated the implementation of features such as farming (running various inputs within a single job), general input parameter optimization, PIMD, or Gibbs-ensemble MC. 

Good performance and massive parallel scalability are key features of \cp. This is achieved using a multi-layer structure of specifically designed parallel algorithms. On the highest level, parallel algorithms are based on message passing with MPI, which is suitable for distributed memory architectures, augmented with shared memory parallelism based on threading and programmed using OpenMP directives. Ongoing work aims at porting the main algorithms of CP2K to accelerators and GPUs, as these energy efficient devices become more standard in supercomputers. At the lowest level, auto-generated and auto-tuned code allows for generating CPU-specific libraries that deliver good performance without a need for dedicated code development.

\subsection{Hardware Acceleration}
\label{subsection_acceleration}
\cp\ provides mature hardware accelerator support for GPUs and emerging support for FPGAs. The most computationally intensive kernels in \cp\ are handled by the \DBCSR library (see Sec.~\ref{sec:DBCSR}), which decomposes sparse matrix operations into BLAS and \LIBXSMM library calls. For GPUs, \DBCSR provides a CUDA driver, which offloads these functions to corresponding libraries for GPUs (\LIBSMMACC and cuBLAS). FFT operations can also be offloaded to GPUs using the cuFFT library.
For FPGAs, however, there exist currently no general and widely used FFT libraries that could be reused. Hence, a dedicated FFT interface has been added to \cp\ and an initial version of an FFT library for offloading complex-valued, single-precision 3D FFTs of sizes $32^3$ to $128^3$ to Intel Arria 10 and Stratix 10 FPGAs has been released~\cite{Ramaswami2019}.

\subsection{Approximate Computing}
\label{subsection_ac}
The AC paradigm is an emerging approach to devise techniques for relaxing the exactness to improve the performance and efficiency of the computation~\cite{Plessl2015}. The most common method of numerical approximation is the use of adequate data-widths in computationally intensive application kernels. Although many scientific applications use double-precision floating-point by default, this accuracy is not always required. Instead, low- and mixed-precision arithmetic has been very effective for the computation of inverse matrix roots~\cite{lass17-esl}, or solving systems of linear equations~\cite{Klavik2014,Bekas,Dongarra2017,Dongarra2018}. 
Driven by the growing popularity of artificial neural networks that can be evaluated and trained with reduced precision, hardware accelerators have gained improved low-precision computing support. For example, NVidia V100 GPU achieves 7~GFlops in double-precision and 15~GFlops single-precision, but up to 125~GFlops tensor performance with half-precision floating point operations.
Using low-precision arithmetic is thus essential for exploiting the performance potential of upcoming hardware accelerators.

However, in scientific computing, where the exactness of all computed results is of paramount importance, attenuating accuracy requirements is not an option. Yet, for specific problem classes it is possible to bound or compensate the error introduced by inaccurate arithmetic. In \cp, we can apply the AC paradigm to the force computation within MD and rigorously compensate the numerical inaccuracies due to low-accuracy arithmetic operations and still obtain exact ensemble-averaged expectation values, as obtained by time averages of a properly modified Langevin equation. 

For that purpose, the notion of the second-generation CPMD method is reversed and we model the nuclear forces as
\begin{equation} \label{fFPGA}
\textbf{F}_{I}^{N} = \textbf{F}_{I} + \mathbf{\Xi }_{I}^{N},
\end{equation}
where $\mathbf{\Xi }_{I}^{N}$ is an additive white noise that is due to a low-precision force computation on an GPU or FPGA-based accelerator. Given that $\mathbf{\Xi }_{I}^{N}$ is unbiased, i.e. %
\begin{equation} \label{CrossCorr}
 \left \langle \textbf{F}_{I}\left ( 0 \right ) \mathbf{\Xi } _{I}^{N}\left ( t \right )\right \rangle \cong  0
\end{equation}
holds, 
it is nevertheless possible to accurately sample the Boltzmann distribution by means of a modified Langevin-type equation~\cite{Richters2014,Karhan,Rengaraj2019}: 
\begin{equation} \label{modLangevin}
M_{I}\ddot{\textbf{R}}_{I} = \textbf{F}_{I}^{N}-\gamma _{N}M_{I}\dot{\textbf{R}}_{I}.
\end{equation}
This is to say that that the noise, as originating from a low-precision computation, can be thought of as an additive white noise channel associated with a hitherto unknown damping coefficient $\gamma_N$, which satisfies the fluctuation-dissipation theorem 
\begin{equation}
\left \langle \mathbf{\Xi }_{I}^{N}\left ( 0 \right ) \mathbf{\Xi }_{I}^{N}\left ( t \right ) \right \rangle \cong  2 \gamma_{N} M_I k_{B} T  \delta \left ( t \right ).
\label{FDT}
\end{equation}
%
%
As before, the specific value of $\gamma_N$ is determined in such a way so as to generate the correct average temperature, as measured by the equipartition theorem of Eq.~\ref{EPT}, by means of the adaptive Langevin technique of Leimkuhler and coworkers~\cite{JonesLeimkuhler2011, Mones2015, LeimkuhlerStoltz2019}.

\subsection{Benchmarking}
\label{subsection_benachmarking}
\begin{figure}
    \centering
    \includegraphics[width=\linewidth]{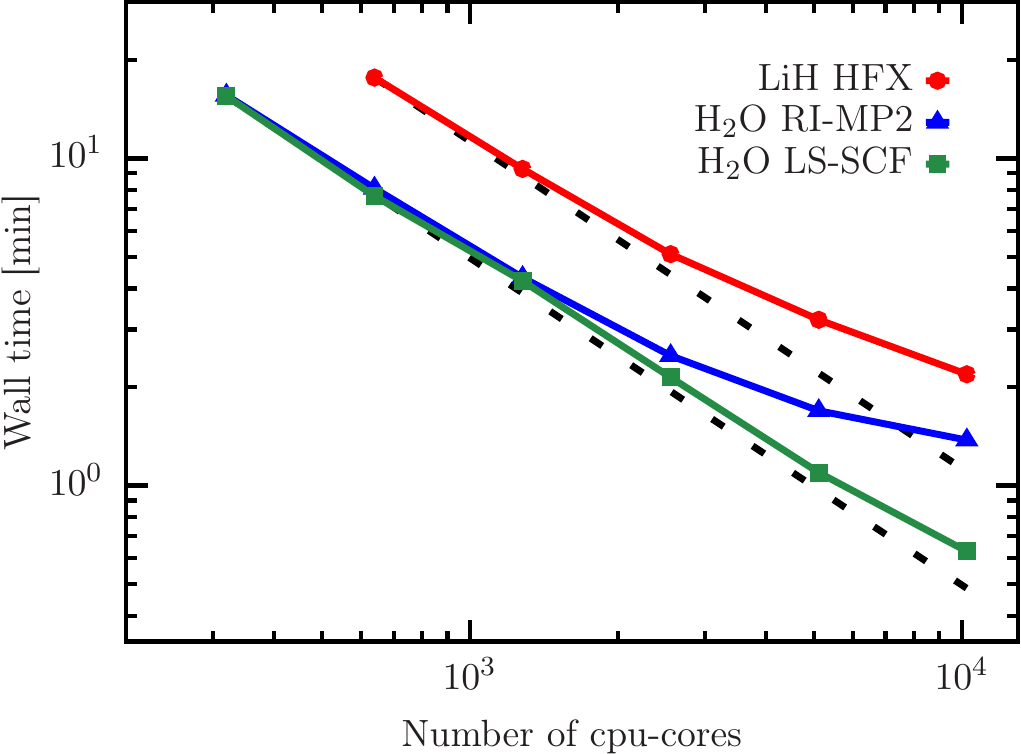}
    \caption{Parallel scaling of \cp\ calculations up to thousands of cpu-cores: single-point energy calculation using GAPW with exact Hartree-Fock exchange for a periodic 216 atom lithium hydride crystal (red line with circles),  single-point energy calculation using RI-MP2 for 64 water molecules with periodic boundary conditions (blue line with triangles) and a single-point energy calculation in linear-scaling DFT for 2048 water molecules in the condensed phase using a DZVP MOLOPT basis set (green line with boxes).
    All calculations have been performed with up to 256 nodes (10240 cpu-cores) of the Noctua system at PC$^2$.}
    \label{fig:HFX_RI-MP2}
\end{figure}
To demonstrate the parallel scalability of the various DFT-based and post-Hartree-Fock electronic structure methods implemented in \cp\, strong-scaling plots with respect to number of cpu-cores are shown in Fig.~\ref{fig:HFX_RI-MP2}. 

%
The benefit of the AC paradigm, described in section~\ref{subsection_ac}, in terms of reduction of wall time for the computation of the STMV virus, which contains more than one million atoms, is illustrated in Fig.~\ref{fig:NS_STMV_epsfilter}. Using the periodic implementation in \cp\ of the GFN2-xTB model~\cite{xtb}, an increase in efficiency of up to one order of magnitude can be observed within the most relevant matrix-sqrt and matrix-sign operations of the sign-method described in section~\ref{subsection_sign-method}. 
\begin{figure}
    \centering
    \includegraphics[width=\linewidth]{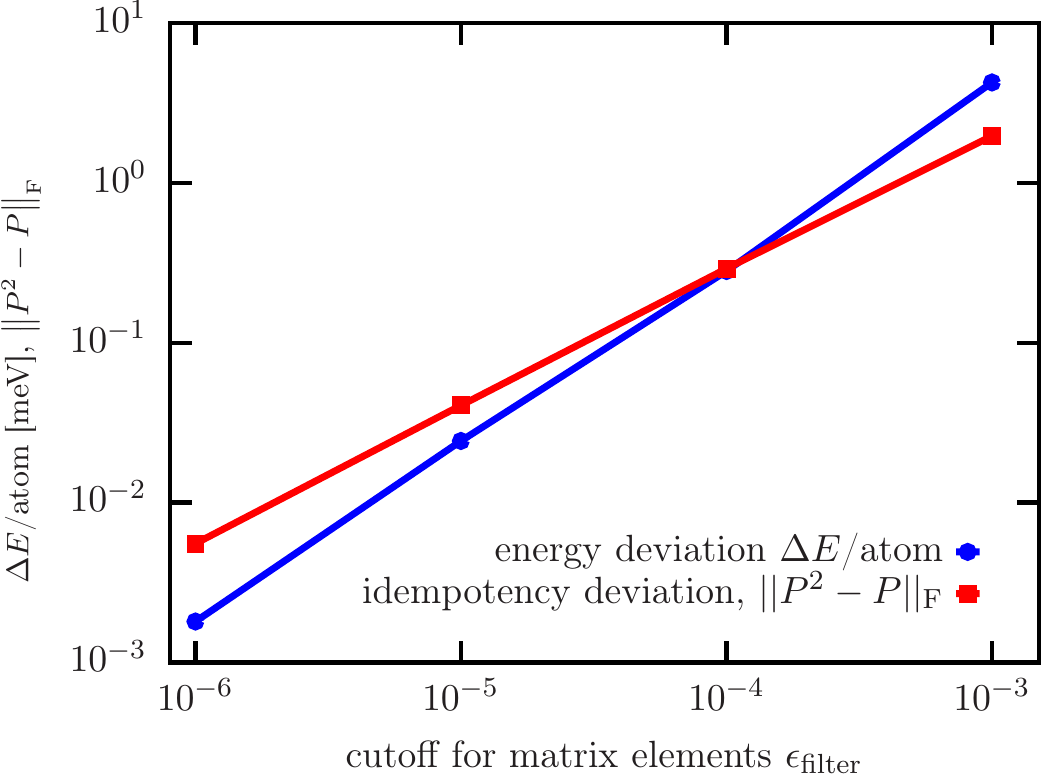}
    \caption{Energy deviation and deviation from the idempotency condition of the density matrix as a function of matrix truncation threshold for the same system and calculation as in Fig.~\ref{fig:NS_STMV_epsfilter}. For the reference energy a calculation with $\epsilon_\mathrm{filter}=10^{-7}$ was used.}
    \label{fig:STMV_NS_Noctua_E}
\end{figure}
The corresponding violation of the idempotency condition of the density matrix and the energy deviation are shown in Fig.~\ref{fig:STMV_NS_Noctua_E}. The resulting error within the nuclear forces can be assumed to be white, and hence can be compensated by means of the modified Langevin equation of Eq.~\ref{modLangevin}. 

\subsection{MOLOPT basis set and deltatest}
\label{subsection_deltatest}
\begin{figure}[htpb]
    \centering
    \includegraphics[width=\linewidth]{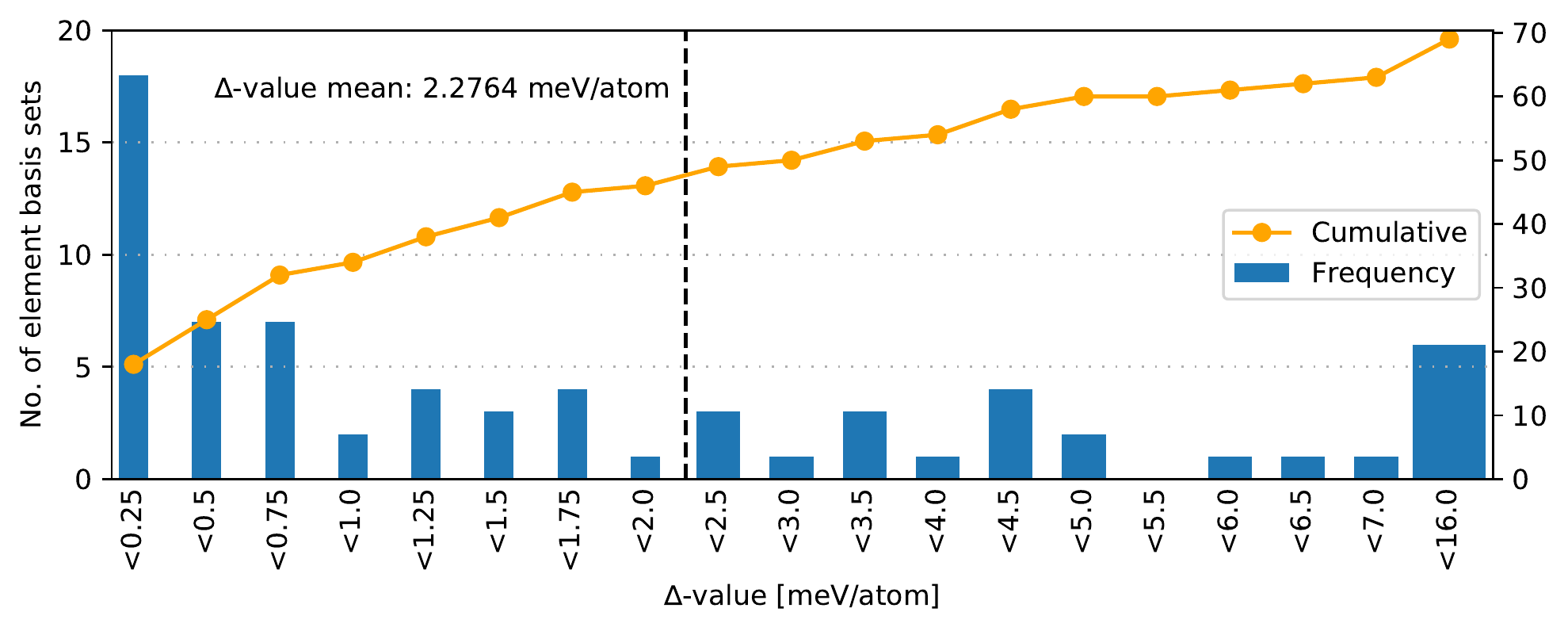}
    \caption{$\Delta$-values for DFT calculations (using the PBE XC functional and GTH pseudopotentials) of 1st to 5th row elements without Co \& Lu, as computed between the best-performing MOLOPT (i.e. DZVP, TZVP or TZV2P(X)) basis sets and WIEN2k reference results~\cite{lejaeghereReproducibilityDensityFunctional2016}. For comparison, the corresponding average for Abinit is $2.1$ meV/atom for the semicore and $6.3$ meV/atom for the regular pseudopotentials, respectively~\cite{deltacodesdft2019}.}%
    \label{fig:deltatest_histogram}
\end{figure}

Even though \cp\ supports different types of Gaussian-type basis sets, the MOLOPT type basis set have been found to perform particularly well for a large number of systems and also targets a wide range of chemical environments, including the gas phase, interfaces, and the condensed phase~\cite{VandeVondele_JCP_2007}. These generally rather contracted basis sets, which include diffuse primitives, are obtained by minimizing a linear combination of the total energy and the condition number of the overlap matrix for a set of molecules with respect to the exponents and contraction coefficients of the full basis. To verify reproducibility of DFT codes in the condensed matter community, the $\Delta$-gauge \eqref{eq:delta-gauge} 
\begin{equation}
    \Delta_i(a,b) = \sqrt{\frac{\displaystyle\int_{0.94V_{0,i}}^{1.06V_{0,i}} \big(E_{b,i}(V) - E_{a,i}(V)\big)^2\; \mathrm{d}V}{0.12V_{0,i}}} \label{eq:delta-gauge}
\end{equation}
based on the Birch-Murnaghan equation of state has been established~\cite{lejaeghereErrorEstimatesSolidState2014}. To gauge our $\Delta$-test values, we have compared them to the ones obtained by plane wave code Abinit~\cite{gonzeRecentDevelopmentsABINIT2016}, where exactly the same dual-space Goedecker-Teter-Hutter (GTH) pseudopotentials as described in section~\ref{subsection_pp} have been employed~\cite{Goedecker1996,Hartwigsen1998,Krack2005}.
As can be seen in Fig.~\ref{fig:deltatest_histogram}, \cp\ generally performs fairly well, with the remaining deviation being attributed to the particular pseudization approach. 

\subsection{CP2K workflows}
\label{subsection_workflows}
With the abundance of computing power, computational scientist are tackling ever more challenging problems.
As a consequence, today's simulations often require complex workflows.
For example, first the initial structure is prepared, then a series of geometry optimizations at increasing levels of theory are performed, before the actual observables can computed, which then finally need to be post-processed and analyzed.
There is a strong desire to automate these workflows, which not only saves time, but also makes them reproducible and shareable.
\cp\ is interfacing with two popular frameworks for automating of such workflows: The Atomic Simulation Environment (ASE)~\cite{ase} and the Automated Interactive Infrastructure and Database for Computational Science (AiiDA)~\cite{aiida}.

The ASE framework is a Python library that is build around the \textit{Atoms} and the \textit{Calculator} classes. An \textit{Atoms} object stores the properties of individual atoms such as their positions and atomic numbers. A \textit{Calculator} object provides a simple unified API to a chemistry code such as \cp.
A calculation is performed by passing an \textit{Atoms} object to a \textit{Calculator}, which then returns energies, nuclear forces, and other observables.
For example, running a \cp\ calculation of a single water molecule requires just a few lines of Python code:
\begin{verbatim}
$ export ASE_CP2K_COMMAND="cp2k_shell.sopt"
$ python
>>> from ase.calculators.cp2k import CP2K
>>> from ase.build import molecule
>>> calc = CP2K()
>>> atoms = molecule('H2O', calculator=calc)
>>> atoms.center(vacuum=2.0)
>>> print(atoms.get_potential_energy())
-467.191035845
\end{verbatim}
Based on these two powerful primitives, the ASE provides a rich library of common building blocks including structure generation, MD, local and global geometry optimizations, transition-state methods, vibration analysis, and many more.
As such, the ASE is an ideal tool for quick prototyping and automatizing a small number of calculations.

The AiiDA framework, however, aims to enable the emerging field of high-throughput computations~\cite{high_throughput}.
Within this approach, databases of candidate materials are automatically screened for the desired target properties.
Since a project typically requires thousands of computations, very robust workflows are needed that can handle also rare failure modes gracefully.
To this end, the AiiDA framework provides a sophisticated event-based workflow engine.
Each workflow step is a functional building block with well defined inputs and outputs.
This design allow the AiiDA engine to trace the data dependencies throughout the workflows and thereby record the provenance graph of every result in its database.

\subsection{GitHub and general tooling}
\label{subsection_tooling}
The \cp\ source code is publicly available and hosted in a Git repository on \url{https://github.com/cp2k/cp2k}. While we still maintain a master repository similar to the previous Subversion repository, the current development process via \emph{pull requests} foster code reviews and discussion, while the integration with a custom \emph{continuous integration} (CI) system based on the Google Cloud Platform ensures the stability of the \emph{master branch} by mandating successful regression tests prior to merging the code change~\cite{Cp2kCp2kci2019}.
The \emph{pull request} based testing is augmented with a large number of additional regression testers running on different supercomputers~\cite{CP2KDashboard2019}, providing over 80\% code coverage across all variants (MPI, OpenMP, CUDA/HIP, FPGA) of \cp~\cite{LCOVCP2KRegtests2019}. These tests are run periodically, rather than triggered live by Git \emph{commits}, to workaround limitations imposed by the different sites and developers are being informed automatically in case of test failures.

Over time additional code analysis tools have been developed to help avoid common pitfalls and maintain consistent code style over the large code base of over 1 millions lines of code. Some of them - like our \emph{fprettify} tool~\cite{Seewald2019} - have now been adopted by other Fortran-based codes~\cite{pizziWannier90CommunityCode2019}.
To simplify the workflow of the developers, all code analysis tools not requiring compilation are now invoked automatically on each Git \emph{commit} if the developer has setup the \emph{pre-commit} hooks for her \emph{clone} of the \cp\ Git repository~\cite{Precommit2019}.
Since \cp\ has a significant number of optional dependencies, a series of \emph{toolchain} scripts have been developed to facilitate the installation, currently providing the reference environment for running the regression tests. Alternatively, \cp\ packages are directly available for the Linux distributions Debian~\cite{DebianPackageCP2K2019}, Fedora~\cite{FedoraPackageCP2K2019}, as well as Arch Linux~\cite{ArchAURCP2K2019}, thanks to the efforts of M. Banck, D. Mierzejewski and A. Kudelin. In addition, direct \cp\ support is provided by the HPC package management tools Spack and EasyBuild~\cite{SpackCP2KPackage2019, EasyBuildCP2K2019}, thanks to M. Culpo and K. Hoste.

\begin{acknowledgments}
TDK has received funding from the European Research Council (ERC) under the European Union's Horizon 2020 research and innovation programme (grant agreement No 716142). J.V. was supported by an ERC starting grant No. 277910. VRR has been supported by the Swiss National Science Foundation in the form of Ambizione grant No. PZ00P2\_174227 and RZK by the Natural Sciences and Engineering Research Council of Canada (NSERC) through Discovery Grants (RGPIN-2016-0505). GKS and CJM are supported by the Department of Energies Basic Energy Sciences, Chemical Sciences, Geosciences, and Biosciences Division. 
UK based work was funded under the embedded CSE programme of the ARCHER UK National Supercomputing Service (http://www.archer.ac.uk), grants eCSE03-011, eCSE06-6, eCSE08-9, eCSE13-17 and the EPSRC (EP/P022235/1) grant ``Surface and Interface Toolkit for the Materials Chemistry Community''. 
The project received funding via the CoE MaX as part of the Horizon 2020 program (grant number No. 824143), the Swiss platform for advanced scientific computing (PASC), and the centre on Computational Design and Discovery of Novel Materials (MARVEL). 
Computational resources were provided by the Swiss National Supercomputing Centre (CSCS) and Compute Canada. The generous allocation of computing time on the FPGA-based supercomputer ``Noctua'' at PC$^2$ is kindly acknowledged.\end{acknowledgments}

\section*{Data Availability Statement}
The data that support the findings of this study are available from the corresponding author upon reasonable request.

%

\end{document}